\pdfoutput=1
\documentclass[10pt,english]{elsart3p}

\usepackage{esaf}
\usepackage{amsmath}
\usepackage{amssymb}
\usepackage[pdftex]{graphicx}  
\usepackage[english]{babel}                                         
\usepackage[T1]{fontenc}
\usepackage[dvipsnames]{xcolor}
\usepackage{xspace}
\usepackage{color}

\begin{document}

\begin{frontmatter}

\title{Full Simulation of Space-Based Extensive Air Showers Detectors with ESAF}
\maketitle

\author[Grenoble]{C. Berat},
\author[FirenzeINFN]{S. Bottai},
\author[Genova,GenovaINFN]{D. De Marco\thanksref{DanielAt}},
\author[Grenoble]{S. Moreggia},
\author[GenovaINFN,FirenzeINFN,Dubna]{D. Naumov},
\author[Genova,GenovaINFN]{M. Pallavicini},
\author[Genova,GenovaINFN]{R. Pesce\corauthref{cor1}},
\author[Genova,GenovaINFN]{A. Petrolini},
\author[Grenoble]{A. Stutz},
\author[Firenze,FirenzeINFN]{E. Taddei},
\author[Genova,GenovaINFN]{A. Thea\corauthref{cor2}\thanksref{AleAt}}

\corauth[cor1]{Corresponding author. E-mail: Roberto.Pesce@ge.infn.it}
\corauth[cor2]{Corresponding author. E-mail: Alessandro.Thea@cern.ch}

\address[Genova]{Dipartimento di Fisica, Universit\`a di Genova,
via Dodecaneso 33, I-16146 Genova, Italy}
\address[GenovaINFN]{Istituto Nazionale di Fisica Nucleare, Sezione di Genova,
via Dodecaneso 33, I-16146 Genova, Italy}

\address[Firenze]{Dipartimento di Fisica, Universit\`a di Firenze and INFN Firenze,
via Sansone 1, I-50019 Sesto Fiorentino (FI), Italy}
\address[FirenzeINFN]{Istituto Nazionale di Fisica Nucleare, Sezione di Firenze,
via Sansone 1, I-50019 Sesto Fiorentino (FI), Italy}

\address[Dubna]{JINR, Joliot-Curie, 6, 141980 Dubna, Moscow Region, Russia}
\address[Grenoble]{LPSC, UJF Grenoble 1, CNRS/IN2P3, INPG, 53 Rue des Martyrs, 38026 Grenoble Cedex, France}

\thanks[AleAt]{Now at ETH, Z\"urich}
\thanks[DanielAt]{Now at Bartol Research Institute, University of Delaware, Newark - DE 19716, USA}

\begin{keyword}
cosmic rays; space detectors; extensive air showers; atmosphere models; 
\\PACS 95.55Vj 95.55Fw
\end{keyword}

\begin{abstract}

Future detection of Extensive Air Showers (EAS) produced by Ultra High Energy Cosmic Particles by
means of space based fluorescence telescopes will open a new window on the universe and allow cosmic ray and
neutrino astronomy at a level that is virtually impossible for ground based detectors. In the context of the Extreme Universe Space Observatory (EUSO) project, an end-to-end simulation of EAS observation with a spatial detector has been designed (EUSO Simulation and Analysis Framework, ESAF). This paper describes the detailed Monte-Carlo developed to simulate all the physical processes involved in the fluorescence detection technique, from the EAS development to the instrument response. Particular emphasis is given to modeling the light propagation in the atmosphere and the effect of clouds. The simulation is used to assess the performances of EAS spatial detection.
Main results on energy threshold and resolution, direction resolution and \Xmax determination are reported. Results
are based on EUSO telescope design, but are also extended to larger and more sensitive detectors.

\end{abstract}

\end{frontmatter}

\section{Introduction}\label{sec:intro}

Ultra-High Energy Cosmic Particles (UHECP) with energies in excess of 
$\simeq$ $10^{19}$  eV hit the Earth with a very low flux which falls down to less of one particle  km$^{-2}$ sr$^{-1}$ century$^{-1}$  for particles with energy E $>$ 10$^{20}$ eV~\cite{bib:auger1}. 

The observation of UHECP and the interpretation of the related phenomenology is one of the most challenging 
topics of modern High-Energy Astro-Particle Physics. Direct detection is impossible at these energies, due to the 
exceedingly low flux, but UHECP can be detected by observing the Extensive Air Showers (EAS) produced by 
the interaction of the primary particle with the Earth atmosphere (see~\cite{bib:eas1,bib:eas2,bib:eas3}, and references therein, for recent reviews on these topics). 

The ground-based Pierre Auger Observatory (PAO)~\cite{bib:auger2,bib:auger4,bib:auger5,bib:auger6} is currently taking data: its south site
(surface of 3000 km$^2$) in Argentina and its forthcoming north site in the US (20000 km$^2$)
will provide, in the next few years, a clear understanding 
of many important topics~\cite{bib:auger1,bib:auger3,bib:auger7,bib:auger8,bib:auger9}. However, a next generation of experiments with sensitivity to UHECP flux
well above the PAO one at the highest energies is essential but it seems to be
difficult to realize by extending the ground-based technique at ever larger arrays. 

Ground-based experiments detect EAS either by sampling charged particles at ground by means of suitable counters and/or by detecting the fluorescence light emitted by the atmospheric
nitrogen when excited by the charged particles in the shower. While the former technique is intrinsically
ground based, the latter is not, being the fluorescence light emission isotropic. However, fluorescence light
detection is limited to moonless nights and is affected by weather conditions. 

In 1979 J. Linsley~\cite{bib:linsley} proposed for the first time the idea of detecting EAS fluorescence light with a 
telescope in orbit around the Earth. While this approach is technically, programmatically and economically
challenging, it is one of the most promising that might allow a big step forward in this field, opening a new window
on the universe.  A space based experiment might observe EAS on an effective surface of a few times 10$^5$ km$^2$,
increasing the event statistics by more than an order of magnitude compared to existing and running ground based experiments, even 
when taking into account that the duty cycle for observing EAS with fluorescence is limited to clear 
and moonless nights. Moreover, the huge target mass of the observed air volume \mbox{(> 10$^{12}$ kg),}
three orders of magnitude more than the current or foreseen projects in the South Pole ice \cite{bib:icecube}
or mediterranean sea \cite{bib:km3net} ) 
may open the era of high energy neutrino astronomy, at least above energies of the order of 
10$^{19}$ eV or so. However, we do not discuss high energy neutrino detection in this paper.

The relevance of this innovative approach is widely recognized by the scientific community. The Extreme 
Universe Space Observatory (EUSO) project~\cite{bib:Euso}, originally proposed in 2000 as response to F2/F3 European 
Space Agency (ESA) call for mission successfully completed the
phase A conceptual study. Nevertheless it did not continue only for programmatic reasons related to International Space 
Station and Space Shuttle availability. The inclusion of UHECP physics in the ESA Cosmic Vision 2015-2025 
program~\cite{bib:CV} provides a suitable framework for the study of future space missions.
An introduction to the basic issues concerning the detection of UHECP with a space detector can be found in~\cite{bib:space}.

In this paper we therefore report the results obtained in the context of the EUSO collaboration by means
of a full simulation and reconstruction software code named ESAF, standing for \emph{EUSO Simulation and Analysis
Framework}.

Particularly,
we report about results concerning energy threshold as function of UHECP direction and detector size and geometry, direction and energy resolution, the dependence on atmospheric conditions and clouds.

This paper is structured as follows: section~\ref{sec:esaf} presents the ESAF design; section~\ref{sec:simu} describes the simulation part, giving details on each modules, from the EAS development to the detector response; section~\ref{sec:reco} contains a description of the reconstruction tasks; 
section \ref{sec:results} presents several results obtained with the ESAF code; the most relevant features are summarized in the conclusions.

\section{The ESAF design}\label{sec:esaf}

ESAF handles the whole simulation and reconstruction chain from primary particle interaction
in the Earth atmosphere until the final reconstruction of the event. The code includes:

\begin{itemize}
\item[-] extensive air shower simulation both by means of internally developed 
algorithms (e.g. Greizen-Ilina-Linsley (GIL) parametrization,~\cite{bib:gil}) and with interfaces to existing widely used codes 
(e.g. CORSIKA,~\cite{bib:corsika});
\item[-] a complete description of the atmosphere, including aerosols\footnote{Aerosols are airborne liquid droplets or solid particles, which may exist at low altitude.} and clouds
\item[-] fluorescence and Cherenkov light production
\item[-] a complete simulation of photon propagation, from production point up to the telescope, including detailed interactions with ground and atmosphere, and a Monte-Carlo code dealing with multiple scattering
\item[-] telescope optics simulation
\item[-] telescope geometry
\item[-] telescope photodetector simulation
\item[-] electronics and trigger simulation
\item[-] background simulation
\item[-] pattern recognition and shower signal identification above background
\item[-] reconstruction of direction, energy, and the so called slant depth of the shower maximum (\Xmax),
i.e. the distance (expressed in \gcmsq\ of traversed material) of the shower maximum from the top of the atmosphere.
\end{itemize}

Although the simulation was done having in mind the specific EUSO design, the flexibility of the ESAF code 
allowed us to obtain results that are quite general and might be useful for any future project. 

\begin{figure}[htb]
\includegraphics[width=0.48\textwidth]{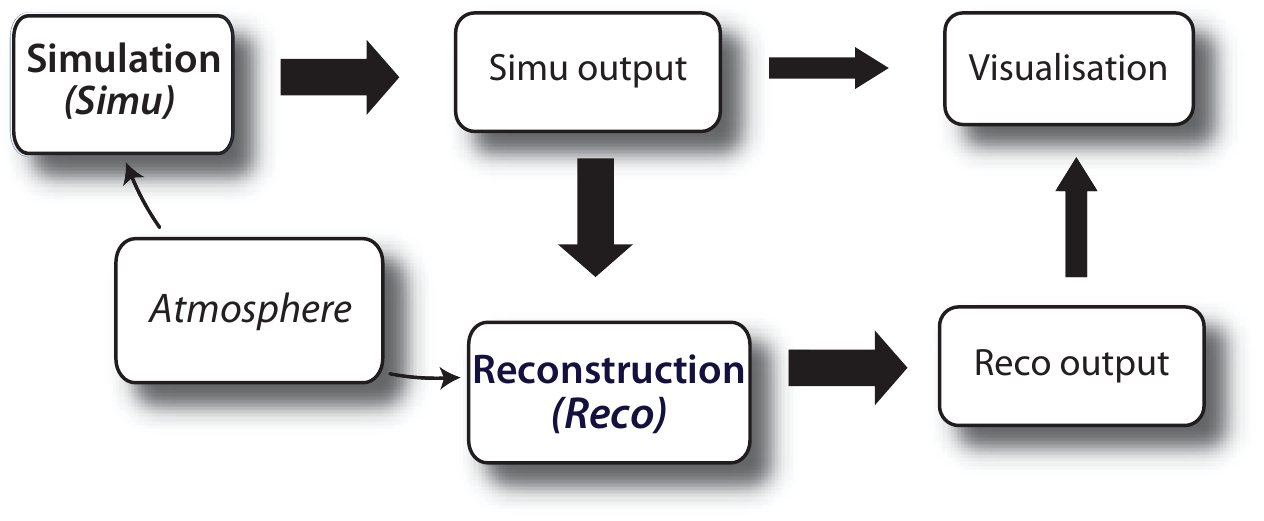}
\caption{The upper level structure of ESAF. The simulation and reconstruction modules are detailed in Fig.~\ref{fig:esaf2}
and Fig.~\ref{fig:esaf3} respectively. }
\label{fig:esaf1}
\end{figure}

\begin{figure}[htb]
\includegraphics[width=0.48\textwidth]{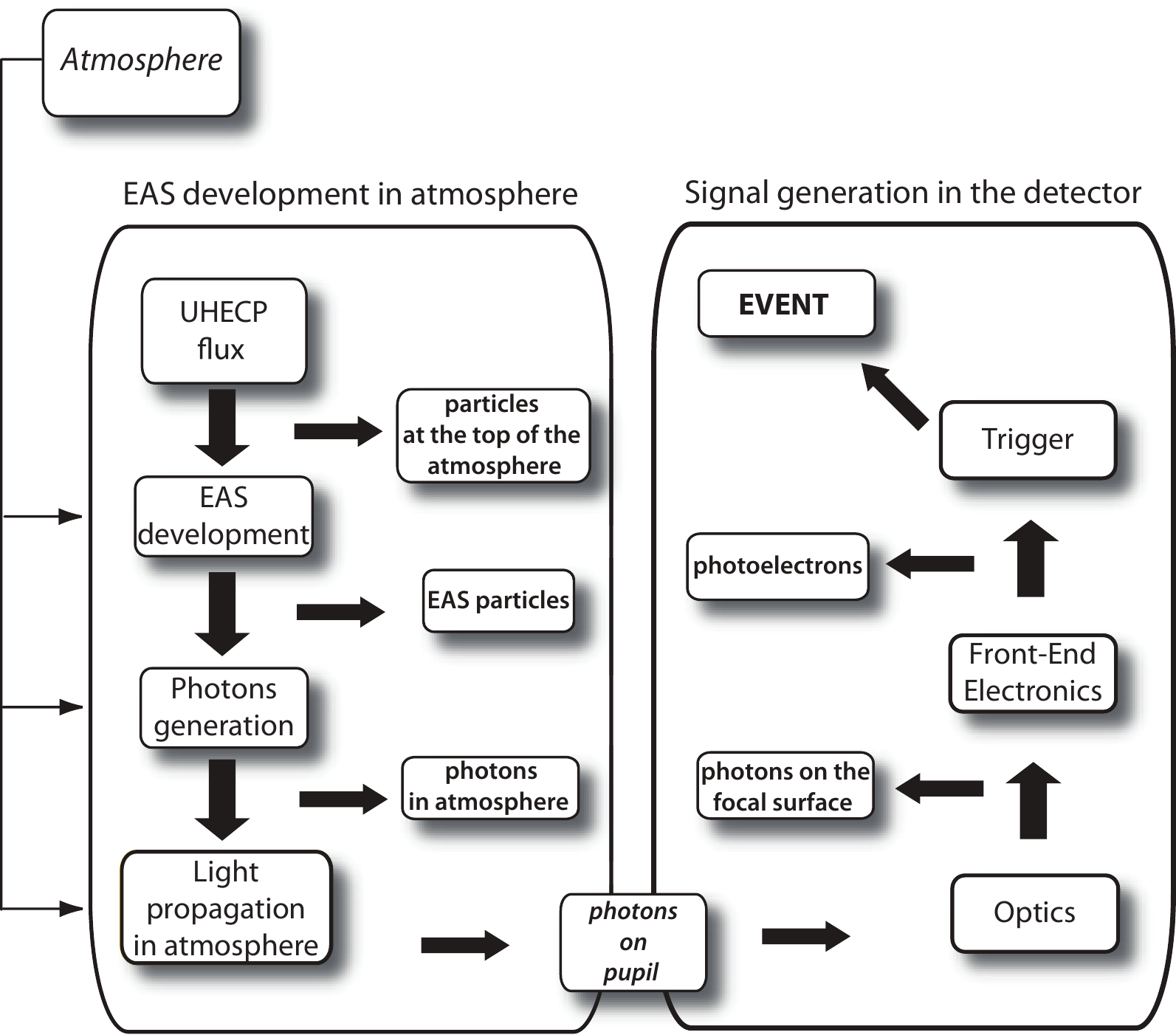}
\caption{ESAF simulation structure.}
\label{fig:esaf2}
\end{figure}

\begin{figure}[htb]
\includegraphics[width=0.48\textwidth]{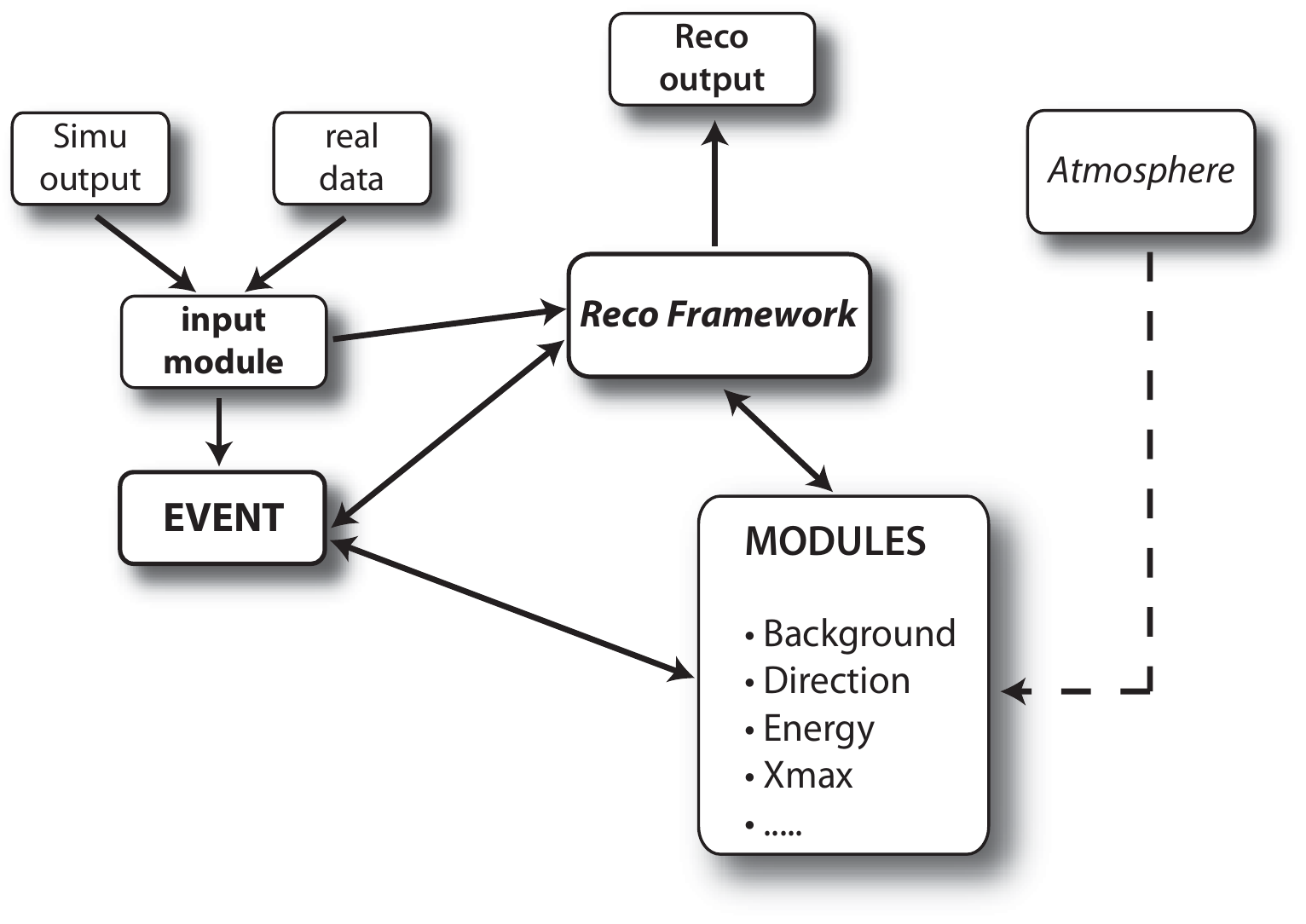}
\caption{ESAF reconstruction structure.}
\label{fig:esaf3}
\end{figure}

\subsection{General structure of the ESAF code}

The ESAF code is written mostly in C++ with a few external Fortran libraries. A complete object-oriented
design is used to allow flexibility, modularity and maintenability of the code. The code is divided in two
independent sections, simulation and reconstruction, that share a common infrastructure and a set of
common libraries. The code is not public, but is available upon request to interested 
users\footnote{http://www.ge.infn.it/$\sim$pesce/esaf.html}. 
The main structure of the program and the interactions between the two main sections are shown in Fig. \ref{fig:esaf1}. 

The simulation is structured in 6 main sections (see Fig.~\ref{fig:esaf2}): shower simulation starting from the expected UHECP flux, fluorescence and Cherenkov photons
production in atmosphere,  light propagation from the production point up to the orbiting telescope with simulation of occurring photon interactions during the propagation, simulation of telescope optics (lenses and focal surfaces),
simulation of front end electronics and trigger. In this paper we do not cover
implementation details that can be found in~\cite{bib:tesi-thea,bib:tesi-pesce,bib:tesi-moreggia}. 

The output of the simulation module is a ROOT file~\cite{bib:root} with a structure that contains 
the same information that is expected from real data plus a set of "Monte Carlo truth" data.
The level of detail is user-configurable, to optimize the output size to the goal of the simulation 
(i.e. physics, optics, trigger, etc.). To provide the user with an easy access to the data, a set of detector 
event viewers is provided with ESAF. They allow a coherent display of the run informations, histograms and 
several 2D and 3D views of the event development in the detector and on the focal surface (some example 
plots are in section~\ref{sec:dete}).

The reconstruction module reads back
the ROOT file representing the fully simulated shower with background and reconstructs the events as if they were
real data events from the telescope. Several modules allow pattern recognition and identification of 
EAS signal above incoherent star light and moonlight background, 3D direction reconstruction, energy
and \Xmax reconstruction. The reconstruction structure is shown in Fig.~\ref{fig:esaf3}.

The reconstruction framework is the main structure that acquires the data (real or simulated) from an input module and builds the chain of modules necessary to reconstruct the events. The event reconstruction is divided into different tasks (e.g. background subtraction, direction reconstruction, $\ldots$) and for each task different modules may be available. The use of such a structure allows to easily replace or exclude a given module. In this way different algorithms or algorithm combinations can be tested and compared on the same events. As in the simulation part, the reconstruction modules are configurable by the user. The event container stores all the relevant informations about the reconstructed event. Every module can access the event container, reads the existing data stored in it and writes its own results.

\subsection{Atmosphere description}\label{subsec:atmodel}

In the simulation process, a description of the atmosphere is required for proper simulation of 
shower development, light production and  photons transfer toward the telescope entrance pupil
(see Fig. \ref{fig:esaf2}). The atmosphere parameters are also required during the reconstruction process.

In ESAF, the atmosphere and Earth surface
are spherically shaped, and several models have been introduced, to form database of atmospheric profiles: 
the US-Standard model and its complements~\cite{bib:USS76} and the {\tt NRLMSIS-00 (MSISE)} empirical model~\cite{bib:MSISE}. 
The formers provide temperature, pressure, refractive index and density profiles,
up to 100~km, as well as  ozone and water vapour profiles. Available data are mean values
over time of atmospheric data at different latitudes.
With {\tt MSISE} model, atmospheric profiles are provided
for given latitude, longitude, day of the year and hour.

Other crucial data concerning  atmosphere description have been implemented in ESAF.
The presence of clouds and aerosols has an impact
on the UHECP detection from space. The cloud and aerosol effects have to be evaluated with the simulation. The cloud mean fraction in the field of view of an orbiting telescope is around 70\%. 
To describe cloud main characteristics, informations collected with the satellite  TOVS 
(TIROS Operational Vertical Sonder) Path-B~\cite{bib:TOVS} are used. 
To describe aerosols ESAF is also supplied with several atmospheric models and parameters which exist in the
{\tt LOWTRAN7}\footnote{{\tt LOWTRAN7} is a standard program used in the atmosphere 
community, to compute 
transmission in atmosphere.} software~\cite{bib:LOWTRAN}.

A set of dedicated configuration files allow to choose one particular atmosphere model and status among all
the possibilities available in ESAF, before runs. The atmospheric parameters are then used at each step
where they are required during the simulation or the reconstruction process.

\section{The simulation code}\label{sec:simu}

\subsection{Shower simulation}\label{sec:showers}
An accurate and physically reliable simulation of EAS is fundamental for the correct 
understanding of the UHECP phenomenology and the possible experimental outcomes. 
During the last 20 years several authors have dealt with the very demanding task of 
preparing detailed UHECP simulation programs. Particularly, CORSIKA~\cite{bib:corsika} and 
AIRES~\cite{bib:Aires} , are at present widely used and represent the 'state of the art' in 
atmospheric shower simulation.  Another shower simulator, UNISIM~\cite{bib:Unisim}, was
developed by some ESAF developers, and has the unique feature of simulating neutrino interactions
in atmosphere.

The basic ESAF approach, therefore, does not uniquely consist in the development of brand 
new shower simulators, but also in having  the capability to handle external shower 
simulation programs. 
This attitude is realized by implementing in ESAF a generic interface by means of a 'Shower Track' object 
which contains the structure for an extremely detailed description of a single shower event, 
a description that is even more accurate than any of the output of the present existing 
simulation programs. 
Using some specific modules, the user can easily write an interface which plugs into the 
'Shower Track' the shower details event by event.

The 'Shower Track' can manage all typical variables (number of charged particles, number of 
electrons, energy loss..) representing the shower during its longitudinal development.
The 'Shower Track' has a block of general information and a list of 'Shower Step' objects along the EAS longitudinal development.

Each 'Shower Step' object contains an information about the shower longitudinal development corresponding to the track length $\diffl{L}$, which may also vary from event to event. For each step there is 
also the possibility to handle histograms representing the local distributions of particles in 
the shower according to their energy, distance from the axis,
direction in space, arrival time, and taking care of preserving some of the most important 
correlation between couples of such physical quantities. 

At present it exists inside ESAF an interface to UNISIM,  a preliminary version of interface to CORSIKA, and an interface to CONEX~\cite{Bergmann:2006yz,Pierog:2004re}. 

In addition to external program interfaces, ESAF can also fill event by event the longitudinal 
description inside the 'shower object' using several parametrizations (Greizen~\cite{bib:Gaisser}, Greizen-Ilina-Linsley (GIL)~\cite{bib:gil}, Gaisser-Hillas~\cite{bib:GaHill}) tuned on data and Monte - Carlo outcomes.
All these parametrizations need the depth of the first interaction point ($X_1$). 
In ESAF it can be fixed or sampled according to Nuclei-Air cross section data~\cite{bib:mielke} or randomly.
Such fast algorithms are very useful during the phase of detector optimization where the 
exact representation of the shower to shower fluctuations is less important.

When some important information is not present in a given external simulation
program, ESAF can decide either to skip or to replace using some of the predefined parametrizations 
implemented in the code. Several different parametrizations are available in ESAF for energy distribution 
of particles inside the shower~\cite{bib:Nerlig,bib:Giller,bib:Hillas}, particles angular
distribution~\cite{bib:Baltru}, and particles lateral distribution~\cite{bib:Gaisser,bib:Dora} as a function of 
the shower age.

Most of the results presented in this paper are achieved using the GIL parametrization for the longitudinal
development of the number of charged particles $N(s)$, as a function of the shower age $s$:

\begin{gather}\label{eq:GIL}
N(s)=\frac{E}{E_1}\exp\left[\xi-\xi_\mathrm{max} - 2\xi\ln s \right] \virgola \\
s = 2\xi/(\xi+\xi_\mathrm{max})\virgola \nonumber \\ 
\xi = (X-X_1)/X_0 \virgola \nonumber \\
\xi_\mathrm{max} = a + b\left[\ln\left(E/E_\mathrm{c}-\ln A\right)\right] \virgola \nonumber
\end{gather}

where $E$ is the energy of the primary particle,
$A$ the atomic number of the primary, $X$ the depth considered,
$X_1$ is the depth of the first interaction, 
$X_0=37.15\,\text{g}\,\text{cm}^{-2}$ (air radiation length), \mbox{$E_\mathrm{c} =81$ MeV} (critical energy), 
$a=1.7$, $b=0.76$ and \mbox{$E_1 =1.45$ GeV}. These values are 
chosen to match CORSIKA-QGSJET results~\cite{bib:qgcjet}.

\subsection{Cherenkov and fluorescence light simulation}\label{sec:light}

Relativistic charged particles (mainly electrons and positrons) in extended air showers  are at the origin of  both fluorescence and Cherenkov light emissions.
Generation of fluorescence and Cherenkov photons has been included in ESAF.  
The shower simulation module provides for each 'Shower Step'
the number of electrons  $N_e(s)$, the electron energy distribution and the lateral distributions. 
The light generation module uses them to compute {\it bunches} of photons, as explained in section~\ref{sec:bunches}.
This approach is dictated by the huge number of photons produced
which makes impossible to follow the fate of each individual photon.

\subsubsection{Fluorescence light simulation}
\label{sec:fluo-light-simu}

The number of photons $\diffl{N}_\mathrm{fluo}$  
created along $\diffl{L}$ by the $N_e(s)$ EAS electrons is:
\begin{eqnarray}\label{eq:Nfluo-a}
\deriv{N_\mathrm{fluo}}{L}   =  &\\ \nonumber
N_e(s)\int\nolimits_{E} &{ \left[  \left(  \frac{1}{N_e} {\deriv{N_e}{E}} (s) \right)   
\int\nolimits_{\lambda}FY_\lambda^L(E,P,T) \diffl{\lambda} \right]  \diffl{E} } \virgola \cr
\end{eqnarray}
where $N_e^{-1} {\deriv{N_e}{E}} (s) $ is the normalized electron energy spectrum, $ FY^L_\lambda(E, P,T)  $ is  the fluorescence yield (in photons/m) at wavelength $\lambda$ for an electron of energy $E$. $P$ and $T$ are the local pressure and temperature considered at the middle of the $\diffl{L}$ track segment.

The fluorescence yield dependence on the local temperature and density conditions is considered according to the empirical relation :
\begin{equation*}
FY^L_\lambda =  \rho \   \frac{A_\lambda}{{1 + B_\lambda \rho \sqrt{T}}} \virgola
\end{equation*}
where $\rho$ is the atmosphere density, in kg$\cdot$m$^{-3}$, and $T$ is in K. The experimental values of the two coefficients $A_\lambda$ and $B_\lambda$ are provided with yield measurements from Nagano measurements~\cite{bib:nagano} or Kakimoto ones~\cite{bib:kakimoto}.
In the Kakimoto case, the fluorescence spectrum is completed with data from Bunner PhD thesis~\cite{bib:bunner}.
In ESAF, the choice to carry out simulations
with one of these two  fluorescence yield measurement data sets is provided by the dedicated configuration file.

The yield is assumed to be proportional to the energy loss $\deriv{E}{X}$ ($X$ is the slant depth, in g$\cdot$cm$^{-2}$) in the atmosphere, thus $FY_\lambda^L$  is given by:

\begin{equation}\label{eq:fluoY}
FY_\lambda^L (E, P,T)  = FY_\lambda^L (E_\mathrm{ref}, P,T)   {\frac{  \deriv{E}{X} (E) }{{\deriv{E}{X} } (E_\mathrm{ref}) }}.
\end{equation}

\begin{figure}[ht]
\begin{center}
\includegraphics[width=0.49\textwidth]{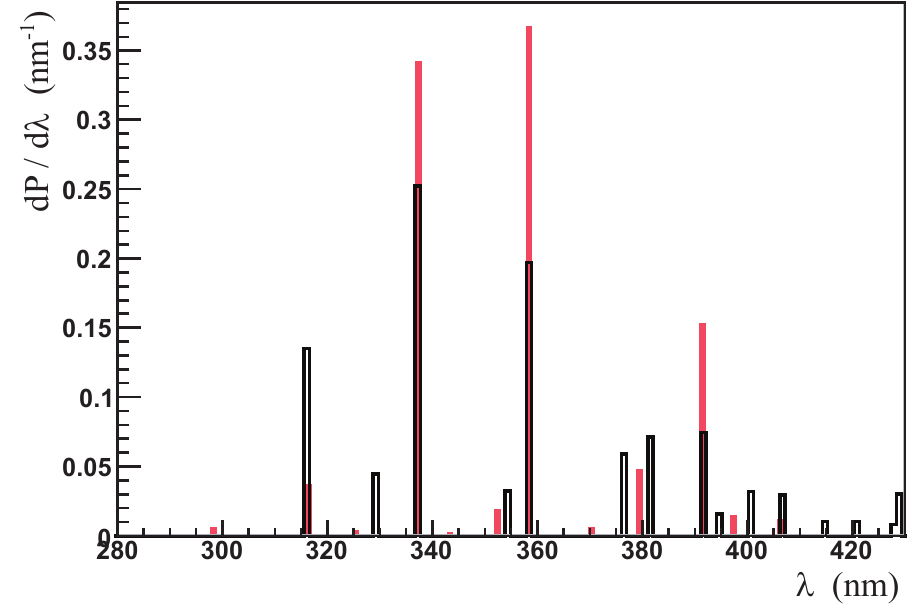}
\caption{Fluorescence emission spectra at \mbox{$P =1013$~hPa} and \mbox{$T =15\degr$C} computed using Nagano measurements~\cite{bib:nagano} (black line histogram) and Kakimoto ones
\cite{bib:kakimoto} completed with Bunner rays~\cite{bib:bunner} (red filled histogram). The main fluorescence rays are between 330 and 400~nm, and the three most  intense rays are at 337, 357 and 391~nm.}
\label{fig:fl-spectre}
\end{center}
\end{figure}

In ESAF, the Berger-Seltzer~\cite{bib:berger} formula is used to compute the energy loss   $ {\deriv{E}{X}} (E) $ 
directly from $E$.\\
Fluorescence yield $ FY^L_\lambda (E_\mathrm{ref}, P,T) $ 
at a fixed energy $E_\mathrm{ref}$ is determined at \mbox{$T = 288$~K} and \mbox{$P= 1013$~hPa} either from~\cite{bib:nagano} or from~\cite{bib:kakimoto,bib:bunner}, according to the chosen configuration, with respectively \mbox{$E_\mathrm{ref}=0.85$~MeV} and \mbox{$E_\mathrm{ref}=1.4$~MeV}. Examples of computed fluorescence yield spectra are presented on Fig.~\ref{fig:fl-spectre}.
In ESAF the normalized electron energy
spectrum is  provided by the shower module (see section \ref{sec:showers}). Using equation (\ref{eq:fluoY}), labeling $FY^L$ the integrated fluorescence yield over the wavelength spectrum,  
and computing the averaged electron energy loss,
\begin{equation*}
\left< {\deriv{E}{X}} \right> _s  = \int\nolimits_{E} {   \left(  \frac{1}{N_e} {\deriv{N_e}{E}} (s) \right)    {\deriv{E}{X}} (E) \diffl{E}} \virgola
\end{equation*}
we can derive from~\eqref{eq:Nfluo-a} the  following formula implemented in ESAF:

\begin{equation} \label{eq:Nfluo}
\deriv{N_\mathrm{fluo}}{L}   =  N_e(s) \  FY^L (E_\mathrm{ref}, P,T) 
\frac{\left< {\deriv{E}{X} } \right> _s}{{\deriv{E}{X}}  (E_\mathrm{ref})}
\end{equation}

\begin{figure}[ht]
\begin{center}
\includegraphics[width=0.48\textwidth]{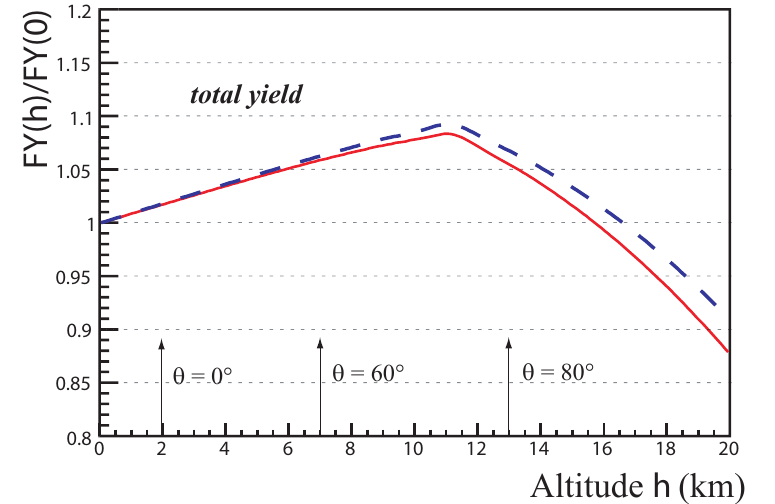}
\caption{{Fluorescence yield $FY^L(h)/FY^L(0)$ integrated over the emission spectrum as a function of the altitude $h$
(dashed blue line: from~\cite{bib:nagano} 
\mbox{$FY^L(0)=3.81$ photons/m}, continuous red line: from~\cite{bib:kakimoto}, \mbox{$FY^L(0)=3.26$ photons/m)}. $P$ and $T$ profiles are obtained from the US Standard atmosphere model (see section~\ref{subsec:atmodel}). The mean altitude of the maximum of shower development is
indicated for 3 shower zenith angles. The weak dependence of the yield with  altitude is observed.}}
\label{fig:fl-yield}
\end{center}
\end{figure}

The resulting total fluorescence yield (Fig.~\ref{fig:fl-yield}) varies weakly along the EAS development (the mean
yield is between 4.1 and 4.7 photons/m). The photons longitudinal distribution along the shower axis is therefore similar to the electrons longitudinal profile. The total number of fluorescence photons isotropically emitted along the shower track is in the order of 10$^{15}$ for a primary energy of 10$^{20}$~eV.  This number depends on the shower  zenith angle, since inclined EAS starting at higher altitude
have a longer path in the atmosphere and thus produce more photons  compared to vertical EAS.

\subsubsection{Cherenkov light simulation}
The algorithm to simulate  Cherenkov photon generation is similar to the fluorescence one.
The number of Cherenkov photons $\diffl{N_\mathrm{ckov}}$ 
originated by $N_e(s)$  electrons along a shower track segment $\diffl{L}$ is:
\begin{eqnarray}\label{eq:Nckov-a}
{\deriv{N_\mathrm{ckov}}{L}}   =  &\\  \nonumber
N_e(s)\int\nolimits_{E_\mathrm{thres}}^{\infty} &{ \left[  \left(  \frac{1}{N_e} {\deriv{N_e}{E}} (s) \right)   CY^L(E,n) \right]  \diffl{E} } \virgola \cr
\end{eqnarray}
where $E_\mathrm{thres}$ is the Cherenkov production energy threshold. 
The term $N_e^{-1} {\deriv{N_e}{E}}(s)  $ describes the normalized energy spectrum of the electrons 
at the emission point. $CY^L (E, n)$ is the total Cherenkov yield integrated over $\left[\lambda_\mathrm{min}, \lambda_\mathrm{max}\right]$. 
The atmosphere refractive index $n$ verifies \mbox{$ n-1  \ll 1$}; it varies very weakly
with $\lambda$ in the UV range of interest, thus only $n$ variation with
atmosphere density is taken into account. Therefore Cherenkov yield can be computed in ESAF using the formula:
\begin{eqnarray}\label{eq:ckovY}
CY^L (E, n)=& \nonumber \\
4 \pi \alpha \ (n-1)  & \left( \frac{1}{\lambda_\mathrm{min}} - \frac{1}{\lambda_\mathrm{max}} \right)   \left(  1 - \frac{E_\mathrm{thres}^2}{E^2}  \right) \;\, ,
\end{eqnarray}
where $\alpha$ is the fine structure constant, and \mbox{$E_\mathrm{thres} \approx \frac{mc^2}{\sqrt{2(n-1)}}$}. Typically $\lambda_\mathrm{min}=300$~nm and $\lambda_\mathrm{max}=450$~nm; these values can be configured.  It has to be noticed that Cherenkov yield depends on
local atmosphere parameters through  the $n-1$ factor present in~\eqref{eq:ckovY} and in the energy threshold expression.
The normalized energy spectrum of the electrons at the
emission point is provided by the shower generation module.

Then, an effective yield  $\left< CY^L \right> _s^\mathrm{eff}(n)$ corresponding to the integral of $CY^L (E, n)$ over the electron energy spectrum
is computed at each step of length $\diffl{L}$ and used to compute the number of Cherenkov photons produced along the step $\diffl{L}$:
\begin{equation}\label{eq:cerenY}
{\deriv{N_\mathrm{ckov}}{L}} = N_e(s) \  \left< CY^L \right> _s^\mathrm{eff}(n) \punto
\end{equation}

\begin{figure}[ht]
\begin{center}
\includegraphics[width=0.5\textwidth]{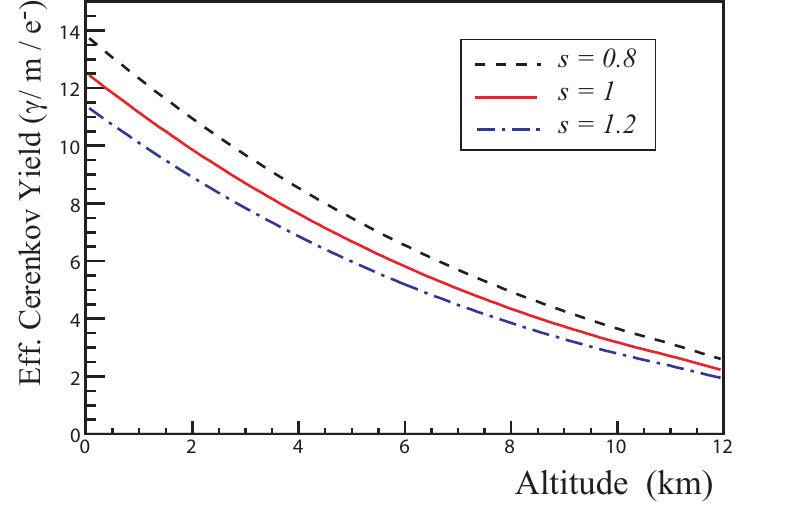}
\caption{{Cherenkov effective yield  $\left< CY^L \right> _s^\mathrm{eff} $  as a function of the altitude, for 3 values of the shower age $s$.}}
\label{fig:ck-yield}
\end{center}
\end{figure}

\begin{figure}[ht]
\begin{center}
\includegraphics[width=0.5\textwidth]{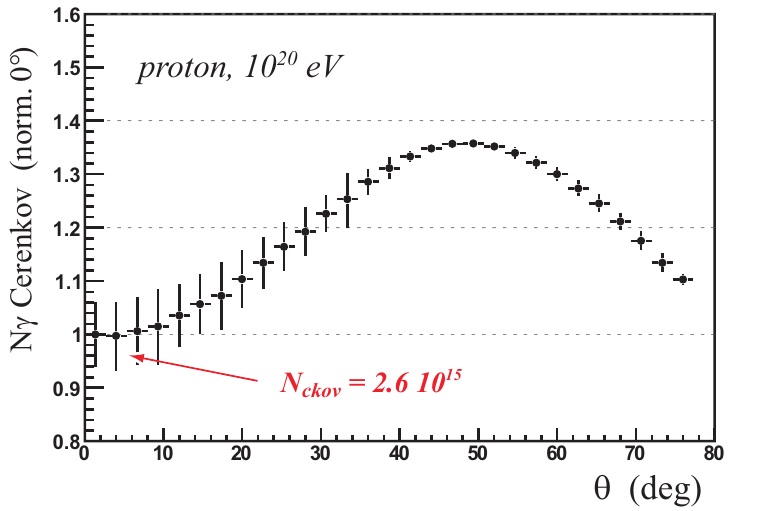}
\caption{Number of Cherenkov photons $N_\mathrm{ckov}(\theta)/ N_\mathrm{ckov}(0\degr)$ produced by an EAS 
(proton, $E=10^{20}$~eV) as a function of the shower zenith angle $\theta$.}
\label{fig:cerenk-theta}
\end{center}
\end{figure}

Contrary to  fluorescence, Cherenkov yield decreases rapidly with altitude~(Fig.~\ref{fig:ck-yield}).
Nevertheless, the number of Cherenkov photons produced all along the shower track
($\sim 3 \cdot 10^{15}$ at 10$^{20}$~eV) is of the same order as the fluorescence one. It varies also with shower zenith
angle (Fig.~\ref{fig:cerenk-theta}). Cherenkov intensity reaches a maximum for shower zenith angle around $50\degr$.
In case of larger zenith angle, development of showers occurs at higher altitudes, where the Cherenkov yield is weaker.
Quasi-vertical showers may reach Earth ground before their development ends, and  Cherenkov photon production is not 
complete.\\

The Cherenkov emission angle $\theta_\mathrm{c}$, as the yield and the energy threshold, is related to the atmospheric refractive index $n$
and therefore correlated with altitude. In the atmosphere and for shower relativistic electrons,  
$\theta_\mathrm{c} \lesssim 1.3\degr$.
In order to obtain the angular distribution of the emitted Cherenkov photons, one has to convolute shower  electron angular distribution 
with $\theta_\mathrm{c}$. In ESAF, the formula established by Elbert et al.~\cite{bib:elbert} is used:
\begin{equation*}
{\deriv{N_\gamma}{\theta_\gamma}} = \frac{C}{\theta_0} \exp\left(-\frac{\theta_\gamma}{\theta_0}\right)
\qquad \text{with} \quad \theta_0 = 0.85\ E_\mathrm{thres}^{-0.66} \virgola
\end{equation*}
where $C$ is a normalization constant such has 
$\int\nolimits_{0^\circ}^{180^\circ}{\deriv{N_\gamma}{\theta_\gamma}}  \diffl\theta_\gamma =1$. 
Photons emitted at high altitude are beamed close to the shower axis: at 12~km, 
$< \theta_\gamma > \simeq 4\degr $, at 4~km, $< \theta_\gamma > \simeq 6\degr $. 

\subsubsection{Bunches of photons}\label{sec:bunches}

The number of photons produced by a typical shower is huge,
of the order of 10$^{15}$. This fact forbids to follow a direct approach, and simulate the fate of each individual
photon. For this reason, the light simulation is done by introducing the concept of {\it bunch}, i.e. an object
dedicated to EAS photon description.

Bunches of photons are created along the EAS axis: the shower longitudinal distribution is 
split in steps of length $\diffl{L}$, set by the user in the steering configuration file; typical values are 
between 1 and 10 g$\cdot$cm$^{-2}$. At each step, two bunches are formed, one for the fluorescence 
emission, the other for the Cherenkov one, according to equations \eqref{eq:Nfluo} and \eqref{eq:cerenY} respectively.
Each bunch is characterized by a mean position value,  a mean direction according to the computed angular distribution, 
and a creation time.  Specific distributions  are associated to each bunch of photons, according to the emission type: 
wavelength spectrum, longitudinal distribution inside the step, lateral and angular distributions (w.r.t the shower axis). 
So, when a photon has to be randomly sampled from a bunch to be propagated, its characteristics can be retrieved
from each distribution associated to the bunch. 

\subsection{Photon propagation from EAS to the space-based detector}\label{sec:atmo}

As mentioned above, more than  10$^{15}$ UV photons are produced by a 10$^{20}$~eV EAS. 
Before reaching the telescope, 
these photons may interact with atmosphere constituents or Earth ground: 
the physics processes to be considered are absorption and scattering. The tracking of such an amount of photons 
through the atmosphere is obviously impossible with ``full'' Monte-Carlo algorithms. 
In ESAF, two different approaches have been designed, each of them
being adapted to  specific simulation needs. 
In the following, the different types of  interactions which can
interfere with photon transfer towards the space-based detector are outlined, as well as
their implementation in ESAF.  Then the two propagation algorithms are described.

\subsubsection{Scattering and absorption processes}\label{subsubsec:scat-abs}

Photons emitted in the direction of the telescope, and that are neither absorbed 
nor scattered during their path  arrive at the entrance pupil. To evaluate this fraction 
of light  straightly transmitted, the transmission $T = e^{-\delta}$ is evaluated from the interaction cross section (or from associated quantities such as the optical depth and the interaction length) related to  the different extinction processes, and taking
into account each photon characteristics (wavelength, position, ...).
EAS photons that are not emitted in the direction of the detector may be scattered later 
in this direction. To quantify this effect, one needs to know as well the angular distributions of the scattering processes,
represented by their phase function. To simulate the radiative transfer, both optical depths and phase 
functions should be computed in ESAF.

When the scattering particle size is much smaller than the photon wavelength, 
Rayleigh theory is used, as for scattering on air molecules.
When the scattering particle size is of the same order or larger than the photon wavelength,
the scattering is described by Mie theory~\cite{bib:mie}. 
In the case of a space-based detector, this latter type of description is required to treat photon propagation in 
clouds and in aerosols.

\paragraph{Rayleigh scattering}

In ESAF, the optical depth $\delta_\mathrm{rayl}$ at a given  wavelength $\lambda$ (in nm)  is computed with the formula used in 
{\tt LOWTRAN7}:
\begin{eqnarray*}
\delta_\mathrm{rayl} = & \frac{X}{\gcmsq[3102]} \left( \frac{400\;\text{nm}}{\lambda}\right ) ^4 \cdot \\ \nonumber  
&\cdot{\left( {1-0.0722 \left( \frac{400\;\text{nm}}{\lambda}\right ) ^2} \right)^{-1}}
\virgola \cr
\end{eqnarray*}
where $X$ is the atmospheric slant depth along the photon path. The $\lambda^{-4}$ dependence, 
at the first order, of 
Rayleigh cross-section  is dominant here.    
The term in $ \lambda^2$ is partly  due to the dependence of the cross-section on the  air index. 
Studies made with ESAF show that vertical transmission
between ground and top of atmosphere is around 50\% at $\lambda = 337\;\text{nm}$, 30\%
at 300~nm, and 80\% at 450~nm. 
Simulations have been performed with several different density profiles:  results show that
air density profile variations have very small impact on associated vertical transmission value (${\delta T / T} \leq 2\% $).

The normalized phase function  $\Phi_\mathrm{rayl}$  of  Rayleigh  scattering is given by:
\begin{equation*}
\Phi_\mathrm{rayl}  (\theta, \varphi) = \frac{3}{16\pi} (1 + \cos^2 \theta).
\end{equation*}
Rayleigh scattering is independent of the azimuth angle $\varphi$ 
and almost isotropic: the phase function varies by only a factor of two between 0\degr and 180\degr,
that is very weak compared to phase functions associated to other scattering processes (see below, and Fig.~\ref{fig:cloud-phase}).

\paragraph{Mie scattering}
Below an altitude of around 8~km, clouds are made of tiny water droplets. Due to the spherical shape 
of droplets, the original Mie theory  can be applied, provided knowledge of the distribution 
of the scattering droplets radius $n(r)$. At a given $\lambda$, $n(r)$ determines the phase function.
In UV range, clouds do not absorb photons, but scatter them. Scattering can be  considered 
as independent of  wavelength in the range 300-450~nm \cite{bib:bunner}.
The same behaviour is observed for high altitude cirrus clouds (cirrostratus) composed of ice crystals.
In ESAF  a cloud is described as a homogeneous layer,  with a finite vertical height (TOVS cloud). 
Cloud databases provide integrated vertical depth $\Delta_\mathrm{vert}$. Laying from 1.2 to 3.3, the mean value of 
$\Delta_\mathrm{vert}$ is around 2. For $\Delta_\mathrm{vert} = 2$, the transmission of a vertical UV  light beam crossing the cloud layer is $\sim15$\%.

Phase function values for low altitude clouds, obtained from one  particular cumulus model~\cite{bib:cumulus1}, 
are stored in tables for $\lambda = 450$~nm, as well as  those for cirrostratus, for  $\lambda = 500$~nm~\cite{bib:cirrus}.
Contrary to Rayleigh scattering,  Mie scattering in clouds is highly anisotropic, and very forward peaked. 
Phase functions are compared in Fig.~\ref{fig:cloud-phase}.

\begin{figure}[ht]
\begin{center}
\includegraphics[width=0.48\textwidth]{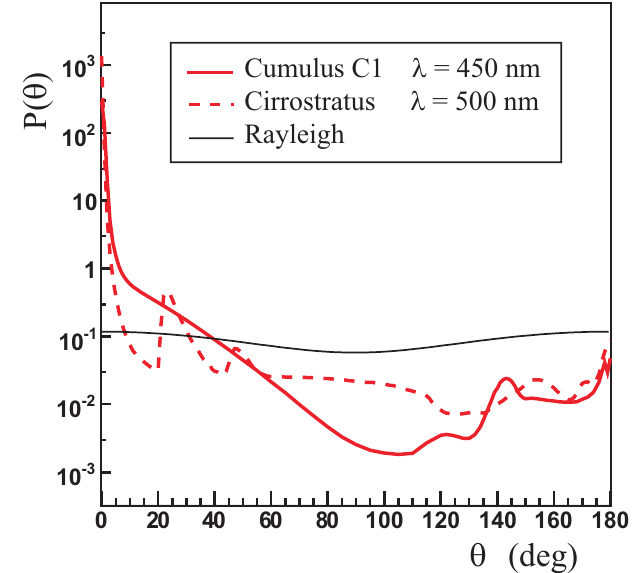}
\caption{Cloud phase functions (cumulus and cirrostratus) compared to Rayleigh one. 
$P(\theta)$ is equal to $ 2 \pi \Phi(\theta, \varphi) =  2 \pi \Phi(\theta)$.}
\label{fig:cloud-phase}
\end{center}
\end{figure}

Concerning aerosols, the three main models (rural, urban, maritime) provided by \texttt{LOWTRAN7} have been introduced in ESAF. These models provide the dependence of the extinction coefficient with altitude and wavelength, specifying the respective part of scattering and absorption processes. These aerosol layers are located between 0 and 2 km altitude. More details can be found in \cite{bib:tesi-moreggia} (in french) or \cite{bib:ramond}. The aerosol layer opacity is specified via the visibility. The smaller is the visibility, the greater is the extinction coefficient. Default visibility values in \texttt{LOWTRAN} are 5 km and 23 km, corresponding to a vertical transmission of 75\% and 20\% respectively.
To describe scattering angular distribution, ESAF is supplied with  phase functions from \texttt{LOWTRAN7} for the three mentioned models.

\paragraph{Ozone absorption}
In the UV range of interest, the molecular absorption in  atmosphere is dominated by ozone (O$_3$) absorption. 
At a fixed altitude $h$, the ozone absorption length $L_{\mathrm{abs,O}_3}$ is expressed by
\begin{equation*}
\frac{1}{L_{\mathrm{abs,O}_3}} = \rho_{\mathrm{O}_3} (h) \   \sigma_{\mathrm{abs,O}_3}(\lambda,h).
\end{equation*}
Ozone absorption cross section ($\sigma_{\mathrm{abs,O}_3}$) data provided by {\tt LOWTRAN}  
have been introduced in ESAF. The dependence of cross-section on local temperature is taken into account, entailing a dependence on altitude. $\rho_{\mathrm{O}_3} (h)$ is determined from available ozone profiles, and 
air density. Ozone exists mainly between 10 and 30~km, with a maximum absorption around 20~km. 
It absorbs the UV light below 330~nm, therefore it  has a weak impact on  fluorescence transmission, since the three main 
fluorescence lines are above the wavelengths affected by ozone; it  affects mainly  Cherenkov light.

\begin{figure}[ht]
\begin{center}
\includegraphics[width=0.48\textwidth]{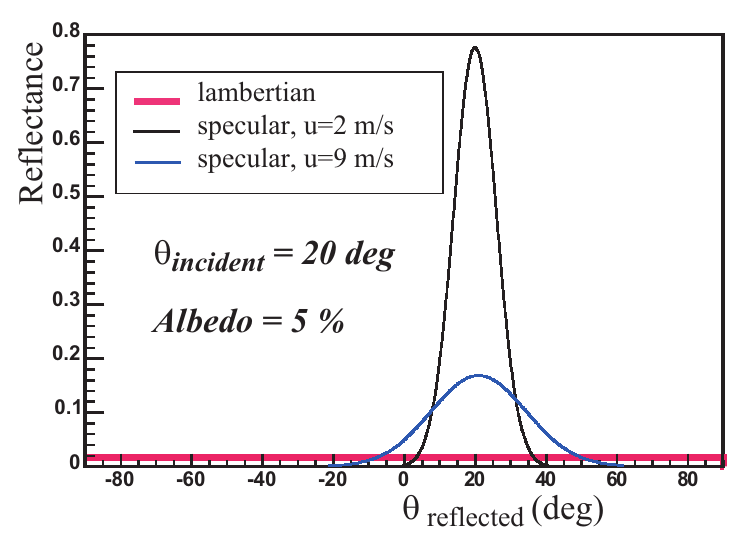}
\caption{Simple model for light reflection on ocean. The zenith angle $\theta_\mathrm{i}$ of the incident light beam is 20\degr. The reflectance is computed
in the plane containing the incident direction,  as a function of the reflected light zenith angle $\theta_\mathrm{r}$, in case of both lambertian or
specular reflection (for 2 wind speed values $u$).}
\label{fig:brdf}
\end{center}
\end{figure}

\paragraph{Ground reflection}
The reflection on the Earth surface is a crucial point for Cherenkov signal detection.
The Earth albedo, in  UV range, has been measured by ULTRA \cite{bib:ultra2} experiment and is less than 10\% \cite{bib:ultra}. Moreover, reflected photon angular distribution
depends on the reflection type, which can be Lambertian (reflected light is scattered isotropically by
a Lambertian surface) or specular (highly anisotropic, as on the ocean surface).  Both types have been implemented in  ESAF.
To describe specular reflection, a simple model, reproducing qualitatively air-born measurements of the ocean 
BRDF ({\it Bidirectional Reflectance Distribution Function})~\cite{bib:brdf} with dependence on wind speed, has been introduced. 
For a wind speed of 9~m/s,  specular peak amplitude is 5 times lower than for a wind speed of 2~m/s.
If specular reflection is dominant, reflected light intensity is not isotropic in $\theta_\mathrm{r}$, and can be zero on 
a large zenith angle range (see Fig.~\ref{fig:brdf}). 

\subsubsection{Propagation algorithms}
Below are described two original propagation algorithms designed and implemented in ESAF.

\paragraph{``Bunch'' algorithm}

The algorithm has been designed to handle mainly three components of the light signal received by the telescope. The first component, called 'Direct Fluorescence', is the fluorescence light reaching directly the lens without undergoing any interaction in the atmosphere. The second component, called 'Reflected Cherenkov', is the ground reflected Cherenkov light, having interacted only once along its path from emission point to the telescope. The third component, called 'Air Scattered Cherenkov', represents Cherenkov photons which have been scattered only once, on air molecules.

The extreme low value of the detector solid angle ($\Omega \sim 10^{-11}\; \text{sr} $ for EUSO telescope
shipped on ISS) is of major importance: when considering photons emitted or scattered within the detector solid angle, 'single' photons can be sampled from the related bunches, and be transmitted one by one to the telescope.

In the case of fluorescence photons directly emitted towards the detector, single photons are sampled from bunches created at fluorescence emission point. Then, they are distributed in time, position and wavelength,  according to distributions associated to the bunch they originate. 
For each photon, the probability that it is transmitted up to the telescope  is computed taking into account all possible interactions  in atmosphere.

Concerning the two other components, they are both simulated by the following algorithm. 
Since Cherenkov radiation is beamed in the forward direction, it is possible to propagate each Cherenkov
photon bunch step by step, until it reaches ground or exits atmosphere. At each step, two operations are performed:
\begin{enumerate} 
\item[(i)] The number of Cherenkov photons scattered once on air molecules, and directed toward the telescope after the interaction, is computed. A corresponding number of single photons are therefore created. Their final transmission to the telescope from their scattering position is then evaluated : if they undergo another interaction along their path to the detector, they are lost.
\item[(ii)] The wavelength spectrum of the propagated bunch is convoluted by transmission values computed along the step. Then, the possibility to pursue the bunch propagation is evaluated.
\end{enumerate}
When a Cherenkov bunch reaches the ground, the number of photons reflected toward the detector is computed. A corresponding number of single photons are created and their transmission to the telescope is evaluated : if they undergo another interaction, they are lost.

In presence of aerosols or clouds, the bunch algorithm takes into account their effects on the transmission values, considering the photons scattered on these media as lost.

\begin{figure}[ht]
\begin{center}
\includegraphics[width=0.48\textwidth]{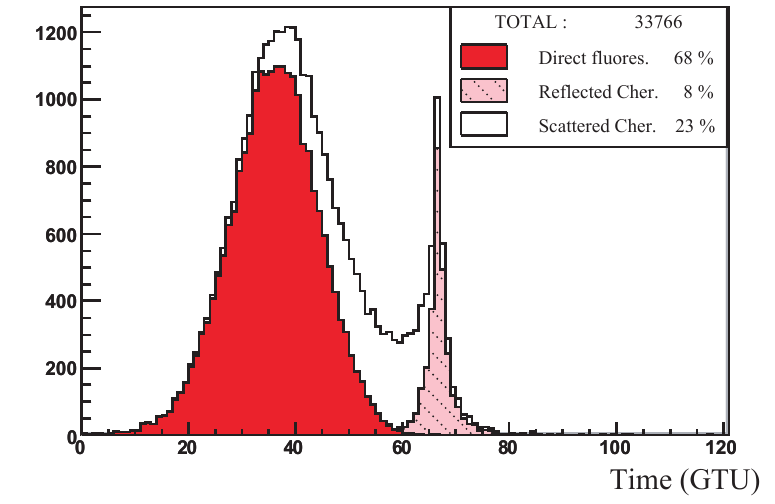} \\
\includegraphics[width=0.45\textwidth]{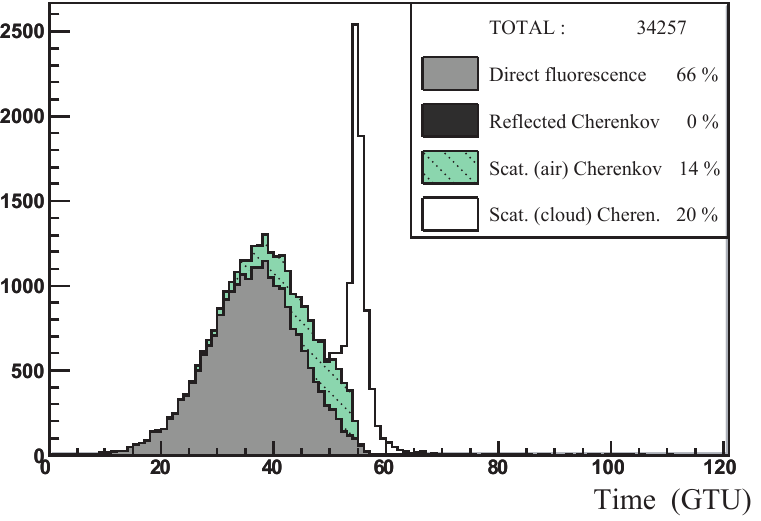} \\
\caption{Examples of time photon distributions, for a shower with $E=10^{20}$~eV, $\theta=60\degr$, and a footprint at the telescope nadir -
The different contributions are superimposed: fluorescence photons transmitted directly to the telescope,  Cherenkov photons scattered once by the atmosphere molecules, and Cherenkov photons reflected by the ground in case of clear sky (top panel) or scattered by the cloud in case of cloudy sky (bottom panel).}
\label{fig:bunch}
\end{center}
\end{figure}

The bunch algorithm is useful to perform fast simulations (see Fig.~\ref{fig:bunch}). But such an algorithm can not handle properly single scattering of fluorescence photons, and multiple scattering of both fluorescence and Cherenkov photons. A mandatory point in EAS analysis is to evaluate the detailed contributions of scattered photons to the detected signal, as this component may introduce systematic uncertainties if not taken into account during reconstruction procedure. To achieve this, a more sophisticated algorithm has been designed to deal with single and multiple scattering of the light produced by EAS.

\begin{figure}[ht] 
\begin{picture}(200,240)
{\small
\put(0,230){\makebox(0,0)[l]{\it  ``full'' MC: $N_0$ photons}}
\put(0,220){\makebox(0,0)[l]{\it  ``reduced'' MC: $N_{red}$ photons}}

\put(24,205){\makebox(0,0){{one photon}}}
\put(24,196){\makebox(0,0){($\lambda, \theta$...)}}
\put(01,190){\line(1,0){48}}
\put(01,210){\line(1,0){48}}
\put(01,210){\line(0,-1){20}}
\put(49,210){\line(0,-1){20}}

\put(49,200){\vector(1,0){23}}
\put(97,200){\makebox(0,0){\framebox(52,20){scattered ?}}}
\put(123,200){\vector(1,0){30}}
\put(138,205){\makebox(0,0){no}} 
\put(188,200){\makebox(0,0){\dashbox{1.5}(70,20){rejected photon}}}

\put(97,190){\vector(0,-1){20}}
\put(105,180){\makebox(0,0){yes}}
\put(65,170){\vector(0,1){25}} 
\put(72,180){\makebox(0,0){no}}
\put(97,160){\makebox(0,0){\framebox(82,20){scattered $n$ times?}}}

\put(97,150){\vector(0,-1){20}}
\put(105,140){\makebox(0,0){yes}}
\put(58,130){\line(1,0){74}}
\put(58,105){\line(1,0){74}}
\put(58,130){\line(0,-1){25}}
\put(132,130){\line(0,-1){25}}
\put(97,122){\makebox(0,0){in the detector}}
\put(97,112){\makebox(0,0){solid angle ?}}
\put(132,118){\vector(1,0){20}}
\put(140,122){\makebox(0,0){no}} 
\put(188,120){\makebox(0,0){\dashbox{1.5}(70,20){rejected photon}}}
\put(50,130){\makebox(0,0)[r]{\it probability $P$}}
\put(50,120){\makebox(0,0)[r]{\it (``full'' MC)}}
\put(50,105){\makebox(0,0)[r]{\it probability $P_1$}}
\put(50,95){\makebox(0,0)[r]{\it (``reduced'' MC)}}

\put(97,105){\vector(0,-1){20}}
\put(105,95){\makebox(0,0){yes}}
\put(58,85){\line(1,0){74}}
\put(58,60){\line(1,0){74}}
\put(58,85){\line(0,-1){25}}
\put(132,85){\line(0,-1){25}}
\put(97,77){\makebox(0,0){transfered up }}
\put(97,67){\makebox(0,0){to detector ?}}
\put(132,73){\vector(1,0){20}}
\put(140,77){\makebox(0,0){no}} 
\put(188,75){\makebox(0,0){\dashbox{1.5}(70,20){rejected photon}}}

\put(97,60){\vector(0,-1){20}}
\put(105,50){\makebox(0,0){yes}}
\put(96,25){\oval(75,30)}
\put(97,32){\makebox(0,0){Photon on the}}
\put(97,22){\makebox(0,0){entrance pupil}}
}
\end{picture}
\caption{Scheme of the Monte-Carlo (MC) algorithm principle to simulate photons scattered exactly $n$ times.}
\label{fig:montecarlo}
\end{figure}
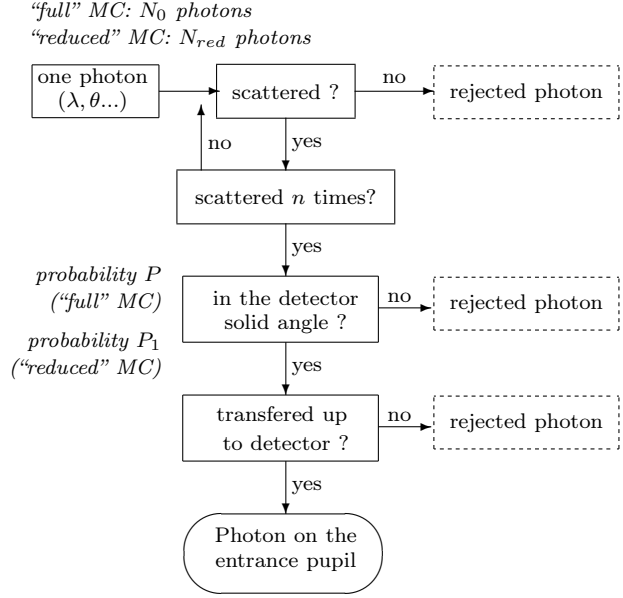

\paragraph{``Reduced'' Monte-Carlo algorithm}
First of all, the simulation of 'Direct Fluorescence' component does not raise any problem, and it is treated in the same way as presented in the previous section. Then, the aim is to simulate fluorescence and Cherenkov light propagation by performing a Monte-Carlo ray-tracing algorithm, propagating photons one by one.
Let us consider photons that reach the telescope after exactly $n$ scattering processes.
To simulate the transfer of each of them, an ideal algorithm has to follow the steps described
below, as illustrated on Fig.~\ref{fig:montecarlo}. 
\begin{enumerate}
\item Photon position, direction and wavelength have to be randomly generated, taking into account related distributions defined during light production process (see section~\ref{sec:light}).
\item  Interaction type and position are randomly determined according to the different physical processes able to alter photon path. 
If the photon is absorbed or if is going out of the atmospheric medium before interacting, it is considered as lost.
\item Previous step is repeated on the scattered photon, until its loss or until $n$-th scattering.
\item At the $n$th scattering,  a random sampling is performed considering the probability :
\begin{equation}
P=\Phi_n (\theta, \varphi) \Omega_\mathrm{det}  \label{eq:proba}
\end{equation}
in order to determine whether the photon is scattered toward the telescope or not.
$\Phi_n (\theta, \varphi)$ is the phase function of the $n$-th scattering, and $\Omega_\mathrm{det} $ the solid angle associated 
to telescope pupil entrance. If the photon is not scattered toward the telescope, it is rejected.
\item Then, transmission from the $n$-th scattering position to the detector is computed. If the photon undergoes another interaction it is lost; if not, it reaches the telescope lens.
\end{enumerate}

As previously stated, with the ideal algorithm described above, it will be unaffordable 
CPU time consuming to propagate each of the $\sim 10^{15}$ UV
photons produced by an EAS initiated by a primary of 10$^{20}$eV  (see section~\ref{sec:light}). 
About 10$^{4}$ photons reach  the telescope.
The high decrease between these two numbers is due to the solid angle. 
If this reducing factor could be applied before photon tracking, a ray-tracing simulation should be possible.
The key point comes from the particular situation of a space-based detector : it stands outside the scattering medium.

\begin{figure}[ht]
\begin{center}
\includegraphics[width=0.47\textwidth]{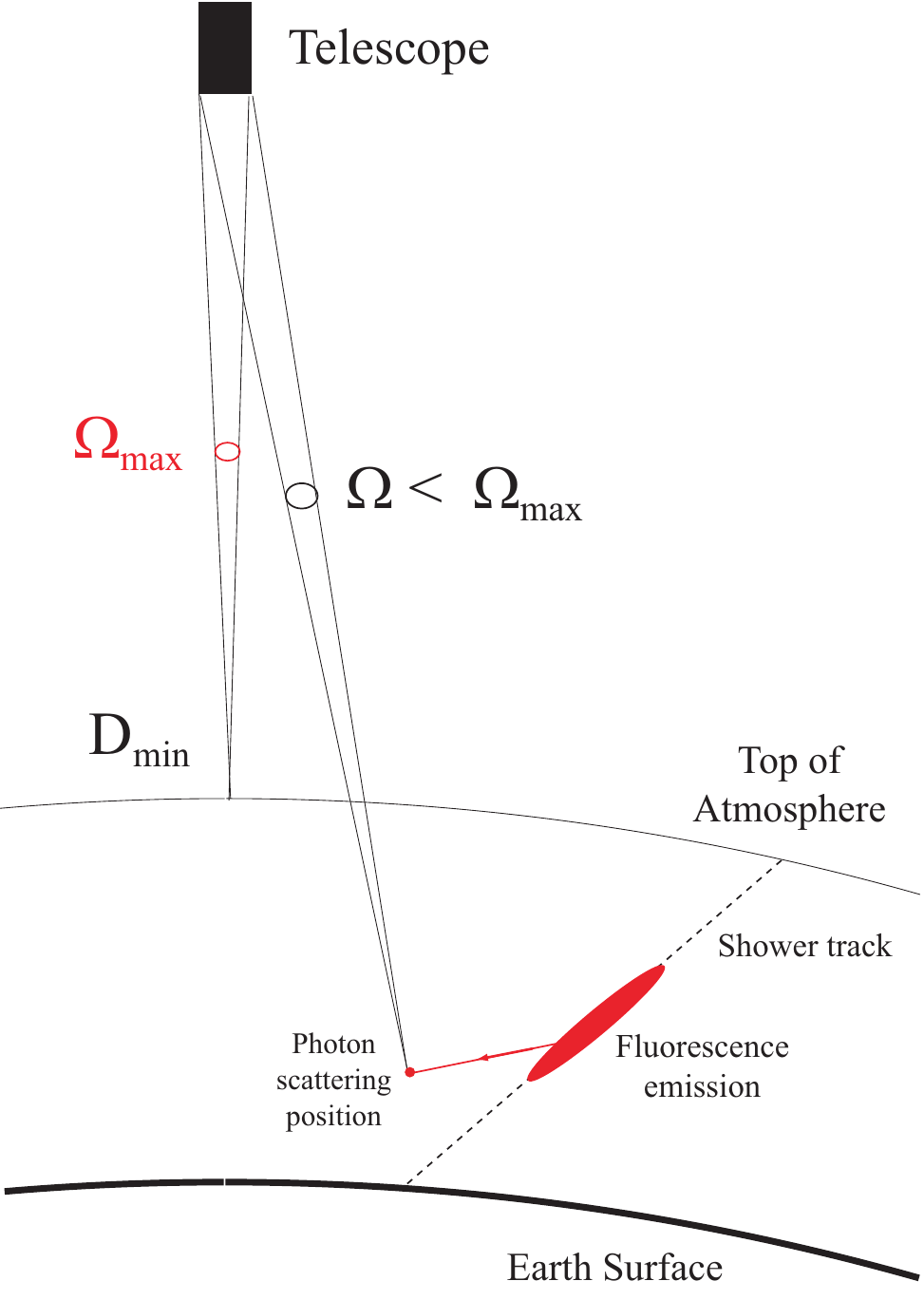}
\caption{Minimum distance $D_\mathrm{min}$ between the top of atmosphere (considered here at an altitude of 30 km) and the telescope and the associated largest solid angle $\Omega_\mathrm{max} $.}
\label{fig:omega-max}
\end{center}
\end{figure}

Above an altitude of 30 km, atmospheric density is very weak, and photon scattering can be neglected. 
As a consequence, one can define a minimum distance $D_\mathrm{min}$ below which photons cannot be scattered. 
This distance and the associated solid angle \mbox{$\Omega_\mathrm{max} = S_\mathrm{det}/D^2_\mathrm{min}$} 
are shown on Fig.~\ref{fig:omega-max}.
In case of a detector shipped on ISS, its altitude is $\sim 430\;\text{km}$, thus $\Omega_\mathrm{max}=3 \cdot 10^{-11}$sr.

The probability $P$ defined by the equation~\eqref{eq:proba} can be written as the product of two
independent factors $ P= P_1 \times  P_2$ defined as:

\begin{equation}
P_1= {\frac{\Phi_n (\theta, \varphi) \Omega_\mathrm{det}}{\Phi_\mathrm{max} \Omega_\mathrm{max} }}
\virgola\qquad
P_2= {\Phi_\mathrm{max}  \Omega_\mathrm{max} }
\end{equation}
$\Phi_\mathrm{max}$ is the maximum over all phase functions involved in simulation. 
By normalising the phase function $ {\Phi_n (\theta, \phi) }$ of the $n$th
scattering (which  type is a priori not known) by $ \Phi_\mathrm{max} $, the
condition $P_1 \leq 1$ is insured. $\Omega_\mathrm{max} $ is small enough to insure $P_2 < 1$.
Random draw of $P$ probability is split into two independent draws of $P_1$ and $P_2$. 
Since $P_2$ has been set up to be independent of all other stochastic processes involved in photon propagation, random computation of $P_2$ can be performed at the very beginning of simulation chain. Moreover, considering that it can be identically applied to all photons, only one
evaluation  is performed, according to a binomial law with ($N_0$, $P_2$) parameters. $N_0$ is the total number of photons produced by the shower, $N_0  \gg 1 \gg P_2$. Then the reduced number
of photons to propagate $N_\mathrm{red}$ when simulating the $n$th scattering, is given by:
\begin{equation}
N_\mathrm{red}={\rm Poisson} (N_0 \Phi_\mathrm{max}  \Omega_\mathrm{max})
\virgola\label{nreduit}
\end{equation}
where the Poisson distribution is the binomial distribution limit, in case of a large
sample size and low probability. 
The described algorithm is then applied, taking probability $P_1$ instead of $P$ in the 4th step. 
All interaction types described in section~\ref{subsubsec:scat-abs} - molecular absorption, scattering on air molecules, aerosols, clouds, and ground - are treated by this algorithm.

The treatment presented above is valid for a fixed scattering order. To obtain the contributions
of different scattering orders $n$ and $p$, the described algorithm has to be performed
for one set of generated photons $N^{(n)}_\mathrm{red}$, until the $n$th scattering, then for another
set $N^{(p)}_\mathrm{red}$, until the $p$th scattering, in a full independent way. Before computation,
a maximum scattering order $n_\mathrm{max}$ has to be fixed. The contribution of the scattering processes to
the signal is the sum of the photons scattered exactly 1 time, 2 times, 3 times, ..., $n_\mathrm{max}$ times.
``Zero'' order corresponds to fluorescence photons produced in the detector solid 
angle, and transmitted without interacting up to the telescope.
Considering the first scattering order in clear sky conditions, the different contributions come from:
\begin{description}
\item [$\bullet$ zero order]~~
\begin{itemize}
\item fluorescence photons straightly transmitted
\end{itemize}
\item [$\bullet$ first order]~~
\begin{itemize}
\item Cherenkov photons reflected on ground
\item Cherenkov photons undergoing molecular scattering
\item fluorescence photons reflected on ground
\item fluorescence photons undergoing molecular scattering.
\end{itemize}
\end{description}

The main advantage of this algorithm is to be able to simulate radiative transfer of photons
emitted by any type of light source,
taking into account the multi-scattered photons at any order;  but its main limitation is the computing time, which can be
high, despite the applied reducing factor.   The computing time
is strongly dependent on  maximum scattering order $n_\mathrm{max}$ , as well as on 
number of photons to propagate $N_\mathrm{red}$. The latter is related to 
the maximum solid angle, fixed by detector setup, to the number of emitted photons, related to shower parameters ($E, \theta$)
and to the maximum value of the phase function. 
$\Phi_\mathrm{max}$ can vary from 0.3 in case of clear sky condition to 1000 in case of cloudy condition.
Simulation runs show that computing time is not affordable in the latter case. 

To bypass this limitation, the cloud phase function involved in the \textit{last} scattering process is approximated. Let us highlight that the cloud phase function involved in previous scattering processes is not approximated ; it concerns only the last scattering process, if it occurs within a cloud. Two kinds of approximation have been studied. The first one uses an analytical form, the Double Henyey-Greenstein function~\cite{bib:DHG}. The second one uses an averaged phase function peak over a limited range of 10\degr. 

In the current state of ESAF development, aerosol phase functions are treated in a simpler way, using the analytical approximation for every scattering process occurring on aerosol layer.

\begin{figure}[ht]
\begin{center}
\includegraphics[width=0.48\textwidth]{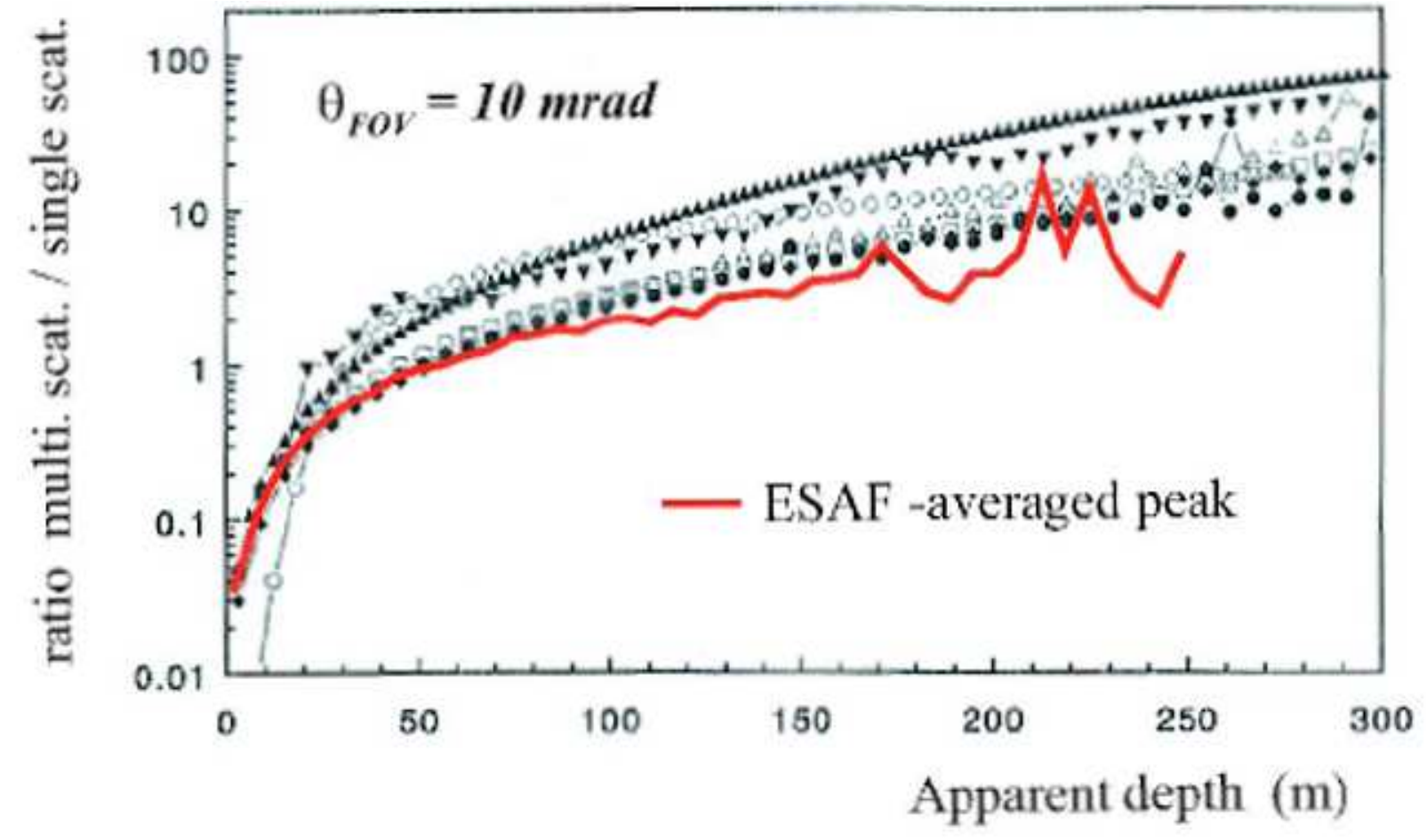} \\
\includegraphics[width=0.48\textwidth]{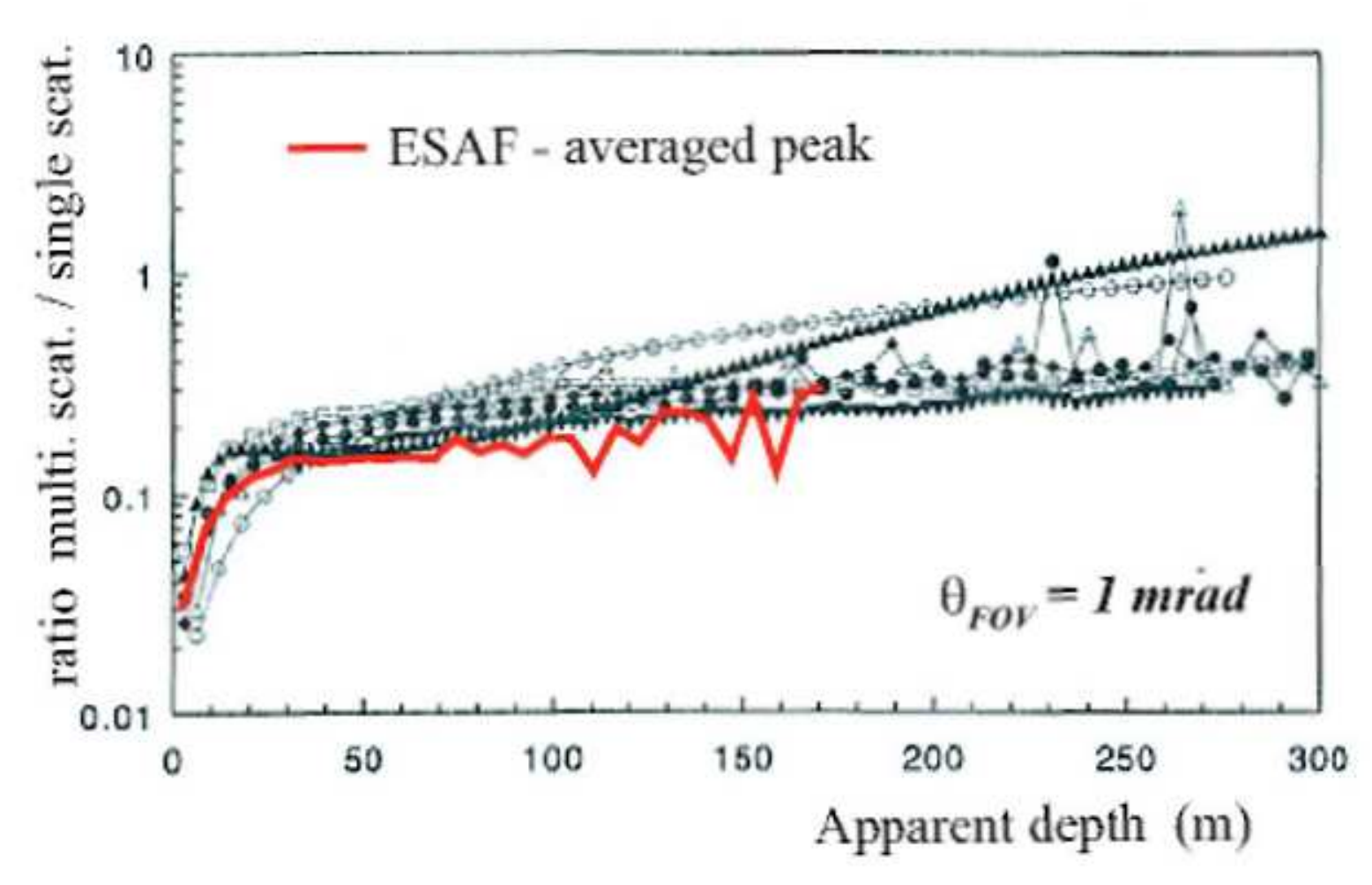}

\caption{LIDAR shot at 1064 nm from ground to 1 km distant cumulus cloud, 300 m deep
and with an extinction coefficient of 17.25 km$^{-1}$ (no atmosphere contribution, cloud only). 
Contribution of the multiple scattering component
for a receiver field of view of 10 mrad (top) and 1 mrad (bottom). 
ESAF results are obtained in simulating scattering process until order 8. 
In both cases the ESAF results are superimposed to MUSCLE ones, showing 
that they are in very good agreement. }
\label{fig:muscle}
\end{center}
\end{figure}

\subsubsection{Validation tests}
To proof that both propagation algorithms are efficient to properly simulate 
radiative transfer, validation tests were carried out.  
First, the propagation in clear sky conditions of photons emitted by a given shower (fixed energy and zenith angles) has 
been simulated by the two designed algorithms. Results concerning Cherenkov photons propagation (molecular single scattering, ground reflection) were compared. The time distribution of photons at telescope entrance pupil and their last scattering position in FoV are similar, clearly proving the consistency of the two algorithms.

In a second step, studies have been undertaken to test the  validity of the most efficient algorithm, 
namely the ``reduced'' Monte-Carlo algorithm. The aim is to compare it  to simulation tools 
developed within the atmosphere physicists community.
One study, handled by MacKee et al.~\cite{bib:mackee}, concerns  cloud albedo change with optical depth, 
predicted by a Monte-Carlo simulation and checked experimentally. Results got with ESAF are in agreement with
those obtained by MacKee et al.
The second set of tests were done to reproduce studies of the MUSCLE (MUltiple SCattering Lidar Experiments) working
group~\cite{bib:muscle}. The aim is to predict LIDAR  shot response after multiple scattering in clouds.
The results obtained with ESAF are compared to the MUSCLE ones: a good agreement has been found (see Fig. 
\ref{fig:muscle} and reference ~\cite{bib:tesi-moreggia} for more details)
that actually proves the validity of the ``reduced'' Monte-Carlo propagation algorithm implemented in ESAF.

Let us specify that the two kinds of cloud phase function approximations have been compared during these tests. It has been shown the use of an averaged peak is the best solution, and will be used further in this paper.

With this propagation package in ESAF, generated fluorescence and Cherenkov
lights are transfered  up to the space-based detector. Thanks to the designed algorithms, 
careful studies of the signal at detector pupil entrance are possible dealing with different
atmospheric situations: clear sky, aerosol layer, clouds.

\subsection{Detector response simulation}\label{sec:dete}

The ESAF detector simulation is the part of the framework which is more tightly connected to the EUSO design. Most of the implementation is based on a monocular cylindrical telescope with a photomultiplier (PMT) based photodetector. The main scheme of the detector is shown in Fig.~\ref{fig:DetScheme}. Nevertheless, the ESAF detector simulation structure retains large degree of flexibility and can be adapted to a different detector concept. 

The telescope simulation part is divided in 2 main modules:

\begin{enumerate}
	\item The Monte-Carlo photon tracking engine through the optical system.
	\item The electronics and trigger response.
\end{enumerate}

\subsubsection{Optics and focal surface}

\begin{figure}[htb]
	\centering
	\includegraphics[width=0.45\textwidth]{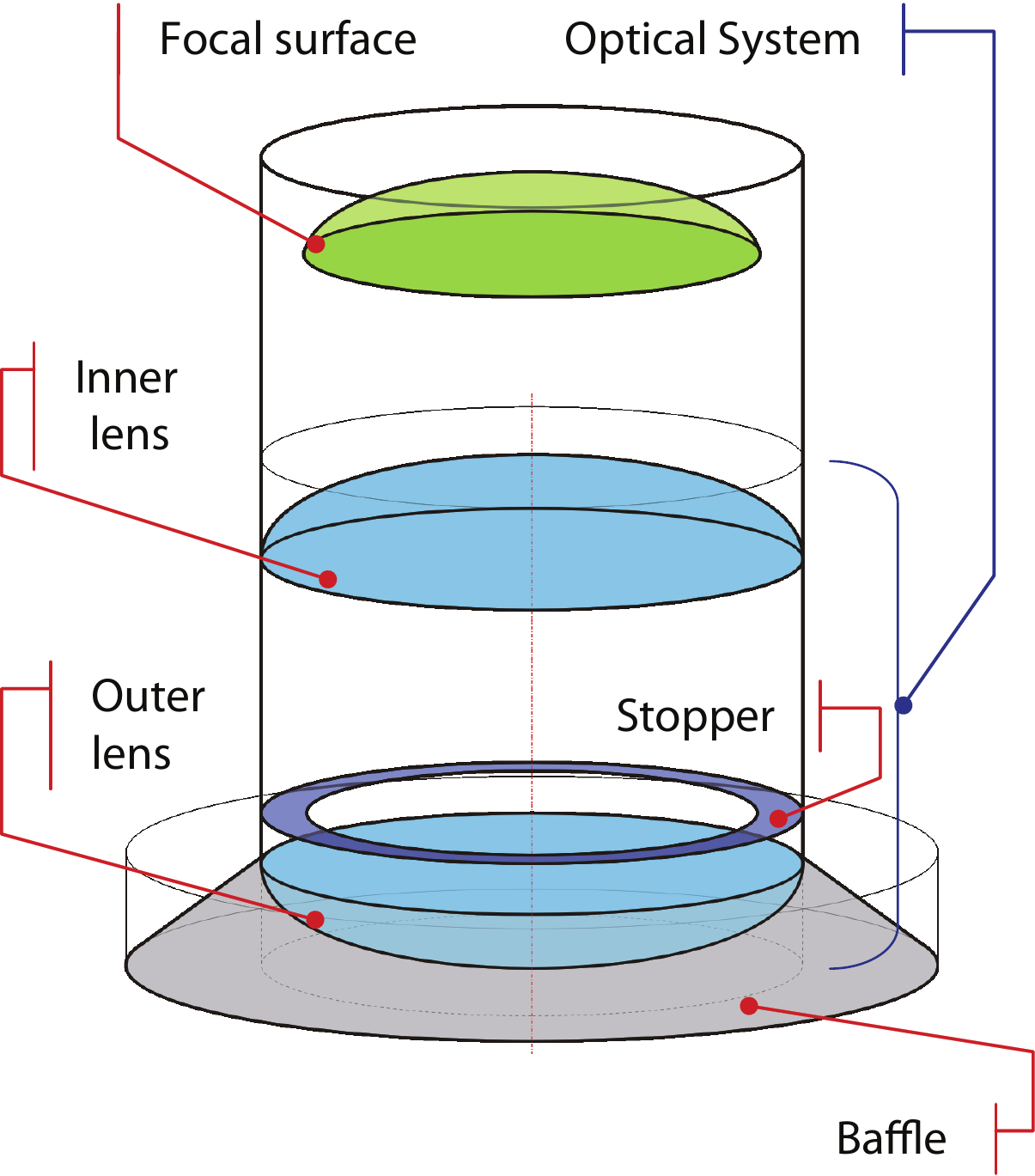}
	\caption{Main scheme of the detector.}
\label{fig:DetScheme}
\end{figure}

\begin{figure}[htb]
	\centering
	\begin{tabular}{c}
	\includegraphics[width=0.45\textwidth]{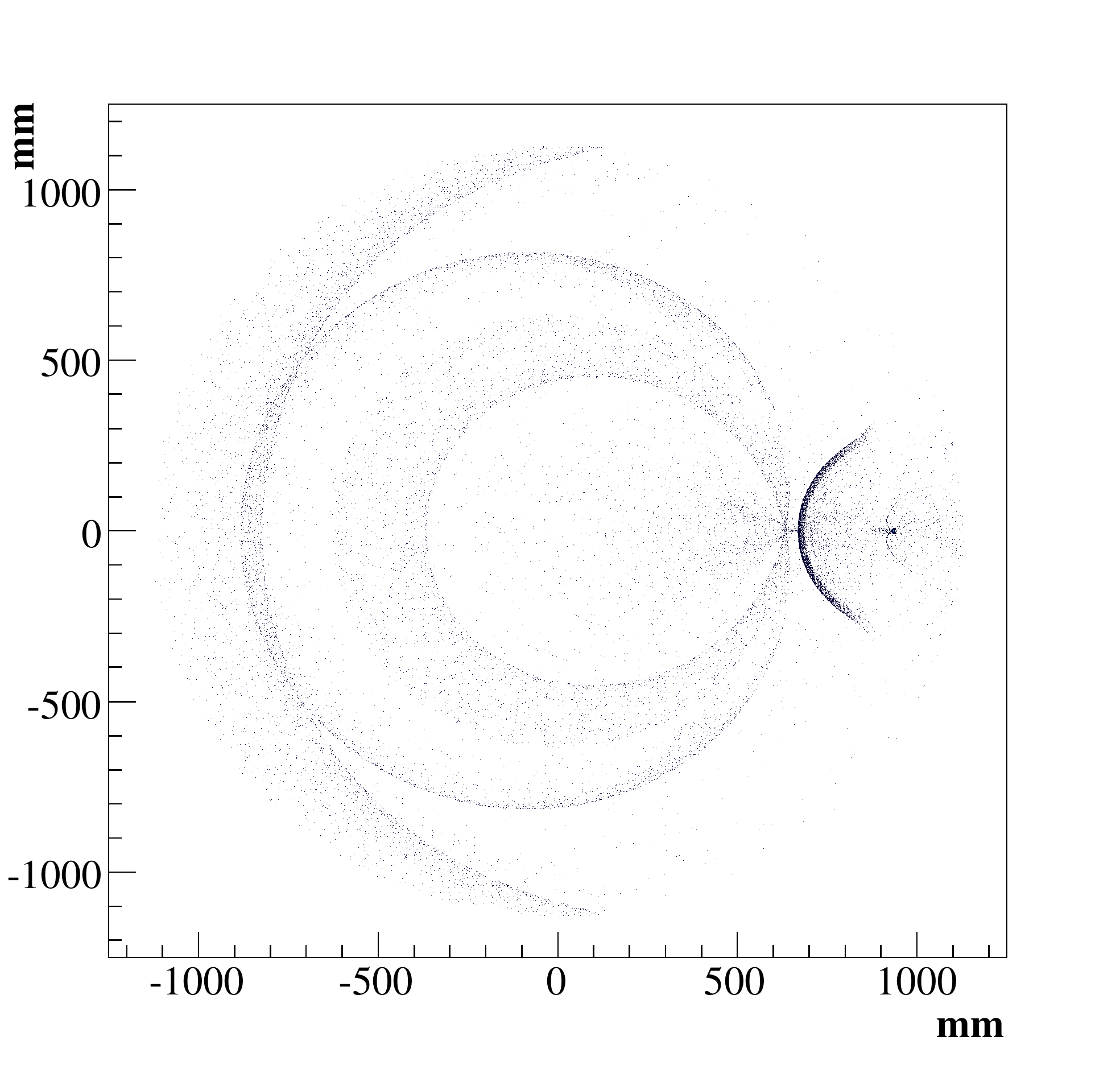} \\
	(a) \\
	\includegraphics[width=0.45\textwidth]{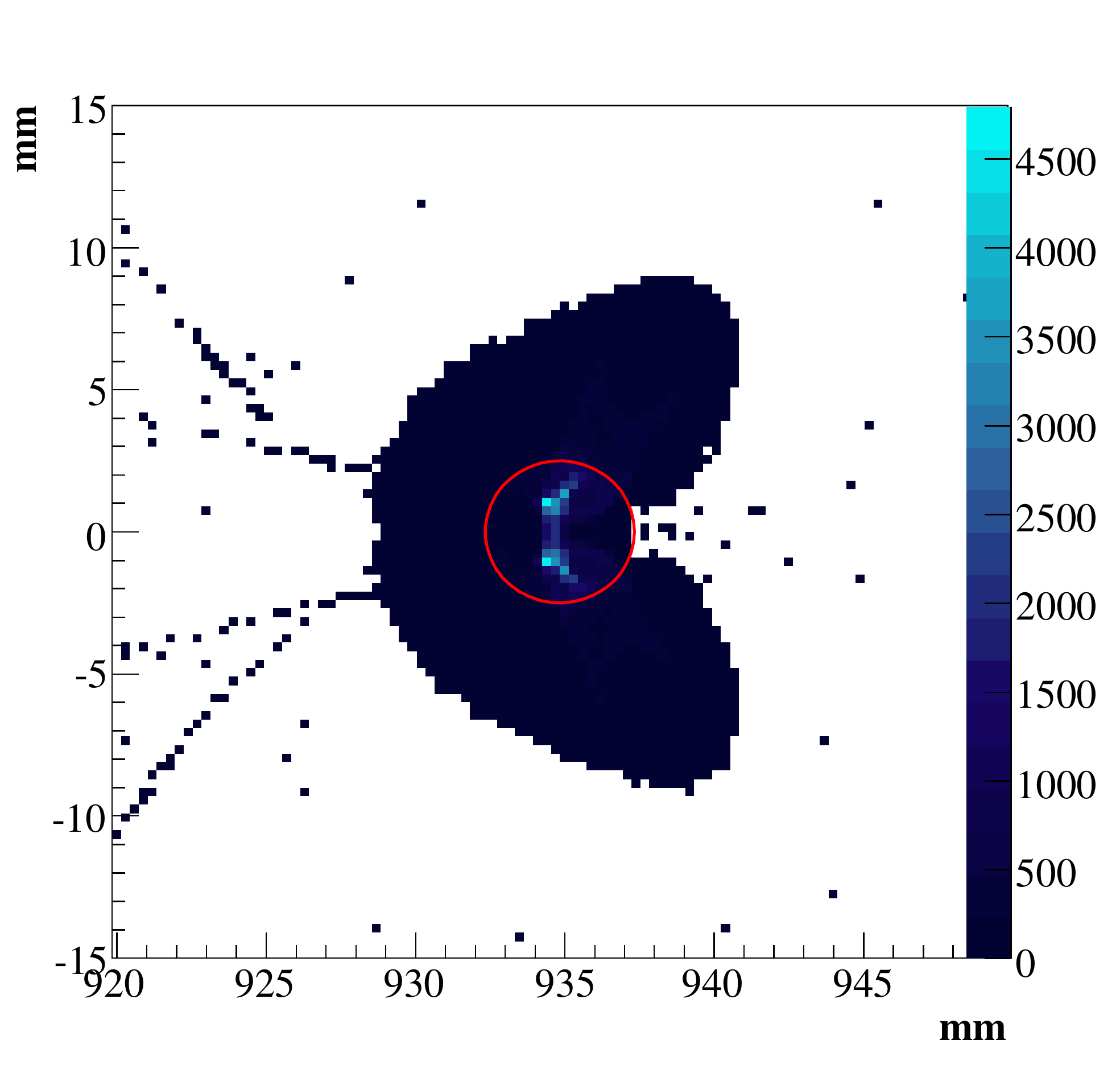} \\
	(b) 
	\end{tabular}	
	\caption{An example of the irregular PSF of the Fresnel optics generated using an
incident flux at an angle $\gamma = 25\deg$ in the \FOV. (a) Distribution of photons on the whole focal surface. (b) Zoom on the spot of the previous picture.
The red circle corresponds to a bucket of \un[5]{mm} of radius.}
\label{fig:PSF}
\end{figure}

\begin{figure}[htb]
\centering
\includegraphics[width=0.5\textwidth]{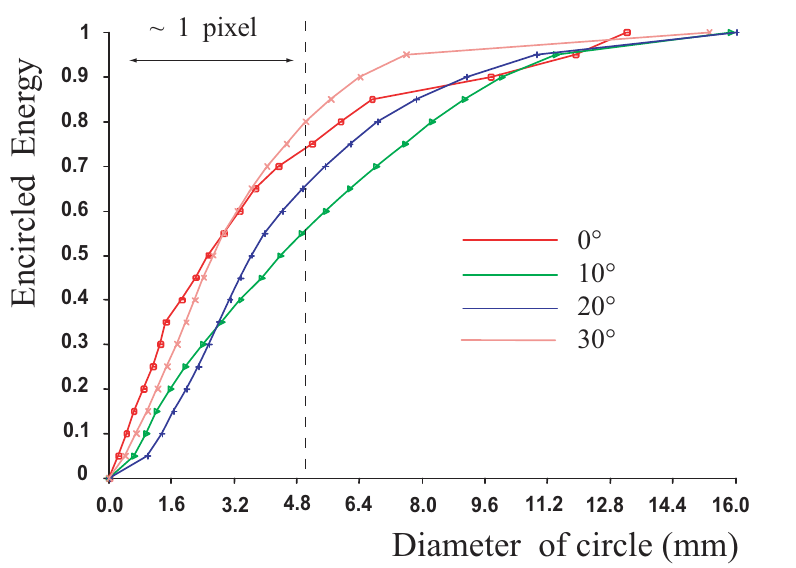}  
\caption{Fraction of energy (fraction of photons) contained in a circle as a function of the circle diameter. }
\label{fig:encyrcled-energy}
\end{figure}

The detector simulation starts from the photons in front of the entrance pupil.
One by one the photons are propagated through the optical elements of the detector; in Fig.~\ref{fig:DetScheme} the baffle, the optical system and the focal surface are visible.
Three simulated optical systems are provided within ESAF:

\begin{enumerate}

\item The EUSO Phase A baseline optics, a double Fresnel system. 
The tracking through this optics is simulated by tracking each individual photon entering the pupil up to the focal surface.  A detailed description of the fine structure of the Fresnel lenses allows to reproduce and study the optics distortions. As shown in Fig.~\ref{fig:PSF}, a large fraction of the light in the Point Spread Function (PSF) of the double-Fresnel optics is spread all over the focal surface: the so-called \emph{stray-light}.
A commercial program (CodeV, \cite{bib:codev}) was used to simulate outside ESAF the exact 
ray tracing properties of the lenses and the result
of this CodeV simulation was used to create an accurate parametrization of the PSF, of the amount of stray light 
and of the aberrations. Within ESAF, this parametrization is used to trace the photons. 
We underline that the light that contributes to the signal and, particularly, to the trigger, is the one that
is contained in a circle of the size of the focal surface pixel. Fig. \ref{fig:encyrcled-energy} shows that 
in a typical pixel of about 5 mm diameter (EUSO design), about 60\% of the photons contribute to the trigger,
while the other 40\% is lost as stray light. Fig. \ref{fig:encyrcled-energy} shows also the (weak) dependence
of this number as a function of the angle that describes the position of the shower in the field of view (see
Fig. \ref{fig:track_view_Cherenkov}). 

\item A second multi-purpose optical system, whose characteristics are implemented through a parametrization of the
optical response (PSF, stray light, aberrations) and are defined by a set of configuration files. Highly flexible, it allows an effective fast simulation without tracking individual photons. It is particularly useful for studies where the performances of the optics have to be tuned according to the overall requirements, or as a reference to compare the performances of the full simulation, or to speed up the simulation for large statistics studies. 
The parametrization and the configuration files were obtained by running the full simulation. 
Most of the work in this paper was obtained with this faster implementation but we have checked that the results obtained with the full simulation are always consistent.

\item A Geant4 based implementation of the optics and of the ray tracing (see next paragraph). This Geant4
implementation is new and was not used in this work.

\end{enumerate}

Ideally the efficiency of the optics scales with the angle in the FoV, $\gamma$, as $\cos\gamma$; for the EUSO optics
	the mean efficiency is of about 45\%. To this number, another  60\% light loss must be added because of the optical aberrations and stray light, as determined from the simulations.

\begin{figure}[htb]
	\centering
	\begin{tabular}{cc}
	\includegraphics[width=0.24\textwidth]{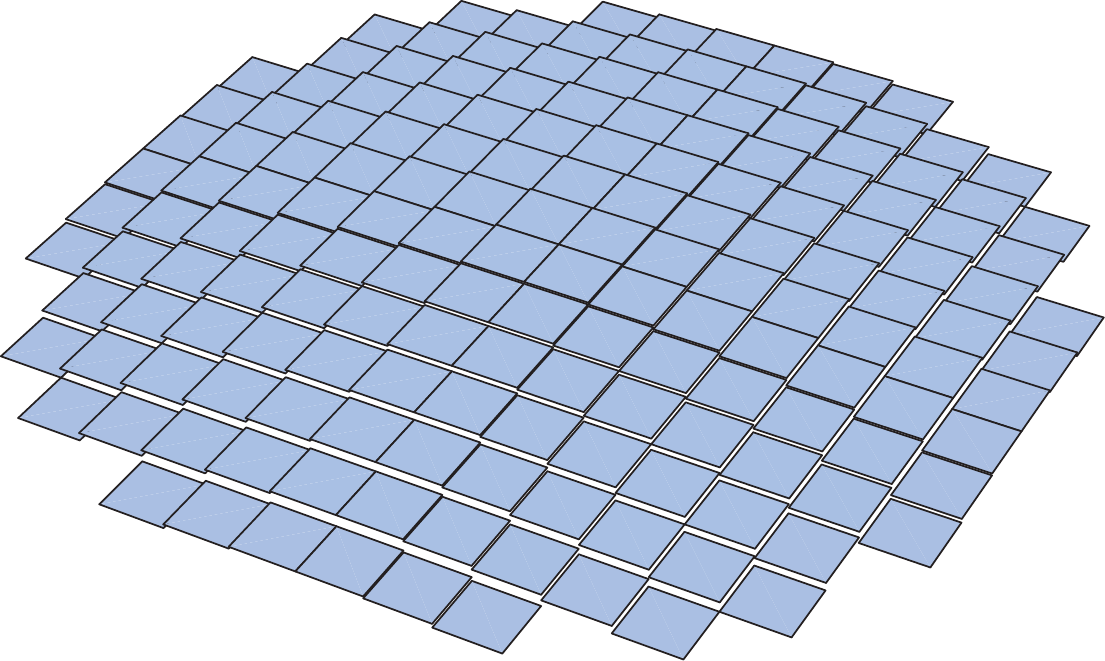} & 
	\includegraphics[width=0.24\textwidth]{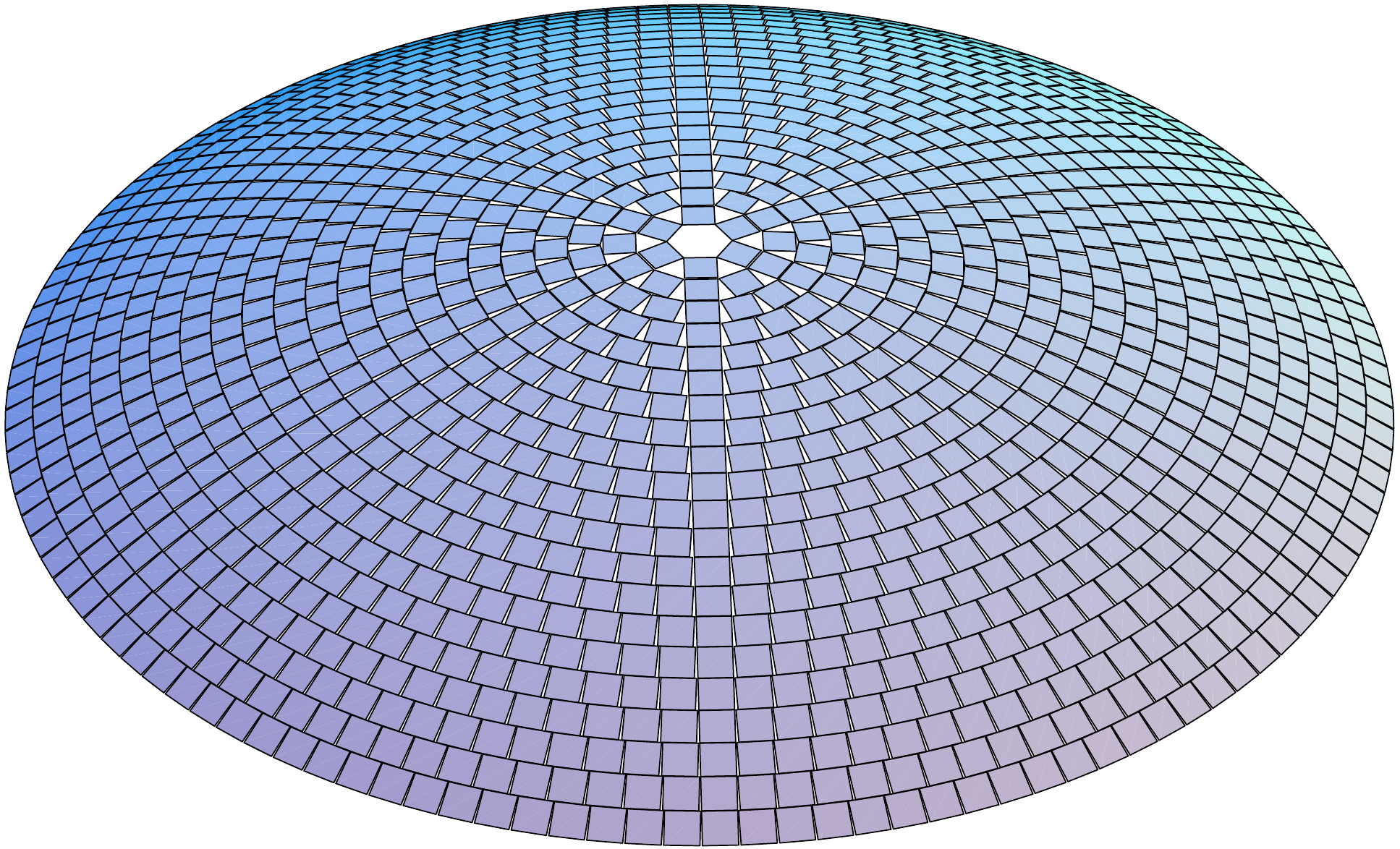} \\
	(a) & (b) \\
	\end{tabular}
	\caption{Focal surface layouts in ESAF. (a) Cartesian layout. (b) Polar layout.}
\label{fig:EusoFS}
\end{figure}

\begin{figure}[htb]
	\centering
	\begin{tabular}{c}
	\includegraphics[width=0.45\textwidth]{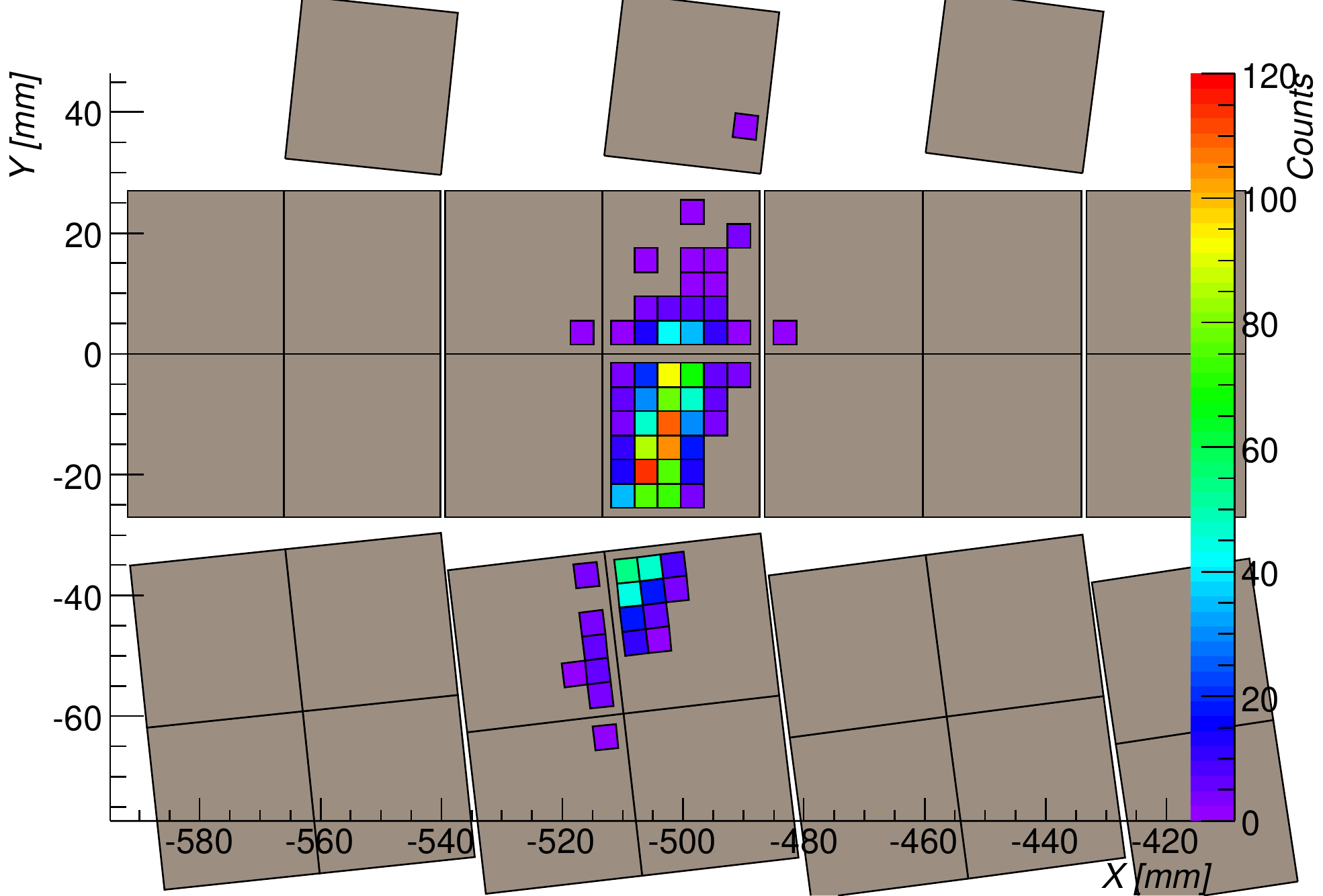}  \\
	 (a) \\ 
	 \includegraphics[width=0.5\textwidth]{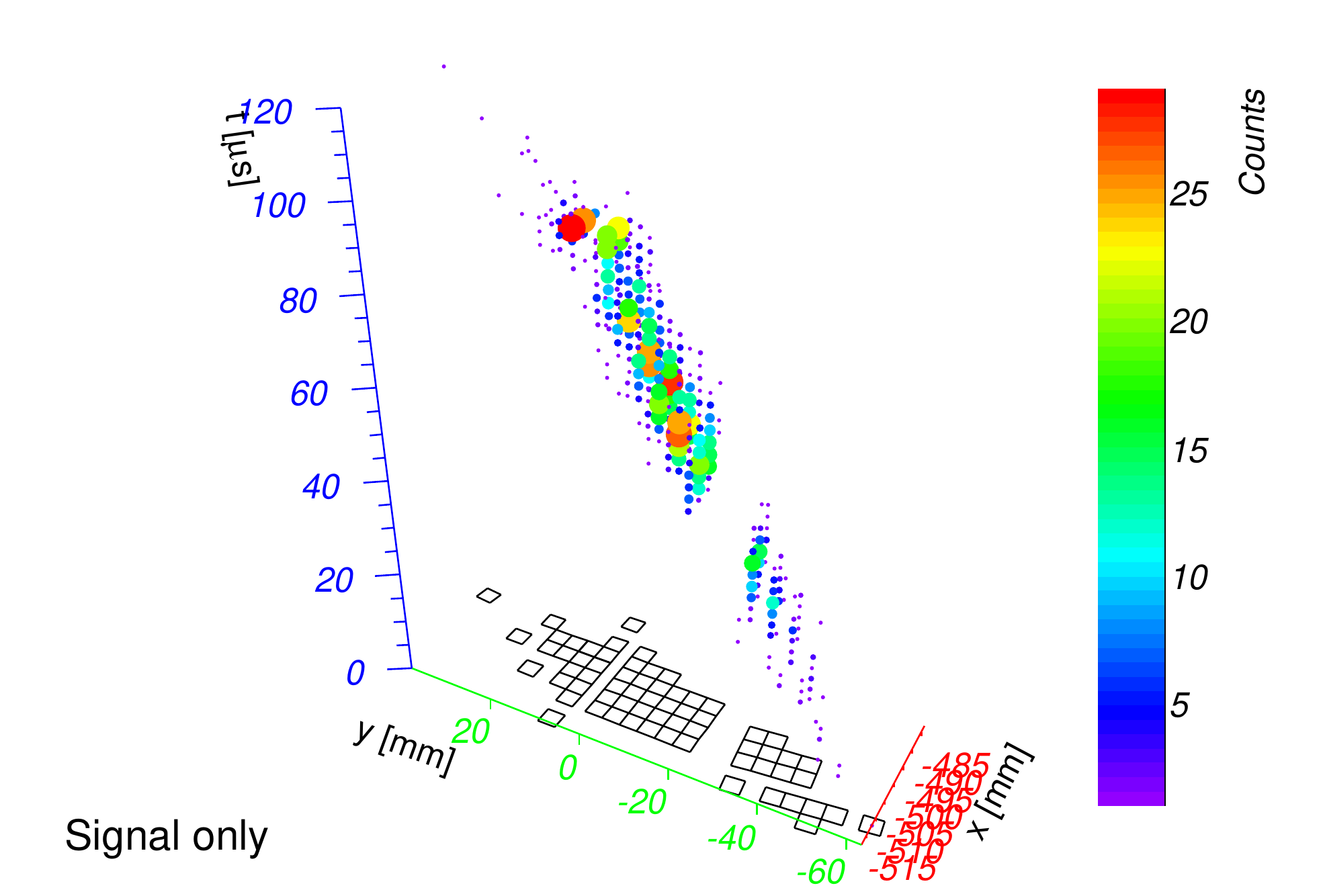} \\
	(b) \\
	\end{tabular}
	\caption{ Image of a \eVsci{1}{20} EAS track ($\theta=45\degr$) on the focal surface. No background is simulated. (a) 2D view. The small squares are the pixels, the big ones the PMT. (b) 3D view including also the time counted in Gate Time Units (GTU) of $2.5\;\mu\text{s}$. The $x$
  and $y$ axis are the $xy$ coordinates of the pixels while the $z$ axis is
  the time from the first signal photon counted (in GTU). On the $xy$-plane is plotted
  the profile of the pixels with signal.}
\label{fig:EViewerExample}
\end{figure}

	The focal surface module implements the geometry of the EUSO photodetector. The two available layouts of the focal surface are illustrated in Fig.~\ref{fig:EusoFS}. The focal surface is based on the modular structure of small autonomous units called elementary cells~\cite{bib:ElemCell}, each one being a matrix $2\times2$ of multi-anode photomultipliers. An optional optical adapter can be placed in front of the PMTs (efficiency of about 95\%). The focal surface elementary cell layout geometry is loaded at runtime from the layout configuration file. The simulations show that the focal surface, in the polar layout, has a ``filing factor'' efficiency of $\sim 90\%$. The tracking engine propagates the photons up to the PMT photocathodes and converts them into an electronic signal, stored for the electronics simulation. 

	Any PMT (or optical adapter) that fulfills the elementary cell specifications and constraints can be easily added to the focal surface simulation. Currently two models\footnote{In the PMT models ESAF takes into account the quantum efficiency as a function of the wavelength, the collection efficiency, the gain, the cross-talk between pixels, the PMT geometry, the time width of the signal and the dark noise rate. The typical PMT efficiency is 25\% (quantum) plus 70\% (collection).} 	are available, the 36-pixel Hamamatsu R8900 and the 64-pixel Hamamatsu R7600. 

In the Fig.~\ref{fig:EViewerExample}a and \ref{fig:EViewerExample}b there is an example of how an EAS appears on the
telescope focal surface.

\subsubsection{GEANT4 simulation of optics in ESAF}

\begin{figure}[p]
\begin{tabular}{c}
\includegraphics[width=0.5\textwidth]{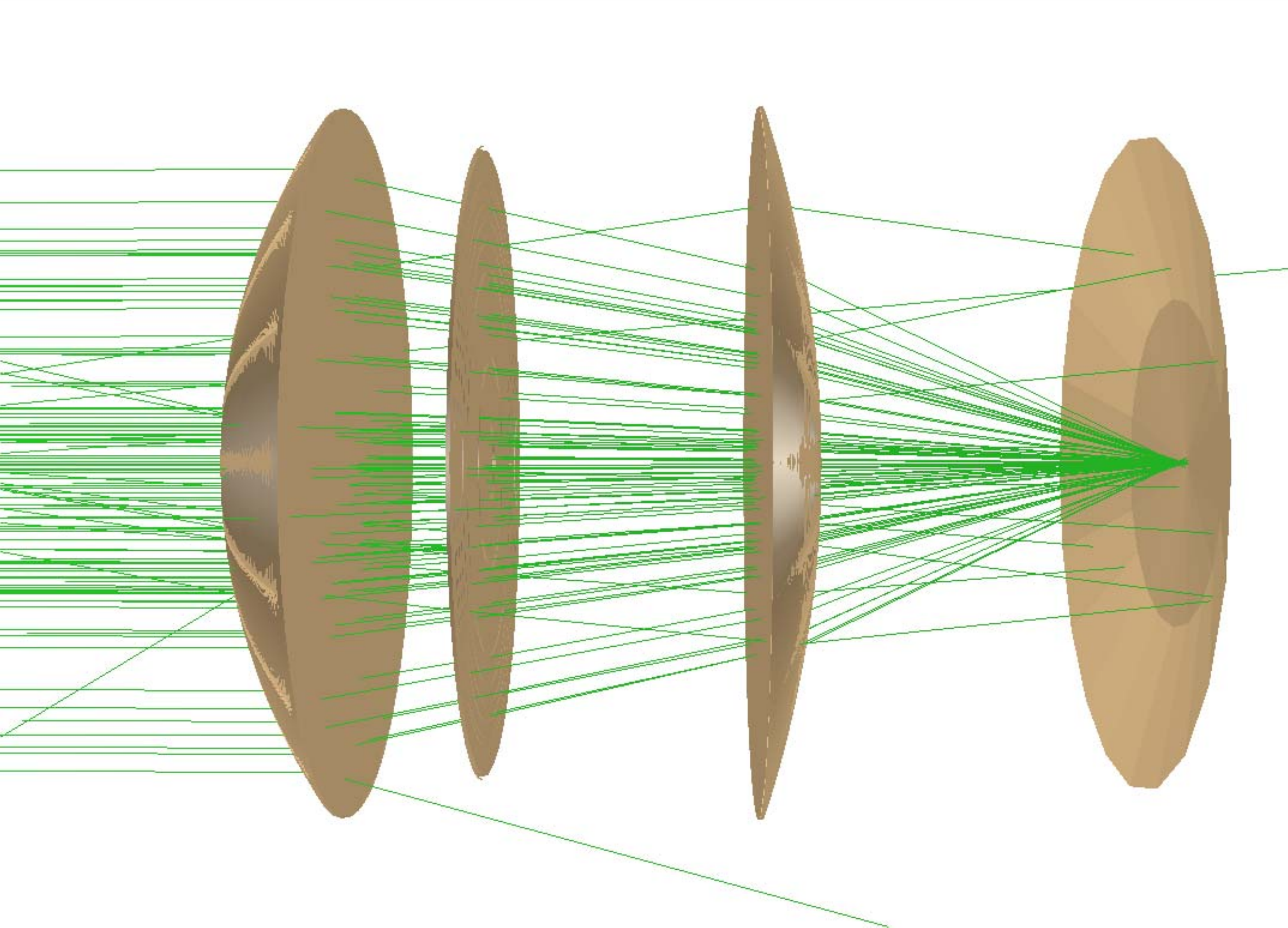} \\
\includegraphics[width=0.5\textwidth]{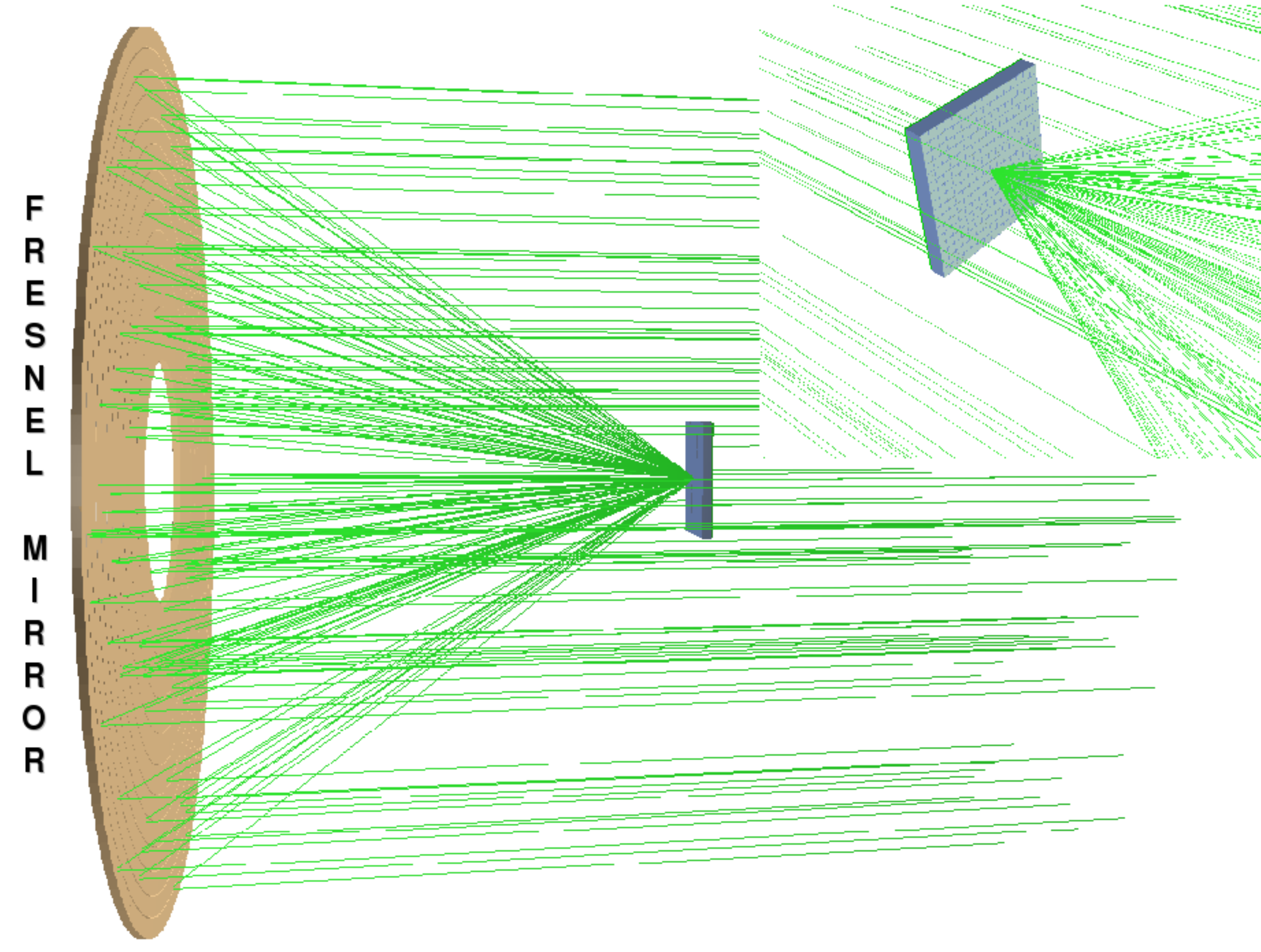}
\end{tabular}
\caption{EUSO Fresnel lens (top) an TUS Fresnel mirror (bottom) detectors built with Geant4. 
Fluxes of photons focused on their respective focal planes are also displayed.} 
\label{fig:g4detectors} 
\end{figure}

A detector simulation chain with the possibility to trace particles through the detector using the GEANT4 package~\cite{bib:geant4} has been implemented in ESAF. We have kept the original way of ray-tracing for checks and compatibility. This upgrade makes ESAF much flexible for a simulation of any detector and not necessarily an EUSO-like detector. 
As two examples we implemented an EUSO like detector with two Fresnel lenses and a diffractive surface between them and a TUS-like detector~\cite{bib:TUS} with a Fresnel mirror. Both examples are shown in Fig.~\ref{fig:g4detectors}.

\begin{figure}[p]
\centering
\includegraphics[width=0.5\textwidth]{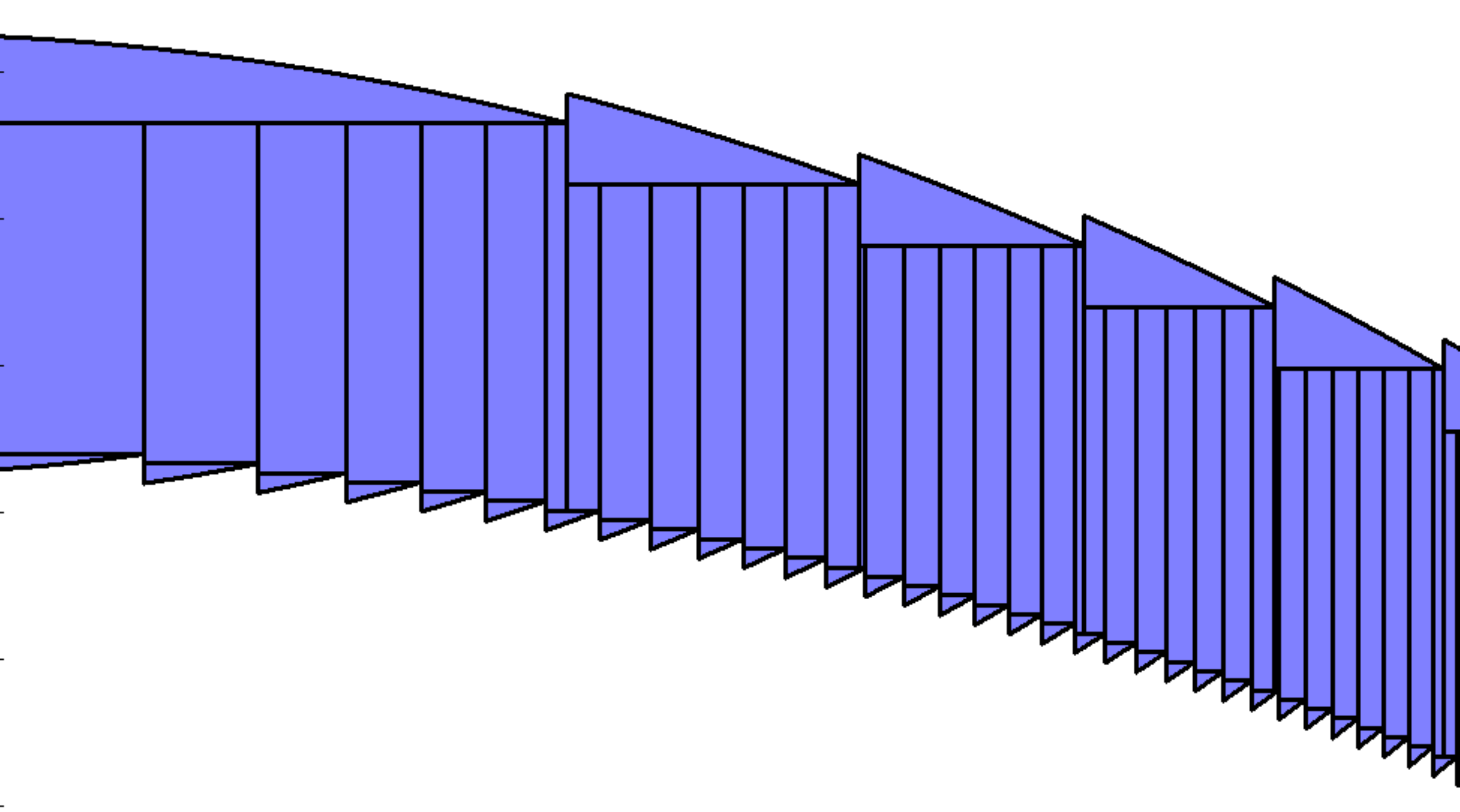} 
\caption{\label{fig:lens_geom1} A piece of a EUSO Fresnel lens simulated with help of GEANT4} 
\end{figure}

Almost all segments of EUSO Fresnel lenses have parabolic surfaces. Every segment is done with boolean operations on the basic shapes (\texttt{G4Tubs} and \texttt{G4Paraboloid}). More complex surface shapes were approximated by cones.
Approximation precision can be chosen on demand. For the current simulation it is equal \mbox{0.1 $\mu$m}. A piece of a EUSO Fresnel lens implemented with GEANT4 is shown in Fig.~\ref{fig:lens_geom1}.

The following standard optical processes are defined in the GEANT4 implementation: Fresnel refraction, 
Fresnel reflection, total internal reflection and absorption. In addition to the standard optical processes, 
ESAF uses a ''special`` process parametrisation: the diffraction on the second Fresnel lens.
In fact, the second lens of the EUSO detector has two surfaces, one with Fresnel structure 
and the second one with a normal diffractive surface. The role of the diffractive
surface is to reduce the chromatic aberrations.
For the simulation of the diffraction process on the second lens we use a parametrization. The momentum direction correction depends on the material refraction index and on transverse component of position on the lens surface. To implement it in the Geant4 simulation we have created a special class that inherits from G4VDiscreteProcess. 
We have added a thin vacuum layer just after the flat lens surface, and have assigned proper material parameters. 
Thus, rays going out of the lens are affected by ordinary boundary processes, and, if not reflected back, they go through that thin vacuum layer and are therefore handled by the new class.

The GEANT4 implementation of the simulation was not used in the work described in this paper. 
For this reason, we do not give a detailed account of GEANT4 - ESAF capabilities. 
However this capability will be very important for future detector simulations and it will be
documented in a future work.

\subsubsection{Electronics and trigger}
\label{sec:elec-trigger}

The electronics simulation converts all the PMT signals into front-end electronics (FEE) counts, simulates the front-end chip response 
and applies the trigger patterns.

In the front-end the  PMT signals are sorted by time, divided in time units called Gate Time Units (GTUs) and 
counted. In the baseline configuration of EUSO the duration of a GTU is $2.5\;\mu\text{s}$. An amplitude threshold and the double hit resolution of the front-end chip are taken into account 
as well, according to the specifications of the EUSO Phase A design report~\cite{bib:Euso}. The electronics has a typical efficiency of about 90\%.

The atmospheric background contribution is added directly after the front-end simulation as additional FEE counts. The motivations for this solution are described in detail in section~\ref{sec:RBSimu}.

The trigger in ESAF is implemented as a set of 'Trigger Engine' objects, one for every trigger algorithm. 
Several engines can be applied to the event, giving the possibility to compare their efficiency on an event-by-event basis.
The specific trigger algorithm used for the studies presented in this paper is described in the section~\ref{sec:CTT}.

\subsubsection{The level-1 trigger algorithm: Tracking}\label{sec:CTT}

The image of a developing EAS on the focal surface is a bright spot moving on a straight line whose brightness rises up to the maximum and gradually fades away. At a given time, the spot is mainly contained in a single pixel, as required by the optics design.
Integrated over the development, the shower appears as a track in space and time whose length and brightness depend on the primary energy and incident angle but also on the EAS position in the \FOV.

In a space-based experiment, only those EAS that produce a track on the focal surface, standing out from the surrounding nightglow background, can be detected and provide any useful information. Without the track structure visible, the EAS signal is not distinguishable from the background fluctuations.

Therefore two conditions can be established for an EAS to be detectable:
\begin{description}
\item[Brightness:] The signal in a given pixel and GTU must be well above the average background counts.
\item[Contiguity:] The bright, or active, pixels must be contiguous in space and time.
\end{description}

At pixel level, to set a ``brightness'' requirement, a background dependent threshold $\mathcal{N}$ can be established: 
\begin{equation}\label{eq:CTTThreshold}
	\mathcal{N}(k) = n_\mathrm{b} + k \sqrt{n_\mathrm{b}} \virgola
\end{equation}
where $n_\mathrm{b}$ is the mean
number of background counts per pixel per GTU, and $k$ a coefficient depending on the requested background
rejection efficiency. Since the random background is a poissonian distribution with mean $n_\mathrm{b}$, $k$ can be regarded as a number of standard deviations. This parametrized threshold is intrinsically adaptable: for $k$ fixed, $n_\mathrm{b}$ can be continuously measured and the threshold can be automatically updated. 

The contiguity criteria requests that the signal must form a track in space and time. Starting from the first active pixel, the next active one in
the next GTU must be the same pixel or one of the nearest neighbors. The minimum number of consecutive active
pixels, the track length $L$, can be assumed as the second requirement a signal must satisfy to be recognized as
a possible EAS candidate.

We have studied the trigger efficiency using the contiguity tracking trigger (CTT) developed during the EUSO
Phase A. It works at the Elementary cell (144 pixels) front-end level, that is an ensemble of 2$\times$2 closely packed multianode
PMT (36 channels each). 
The CTT is a simple algorithm that checks the signal contiguity in space and time, as explained before. Thought to be integrated in the front-end chip as first level trigger, it can be implemented using purely
combinatorial circuits with low cost in term of power and mass budget.
Due to the large background non-uniformities, $\mathcal{N}$ cannot have the same value for every channel on the focal
surface. Instead, we used $k$ as one of the variables of the trigger. In the following we shall denote a particular CTT
configuration with the notation $[L;k]$.

The rate of fake triggers (over the whole detector) due to the random background is easy to compute~\cite{bib:tesi-thea,bib:tesi-pesce}: 
\begin{equation}\label{eq:OverallFTR}
	\nu_\mathrm{FTR}\left[L;k\right] \approx \frac{N_\mathrm{cell}P(L;n_\mathrm{b};\mathcal{N}(k)) }{t_\mathrm{GTU}} \virgola
\end{equation}
where $N_\mathrm{cell}$ is the number of elementary cells (1386 in the EUSO Phase A configuration), $t_\mathrm{GTU}$ is the duration of a gate time unit and 
\[
P(L;n_\mathrm{b};\mathcal{N}(k))\approx 9^{(L-1)}\cdot \left[ 1 - \sum_{i=0}^{\mathcal{N}(k)-1}{\frac{\mathrm{e}^{-n_\mathrm{b}}n_\mathrm{b}^i}{i!}} \right]^L
\]
is the poissonian probability to have a random track of length $L$ with a pixel threshold $\mathcal{N}(k)$.

The maximum acceptable overall fake trigger rate is constrained mainly by the available telemetry resources.
Thus, the maximum fake CTT rate is then limited by the time the second level trigger further processes the data.
The final number depends of course on the details on the second level trigger, but a realistic estimation, based on the
current technology, sets this limit to $\approx\un[100]{Hz}$.

\subsection{Simulation of the random background}\label{sec:RBSimu}

The main source of background to the EAS signal is the atmospheric nightglow (see~\cite{bib:nightglow} and references therein).
Other sources like moon light, lightenings and man-made lights  are not constant in time. Currently, ESAF is able to simulate only the effect of an uniform background. This is not a major issue since the main interest is to study the detector capabilities with various levels of average background. The mean value of the background radiance used in this work is \mbox{$ B \approx \sci{ 0.5^{+0.5}_{-0.2} }{12}\; \un{photons \; m^{-2} \; s^{-1} \; sr^{-1}} $}, as measured by different experiments (e.g. BABY~\cite{bib:baby}, TATIANA~\cite{bib:tatiana}).

Assumed uniform and isotropic on the EAS development timescale, it floods the entrance pupil with a constant flux of photons. The resulting flux on the focal surface is not uniform, because of the optics effects. The average background counts per pixel per GTU, \BkRatePix, varies from the focal surface center to the border with a profile that depends on the optics used. 
	\BkRatePix sums up the contributions of photons from all incoming directions and therefore the stray-light (see Fig.~\ref{fig:PSF}) cannot be neglected.
	\begin{equation}\label{eq:BRatePixMC}
		\BkRatePix \left( \vec{x} \right) = \rho_\mathrm{b} \left( \vec{x} \right) \PDE d^2 \virgola
	\end{equation}
	where $\vec{x}$ is the pixel position, $\rho_\mathrm{b}$ the number of background photons per surface unit, \PDE the photo-detection efficiency and $d$ the pixel side.
	The distribution $\rho_\mathrm{b}\left( \vec{x} \right)$ is obtained by simulating several million photons with an uniform incident direction distribution, according to the background radiance. For such a purpose ESAF has a special running mode where only the optics is loaded, optimizing the memory and CPU consumption.

Once this has been done, and accordingly with the user choice, ESAF can simulate the random background over the whole focal surface only in those elementary cells that have at least one pixel with signal. In each pixel the simulation adds a random number of background counts for every GTU, following the Poisson distribution with mean \mbox{$n_\mathrm{b} = \BkRatePix t_\mathrm{GTU}$}, where $t_\mathrm{GTU}$ is the GTU time length.

\section{Shower reconstruction}\label{sec:reco}

After the observation of one event, the following questions should be addressed: what are the direction, the energy and the nature (nucleus, gamma, neutrino, $\ldots$) of the cosmic ray at the origin of the detected EAS ?

ESAF has a reconstruction package which consists of special modules aiming to recover these informations from an event. Let us consider below the underlying principles of the reconstruction software which consists essentially of the following modules:
\begin{description}

\item[Pattern recognition:] the goal of this module is to find the ``shower track'', i.e. a set of spatially 
correlated pixels on the focal surface that describe a moving point as a function of time (the projection 
of the EAS development on the focal surface).  
The main problem here is to disentangle the pixels that contain some signal to those that are pure 
background.

\item[3D Direction reconstruction:]  the goal of this module is to find the direction of the primary 
particle in space.
The 3-dimensional reconstruction is possible by using both the direction of the projected track on the
focal surface and the arrival time of the photons. 

\item[$\boldsymbol{\Xmax}$ and Energy reconstruction:] the goal of this module is to find the altitude in the atmosphere where the EAS reaches its maximum of development and the corresponding atmospheric slant depth, and to reconstruct the energy of the primary particle. 
The fluorescence light emitted by an EAS is proportional to the deposited energy. However, the collected light is not 
because there are  important corrections that depend on direction, position on the focal surface, 
and detector efficiency. 
The \Xmax and its variation with the energy are very important for primary particle identification.
Since the energy and \Xmax are correlated, it is not possible to infer a precise determination of each variable
in an independent way.
Therefore, this module reconstructs the best estimate of the energy and of the \Xmax.

\end{description}

\subsection{Pattern recognition}\label{Sec:PattReco}
Once a shower track has been triggered by the apparatus, the pattern recognition
module selects, within every considered time interval 
(GTU), the
pixels containing signal from the pixels containing background only.
Since the background light is expected to be randomly distributed both in direction and time,
the space-time correlation in the focal plane is the key tool to select clusters of pixels which 
belong to the shower track on the focal surface.

The pixels selection is implemented in ESAF by a simple multi-step analysis which, within the focal plane,
searches for clusters of pixels in space and time separately. A following module also analyzes the
X versus time and Y versus time correlations inside the focal plane by means of a linear fit, followed
by the exclusion of the points far away from the fitted line.
A more sophisticated method using similar concepts, but based on the Hough transform 
technique~\cite{bib:hough-transf} is also successfully implemented in ESAF.

\subsection{Direction reconstruction}\label{sec:DirectionReco}

Let us introduce a standard reference system (right-handed) with the origin in the center of the detector focal surface
and the $z$-axis pointing towards the nadir of the instrument (see Fig.~\ref{fig:track_view_Cherenkov}).
\footnote{We are assuming that there is no tilt between the nadir and the orientation of the instrument optical axis.}

The EAS axis is described in this reference system by an unit vector $\boldsymbol{\Omega}$ in terms of
two polar angles $\theta$ and $\varphi$:

\begin{equation}\label{eq:EASdirection}
	\boldsymbol{\Omega} = (\sin\theta \cos\phi, \sin\theta\sin\phi, \cos\theta) \punto
\end{equation}
In an analogous way we define the unit vector of the direction that is seen by the $i$-th front-end pixel in the FoV:
\begin{equation}\label{eq:FOVdirection}
	\mathbf{n_i} = (\sin\theta^i_\mathrm{fov} \cos\varphi^i_\mathrm{fov}, 
	\sin\theta^i_\mathrm{fov}\sin\varphi^i_\mathrm{fov}, \cos\theta^i_\mathrm{fov}) \punto
\end{equation}

Therefore, for each selected pixel on the focal plane there is an associated direction in space where the
fluorescence light activating such pixel is expected to come from.\footnote{For any detector configuration, a map that associates each pixel with a unique direction in the FoV is built simulating 
an uniform photon flux in the FoV through the detector.} 

Such directions in space are expected to
belong, within 
instrument resolution and neglecting the shower lateral extension, to the plane
containing the shower axis and the detector itself. Using the information of the light arrival
time in each pixel, it is then possible to look for the unique track in space associated to the
measured direction-time pattern. Such pattern depends strongly on the direction of the shower
axis in the detector-shower plane, but very weakly on the distance of the shower maximum to the
detector because of non-proximity effect. This is one of the advantages of space-based observation: 
the altitude of the shower maximum may vary up to 20 km, depending on the shower inclination, but
it is always a small distance compared to the satellite typical altitude of 400 km, which means that the 
collected light changes at most by 10\%. 

With a monocular detector, it is not possible to determine by means of
a fitting procedure the detector-shower distance, unless the resolution is so good to allow
a measurement of the spatial width of the focal surface track. 
The shower detector distance can be  simply fixed to the expected average value for each zenith angle of the incoming shower.
However, such an information can be inferred by the arrival time of the Cherenkov light reflected on 
the ground, if it is available, since it is strongly correlated to the distance of the shower from Earth and 
detector. 
With the detector-shower distance inferred by such procedure, the fit of the direction-time pattern yields a better determination of the shower-axis direction inside the shower-detector plane. 
In case the Cherenkov light is not detected, the detector-shower distance can be inferred by the spatial 
width of the fluorescence light profile (see section~\ref{sec:hmax}). In both cases, the determination of the detector-shower distances and of the zenith angle is made more precise using an iterative reconstruction process.

The first step that leads to the determination of the EAS direction is the reconstruction of the \emph{Shower-Detector Plane} (SDP), i.e. the ideal plane that contains both the EAS track and the detector. The SDP can be reconstructed by ESAF fitting the ``moving'' EAS image on the planes $x$-time and $y$-time. In this way one obtains a projection of the shower track on the plane
tangent to the Earth surface at the telescope nadir point. The SDP is then the plane that contains this projection and the detector itself.

The second step is the fit of the track on the SDP in order to infer the direction of the shower axis.
In ESAF two different classes of track fitting are implemented~\cite{bib:taddei}. 

\begin{description}
\item[Numerical Exact (NE)] For each signal FEE count we compute its expected arrival time, 
$\hat{t}_i(\mathbf{n_i},\boldsymbol{\Omega},\boldsymbol{\alpha})$, 
as a function of the EAS axis direction (unknown), the pixel direction in the FoV (see above) and other 
free parameters\footnote{These free parameters depend on the particular track fitting algorithm.
They can be, for example, the position in FoV, the time and the distance from the detector of the EAS maximum.}, $\boldsymbol{\alpha}$. 
These times can be exactly calculated using only geometrical and physical relations. 
Since we know the experimental arrival times, $t_i$, we can minimize the $\chi^2$ function
\[
   \chi^2 \left( \boldsymbol{\Omega},\boldsymbol{\alpha} \right) = 
   \sum_i \frac{\left(t_i - \hat{t}_i(\mathbf{n_i},\boldsymbol{\Omega},\boldsymbol{\alpha}) \right)^2}{\sigma^2_i}  \virgola
\]
where the $\sigma_i$ are the uncertainties on the measured arrival times, via numerical minimization methods.
From the minimization results we know the EAS direction.
	\item[Analytic Approximate (AA)] In this case we use approximate formulas, e.g. assuming that the EAS track has a constant angular speed on the focal surface, that can be analytically solved using linear fit procedures.
\end{description}

\subsection{ $X_\mathrm{max}$ and Energy reconstruction}\label{sec:hmax}
The principle of the method described in this section is to assume a functional form $F_\mathrm{long}$ for the longitudinal development of the shower, to calculate the expected number of photoelectrons at the telescope as a function of time and to vary the parameters of the shower function $F_\mathrm{long}$ until an agreement with the measured signal is obtained.
From these parameters, one can deduce the energy and \Xmax of the EAS.

To calculate the number of photons emitted along the track and the fraction of them that reach the telescope, one has to know the shower track position in space and the local atmospheric parameters.
Moreover to obtain the expected signal, one has to estimate the detector efficiency. 

\begin{figure}[htb]
\centering
\includegraphics[height=0.5\textwidth,angle=+90]{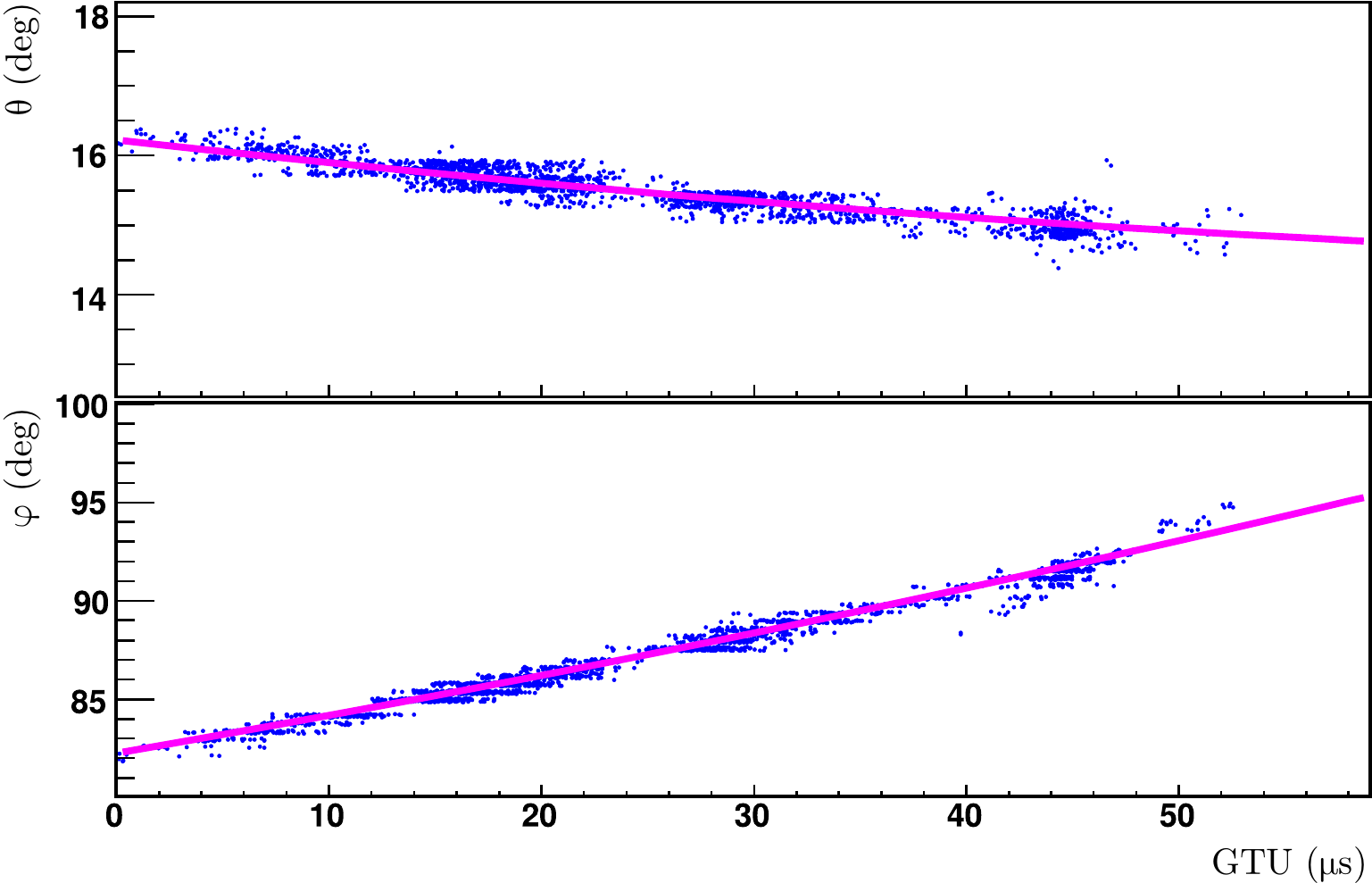} 
\caption{\label{fig:theta-phi-vs-time} An example: $\theta$ vs time (upper plot) and $\phi$ vs time (bottom plot)
(each GTU is 2.5 $\mu$s).} 
\end{figure}

In what follows $\theta_\mathrm{fov}$ and $\phi_\mathrm{fov}$ as functions of time are needed to fix the direction in space where the fluorescence light is expected to come from as a function of time. These functions are obtained by fitting $\theta_\mathrm{fov}$ and $\phi_\mathrm{fov}$ of each active pixel with presumably a signal by a linear function of time. An example of such fit is shown in Fig.~\ref{fig:theta-phi-vs-time}.

\subsubsection{Background and signal estimations}\label{sec:background}
A cluster of pixels  belonging to the shower track is taken from the pattern recognition module to assign for each time stamp a list of pixels presumably enriched by the shower signal. 
For the background study, a similar second assigning map is created which contains the pixels around the {\em signal cluster} within a fixed margin, corresponding, in an EUSO like detector, to the size of two adjacent pixels.
For each pixel of the track, the background can be estimated in two complementary ways: either by looking at the light collected by the adjacent pixels at the same time, or by looking at the light collected by the very same track pixels before and after the shower signal. 
One obtains the observed time profile of the signal by adding at each time stamp the signal from the shower pixels after subtraction of the background. 

\subsubsection{Optics efficiency}\label{sec:efficiency}
\begin{figure}[htb]
\begin{tabular}{c}
\includegraphics[height=0.5\textwidth,angle=-90]{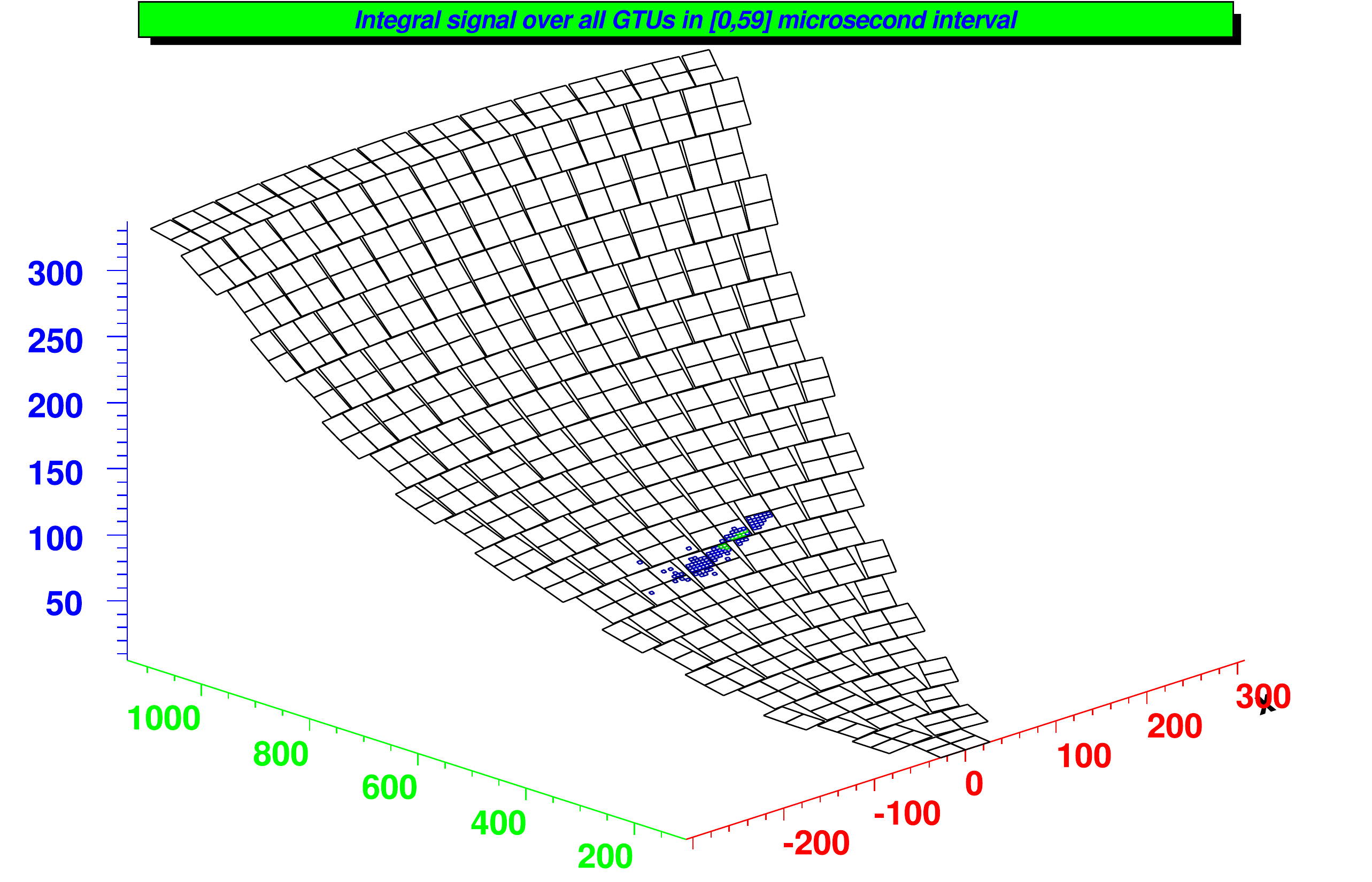} \\
\includegraphics[height=0.5\textwidth,angle=-90]{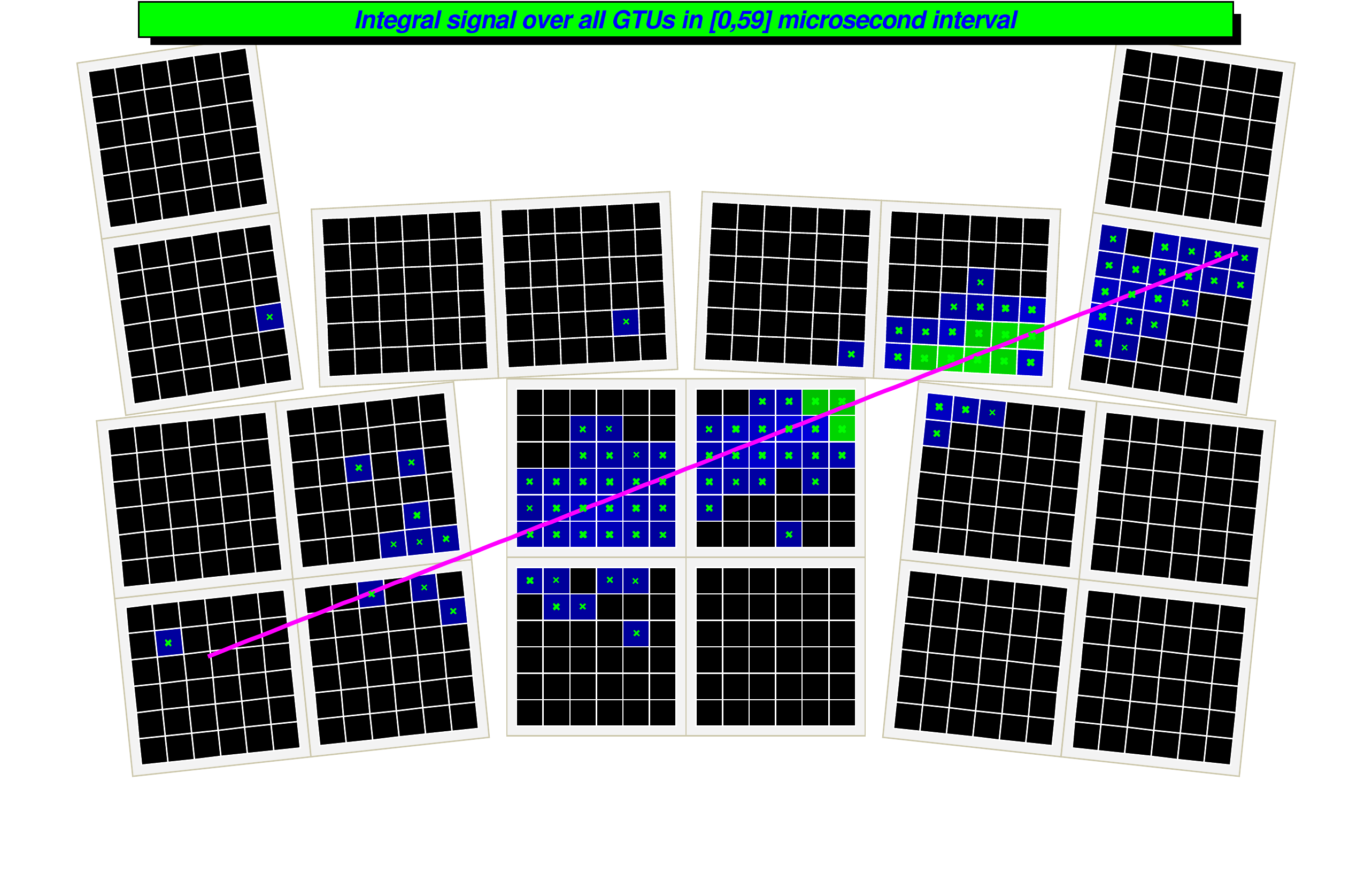} \\
\end{tabular}
\caption{\label{fig:focal_surface} An example of distribution of photoelectrons on the focal surface for a shower energy of 5. 10$^{20}$ eV and zenith angle $\theta$=60$^\circ$. Upper plot: large scale; bottom plot: zoomed picture, in a region where the dead space among PMTs modules is not negligible.} 
\end{figure}

The focal surface may have a significant amount of holes which distort the detected profile as it is e.g. the case for the polar design (see Fig.~\ref{fig:EusoFS}b) of an EUSO like detector. Fig.~\ref{fig:focal_surface} demonstrates an example of how the shower signal looks like on this type of focal surface; the corresponding distribution of numbers of photoelectrons as a function of time is shown in Fig.~\ref{fig:fit_distr}.  One can notice at least three relatively large and rapid losses of detected photoelectrons (around times 13, 24 and 40 GTU) which correspond to holes between macrocells. The dead spaces between PMTs in one macrocell are smaller in size but also produce distortions in the profile. 

These distortions are taken into account in the shower reconstruction process by building for each arrival time $t_i$ an effective optics efficiency $\varepsilon_\mathrm{eff}(t_i)$. 
The previously reconstructed $\theta_\mathrm{fov}(t_i)$ and $\phi_\mathrm{fov}(t_i)$ of the track are used to compute:

\begin{equation} \label{eq:efficiency_eq}
\varepsilon_\mathrm{eff}(t_i) = \int \varepsilon(x,y|\theta_\mathrm{fov}(t_i), \phi_\mathrm{fov}(t_i)) \Delta(x,y) \diffl{x}\diffl{y} \virgola
\end{equation}

where $\Delta(x,y)$ is the function returning 1 if the considered point with coordinates $x,y$ is inside a pixel, and 0 otherwise. $\varepsilon(x,y|\theta_\mathrm{fov}, \phi_\mathrm{fov})$ is the optics point spread efficacy function on the focal surface at position $x,y$ for photons incident direction determined by $\theta_\mathrm{fov}, \phi_\mathrm{fov}$. The PSF and the optics transmission are taken into account to build the optics efficacy function. The PMT efficiencies are carried separately.

\subsubsection{\Hmax reconstruction}\label{sec:hmax_reco}
As already mentioned in section~\ref{sec:DirectionReco}, the determination of the detector-shower distance  can be estimated in two ways: either by using the Cherenkov echo and the information encoded in the fluorescence signal profile, or, 
when the former is not available, by using  fluorescence signal profile only. The 
latter case is, unfortunately, the most common (see section~\ref{sec:results}).
In both methods, all contributions to the signal other than direct fluorescence and reflected Cherenkov lights are neglected. This allows to obtain a first estimation of the altitude of the fluorescence signal maximum \Hmax. 
In what follows the measured signal profile is corrected at each arrival time $t_i$ by $\varepsilon_\mathrm{eff}(t_i)$ for optical loss and focal surface inefficiencies. 

\paragraph{\Hmax reconstruction with the Cherenkov peak}\label{sec:hmax_with_cer}
This method needs to infer from the time profile of the signal the difference ($\Delta T$) between the arrival time of the Cherenkov peak and that of the maximum of the fluorescence light $t_\mathrm{max}$.  $\Delta T$ is used to estimate $H_\mathrm{max}$.

\begin{figure}[htb]
\centering
\includegraphics[width=0.5\textwidth]{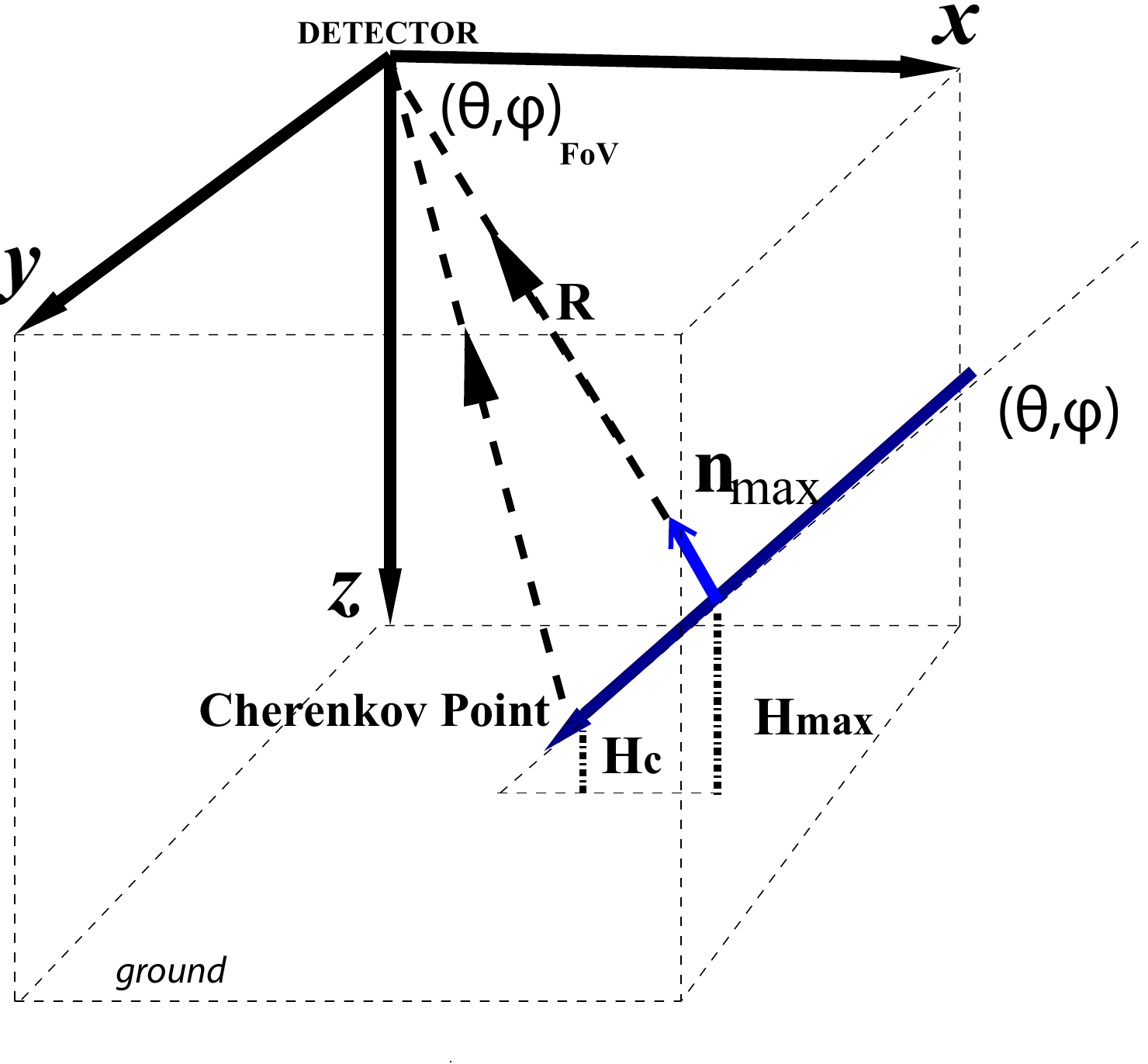} 
\caption{\label{fig:track_view_Cherenkov} Geometry of the Extensive Air Shower seen by the 
detector. The Cherenkov point, when available, gives the position on the impact point
of the shower on the reflecting surface (ground or cloud). } 
\end{figure}

From  Fig.~\ref{fig:track_view_Cherenkov}, one can find that:
\begin{equation}\label{eq:HmaxGolden}
\frac{\Delta \Hmax^\mathrm{C}}{\cos\theta} = \frac{c\Delta T}{2}\frac{c\Delta T + 2 R}{c\Delta T+R(1-\mathbf{n}_\mathrm{max}\cdot\boldsymbol{\Omega})} \virgola
\end{equation}
where:
$$\Delta \Hmax^\mathrm{C} = \Hmax - H_\mathrm{C},$$ and $$R = \frac{H-\Hmax}{\cos(\theta_\mathrm{fov}(t_\mathrm{max}))}$$
with $H_\mathrm{C}$ the known altitude of the reflected Cherenkov surface, $H$ the telescope altitude, 
and the previously reconstructed parameters:  $\boldsymbol{\Omega}$ the shower direction (with zenith angle $\theta$) and $\mathbf{n_\mathrm{max}}$ the unit vector pointing to the telescope and defined by $\theta_\mathrm{fov}(t_\mathrm{max})$ and $ \phi_\mathrm{fov}(t_\mathrm{max})$.

\paragraph{\Hmax reconstruction without the Cherenkov peak}\label{sec:hmax_without_cer}
The altitude of the EAS maximum can be reconstructed relying on the fluorescence light information only. The idea of the method~\cite{Shape1}\cite{Shape2} is based on the simple observation that the time width of the EAS development strongly depends on air density (and thus on the altitude) in the atmosphere at which EAS develops.

A longitudinal profile of the number of shower electrons at given atmosphere depth $X$ can be 
conveniently described by an analytic function $f(\xi)$ which depends on a dimensionless parameter
$ \xi = \frac{X-X1}{X_0}$ as 

$$
\frac{n_e(\xi)}{n_{e,\mathrm{max}}} = e^{f(\xi)}
$$                          

where $X_1$ is the depth of the atmosphere at which EAS starts to develop and $X_0$  is the air 
radiation length. An example of $f(\xi)$ is given in section~\ref{sec:showers} for the GIL parametrization, while the 
consideration below is more general. The EAS profile can be expressed as a function of detection 
time $t$. Around the EAS maximum one can write

\begin{align*}
f(\xi) & = \frac{f^{\prime\prime}(\xi_\mathrm{max})}{2\!}\left(\xi-\xi_\mathrm{max}\right)^2 \\
    & = \frac{f^{\prime\prime}(\xi_\mathrm{max})\xi^{\prime 2}(t_\mathrm{max})}{2\!}\left(t-t_\mathrm{max}\right)^2 \\
   & = -\frac{\left(t-t_\mathrm{max}\right)^2}{2\sigma^2}.
\end{align*}

In this way we have obtained a gaussian distribution of times around $t_\mathrm{max}$ with standard deviation $\sigma$.
In order to calculate $\sigma$, we first obtain $\diffl{\xi}/\diffl{t}$ taking into account the following:
$$
\diffl{\xi} = \frac{\diffl{X}}{X_0} = \frac{\rho(h) \diffl{l}}{X_0} = \frac{\rho(h)}{X_0}\frac{c \diffl{t}}{1-\mathbf{n}\cdot\boldsymbol{\Omega}}
$$

Thus, $\sigma$ is expressed by: 

\begin{equation}
\label{eq:sigma_shapemethod}
\sigma = \frac{X_0\left(1-\mathbf{n_\mathrm{max}}\cdot\boldsymbol{\Omega}\right)}{\rho(\Hmax^\mathrm{fluo}) c \sqrt{-f^{\prime\prime}(\xi_\mathrm{max})}}
\end{equation}
This general equation can be simplified further for the case of GIL parametrization for which \mbox{$f^{\prime\prime}(\xi_\mathrm{max})= -1/2\,\xi_\mathrm{max}$} and the time width becomes
\begin{equation}
\label{eq:sigma_shapemethod_GIL}
\sigma = \sqrt{2 \xi_\mathrm{max}}\frac{X_0\left(1-\mathbf{n_\mathrm{max}}\cdot\boldsymbol{\Omega}\right)}{\rho(\Hmax^\mathrm{fluo}) c}
\end{equation}
with $\xi_\mathrm{max}$ only logarithmically depending on UHECP energy which can be easily taken into account in an iterative procedure or fixed to a mean expected value. 

Because the fluorescence yield is almost constant between 0 and 20 km 
(it varies between 4.1 and 4.7 photons/m, see \ref{sec:fluo-light-simu}) and the atmospheric 
attenuation is a slowly varying function along the shower track, the shape of the shower 
electrons time profile is at first order similar to that of the total collected signal. This way one 
could measure $\rho(\Hmax^\mathrm{fluo})$ and thus $\Hmax^\mathrm{fluo}$ of the shower maximum using observable informations $\sigma, \mathbf{n_\mathrm{max}}$ and reconstructed $\boldsymbol{\Omega}$. 

In principle, when the altitude of the Cherenkov scattering surface is not known, this second method may allow to recover it just as:  
$$
H_\mathrm{C} = \Hmax^\mathrm{fluo} - \Delta \Hmax^\mathrm{C}
$$

\subsubsection{Profile reconstruction}\label{sec:prof_reco}
The aim of this part of the reconstruction is to estimate the atmospheric depth profile of the shower electrons from the light flux observed by the telescope. 

A functional form $F_\mathrm{long}$ for this profile is assumed, a possible choice is the GIL function~\eqref{eq:GIL}, which has three free parameters: E the energy of the shower, \Xmax the atmospheric depth where the number of shower electrons reach its maximum and $X_1$ the depth of the first interaction. $F_\mathrm{long}$ gives the $N_i^{e}$ electrons of the shower as a function of the slant depth along the shower track. In a first estimation of  $E$, $X_1$ and $X_\mathrm{max} $ can be set to their mean expected value.

For each arrival time $t_i$ the expected signal at telescope  $ \hat{S}_i $ is computed as a function of the number of shower electrons (unknown) 
to minimize the following $ \chi^2$ function via numerical minimization methods.
$$
   \chi^2 \left(  E,X_\mathrm{max},X_1 \right) = 
       \sum_i \frac{\left(S_i - \hat{S}_i(E,X_\mathrm{max},X_1) \right)^2}{\sigma^2_i}  \virgola
$$ 
where $S_i$ is the measured signal (background subtracted) at time $t_i$ and $\sigma_i$ is the uncertainty on it.

Direct fluorescence signal observed at time $t_i$ is produced at a slant depth $X_i$ in the atmosphere. Once a reference in altitude and $\boldsymbol{\Omega}$ are reconstructed, the EAS track in the atmosphere can be found unambiguously and $X_i$ computed as an integral along the EAS track from $P_i$, the shower position at time $t_i$, to top of atmosphere: 
$$
X_i = \int \nolimits_{P_i}^\mathrm{t.o.a.} \rho(h)\diffl{l},
$$
this requires to know the precise density profile of the atmosphere for each event. 

For each time stamp the number of  fluorescence photons $N_i^f$ produced in a slant depth interval $\Delta X_i$ by the $N_i^{e}$ electrons of the shower is computed using the following formula:    

$$
N_i^f(E,X_\mathrm{max},X_1) = 
$$
$$
N_i^{e}(E,X_\mathrm{max},X_1) Y_i  \alpha_i(X_\mathrm{max},X_1) \Delta X_i.
$$
with 
$$
\Delta X_i=\frac{\rho_i c  t_\mathrm{GTU}}{1-\mathbf{n_i}\cdot\boldsymbol{\Omega}}
$$
where $\rho_i$ is the atmospheric density at position $P_i$, $Y_i$ is the local fluorescence yield (number of photons per unit of deposited energy) for atmospheric conditions at point $P_i$, $\alpha_i$ is the average energy deposit per unit depth per electron at shower age $s_i(X_\mathrm{max},X_1)$ (a parametrization of $\alpha_i(s)$ can be found in~\cite{bib:Nerlig} or set to its mean value of 2.2 MeV/(\gcmsq) ), and $t_\mathrm{GTU}$ is the length of the time unit. 

The fraction $T_i$ of photons produced at $P_i$  that reach the telescope due to atmospheric attenuation and the distance $R_i$ between the shower point and the telescope are also computed. Thus the expected direct fluorescence signal is expressed by:
$$
\hat{S}_i^f(E,X_\mathrm{max},X_1)=
$$
$$
N_i^f(E,X_\mathrm{max},X_1) T_i \varepsilon_\mathrm{eff}(t_i) \frac{A}{4\pi R_i^2},
$$
where $\varepsilon_\mathrm{eff}(t_i)$ is the previously computed detector efficiency at time $t_i$ defined in~\eqref{eq:efficiency_eq} (see section~\ref{sec:efficiency})
and A the telescope aperture. 

Although this has not yet been implemented inside ESAF a more precise reconstruction process would require to compute the expected reflected Cherenkov signal, the Cherenkov scattered signal and the fluorescence scattered signal, but at first order the expected total signal can only be expressed by $\hat{S}_i^f(E,X_\mathrm{max},X_1)$.

Since the reconstructed shower track depends on the reference in altitude of the shower 
maximum, the fit is done iteratively until a convergence is obtained. Typically 2-3 iterations are enough to get a stable fit result.

\begin{figure}[htb]
\centering
\includegraphics[height=0.45\textwidth,angle=90]{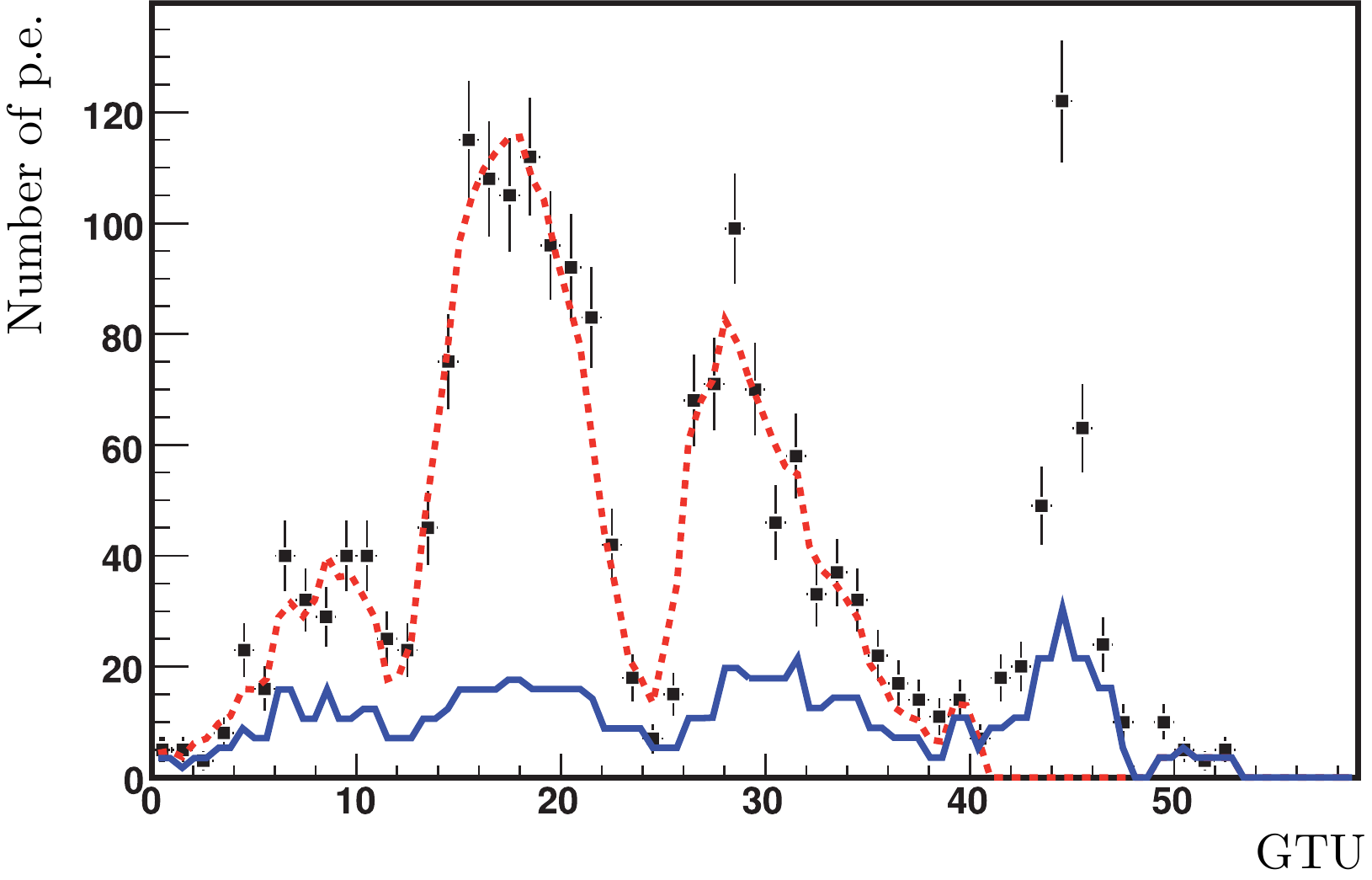} 
\caption{ An example: Fit of temporal distribution. Points with error bars are observed number of p.e., blue curve - estimated background, red curve is the sum of fluorescence and background. The Cherenkov echo seen in the last bins of the temporal distribution was identified by the code and excluded from the fit. Energy and direction are the same of Fig. \ref{fig:focal_surface}} 
\label{fig:fit_distr}
\end{figure}

An example of the fit is shown in Fig.~\ref{fig:fit_distr}.

\section{Simulation results}\label{sec:results}

All the simulations discussed in the present section
are based on a standard configuration which derives
from the EUSO Phase A design, i.e. a 30$^\circ$ FoV space-born telescope
installed on board of the International Space Station at a typical orbital height ranging
from 360 to 450 km. 

However, depending on particular cases, the diameter of the entrance pupil is varied.
In order to study the light flux at the pupil, independently of the detector, the size is fixed at 5 m.

When the performances of an EUSO-like detector are involved, the pupil diameter is fixed at the standard EUSO diameter of 2.5 m. In the trigger efficiency studies different pupil diameters have been tested.

\subsection{Light flux at the pupil in clear sky conditions}\label{sec:pupil}
Results presented in this section concern light flux arriving on a 5 m pupil diameter. 

Simulation results were obtained from showers parameterized with the GIL formula and a fluorescence yield based on the Kakimoto model completed with Bunner rays. The reference detector altitude is assumed to be 430 km and the atmosphere model is a US-Standard clear sky. The Earth ground is assumed to be a lambertian surface, with an albedo of 8\%. Propagation of light is performed with the ``Reduced" Monte-Carlo algorithm.

\begin{figure}[ht]
\begin{center}
\includegraphics[width=0.5\textwidth]{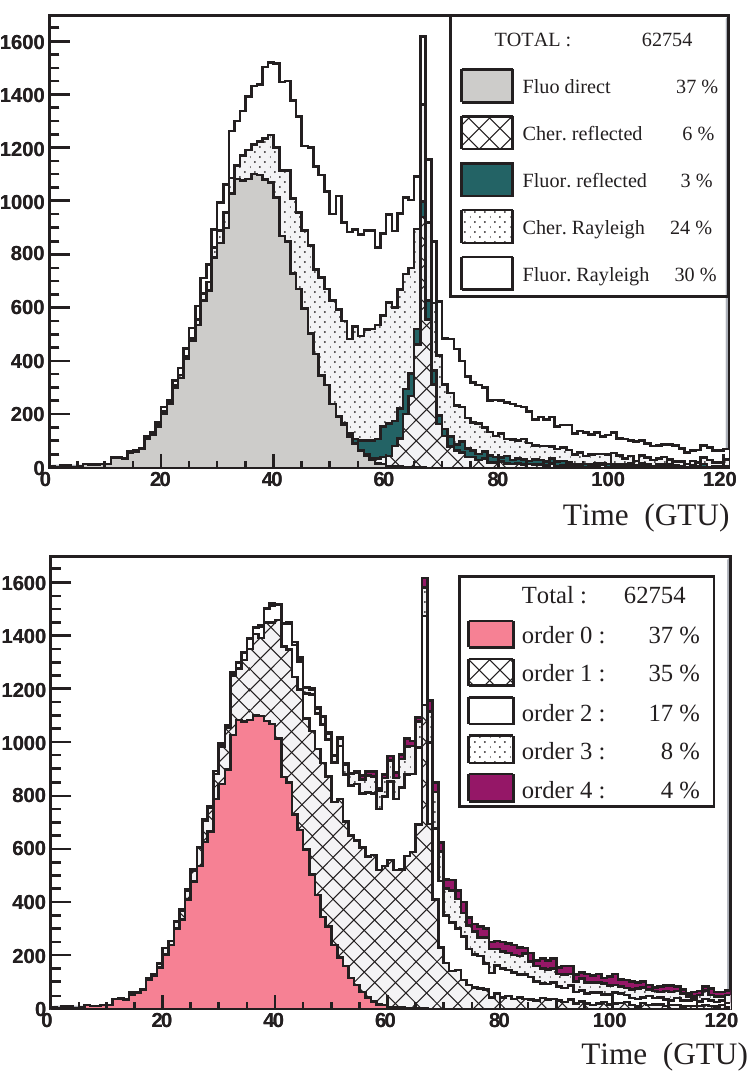}
\caption{Number of photons reaching the pupil as a function of time. The signal is integrated within the whole FoV.
Top histograms refer to the last photon interaction and bottom histograms refer to the number of scattering interactions before reaching the pupil. The simulation has been carried out with 4 orders of scattering in a US-Standard clear sky atmosphere. The energy of the shower is 10$^{20}$~eV and the zenith angle is 60\degr.}
\label{fig:ClearSkySignalOrder4-time}
\end{center}
\end{figure}
\begin{figure}[ht]
\begin{center}
\includegraphics[width=0.5\textwidth]{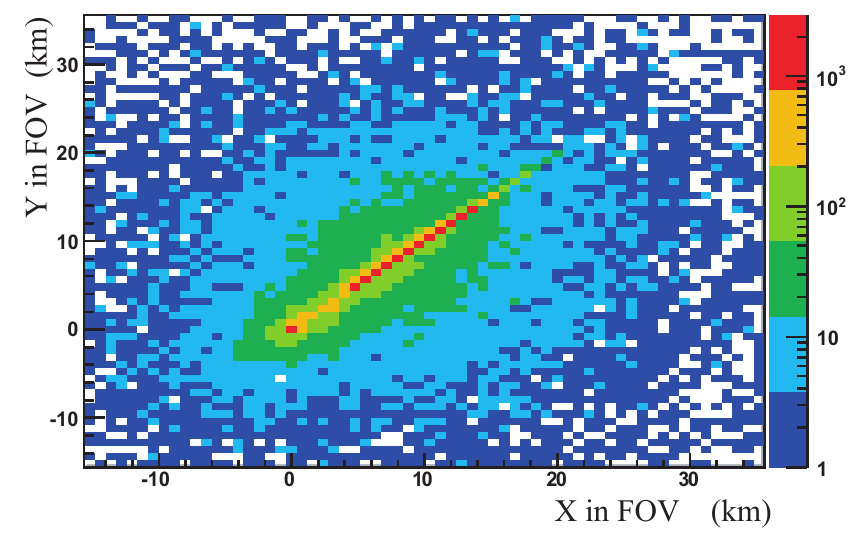}
\caption{Distribution of collected photons positions in the FoV. The simulation has been performed with 4 orders of scattering in a US-Standard clear sky atmosphere. The energy of the shower is 10$^{20}$~eV and the zenith angle is 60\degr.}
\label{fig:ClearSkySignalOrder4-FOV}
\end{center}
\end{figure}

Fig.~\ref{fig:ClearSkySignalOrder4-time} and \ref{fig:ClearSkySignalOrder4-FOV} show an example of the simulated light at the pupil  with the different light components: direct fluorescence, ground reflected lights (fluorescence and Cherenkov), Rayleigh scattered fluorescence and Cherenkov lights. The contribution of the first 4 orders of scattering are also shown. The signal as a function of time has been integrated within the whole FoV. As one can see from Fig.~\ref{fig:ClearSkySignalOrder4-time} the direct fluorescence light contributes only to 37\% of the total signal. This result shows the importance to include the scattered light contamination in the simulation program. The contribution of scattered light decreases by a factor of 2 at each scattering order from 35\% at order 1 to only 4\% at order 4. Although scattered photons are numerous, their positions in the FoV constitute a large halo around the main track as shown in Fig.~\ref{fig:ClearSkySignalOrder4-FOV}.  Besides this, they reach the pupil later than direct signal and thus do not affect the first part of the profile. As a consequence, during the reconstruction process, appropriate pixels selection around the main track in space $x$-$y$ and time will reduce the scattered light contribution as it will be explained later in section~\ref{sec:pupil-diff}.

\subsubsection{Direct fluorescence light}\label{sec:pupil-fluo}
Fluorescence light is related to the shower development in atmosphere, therefore direct fluorescence light is the signal from which the energy and the \Xmax are derived. We report here its main characteristics.

\begin{figure}[ht]
\begin{center}
\includegraphics[width=0.5\textwidth]{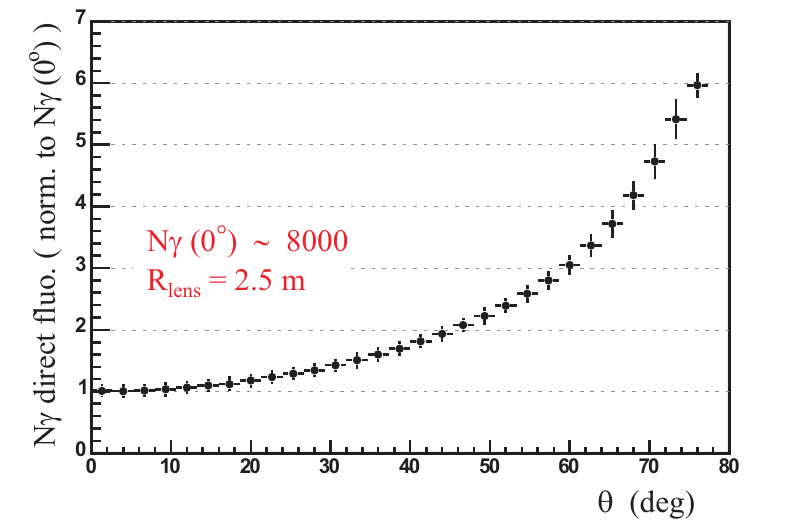}
\caption{Number of direct fluorescence photons at pupil as a function of shower zenith angle for showers at the telescope nadir. Numbers of photons are normalized to the number of photons for a vertical shower, which is around 8000 for a primary energy of 10$^{20}$~eV.}
\label{fig:fluovstheta}
\end{center}
\end{figure}

As underlined in section~\ref{sec:light}, inclined showers produce more fluorescence photons compared to vertical ones. Traveling in a less dense atmosphere these photons are moreover less scattered or absorbed, therefore direct fluorescence light flux at pupil depends strongly on the shower zenith angle. Fig.~\ref{fig:fluovstheta} shows this dependence for showers impacting at nadir. The amount of direct fluorescence light increases by more than a factor 6 between vertical showers and horizontal ones (zenith angle above 75 degrees).

At fixed zenith angle, direct fluorescence light reaching the telescope depends only slightly on the shower position in the FoV.
This weak dependence of signal intensity within the FoV is a specific feature of a space-based detector, due to the large distance between showers and telescope.
Specific studies carried out with ESAF  prove that the  signal from showers at the edge of the FoV is reduced by 40\% compared to the nadir case.

\begin{figure}[ht]
\begin{center}
\includegraphics[width=0.5\textwidth]{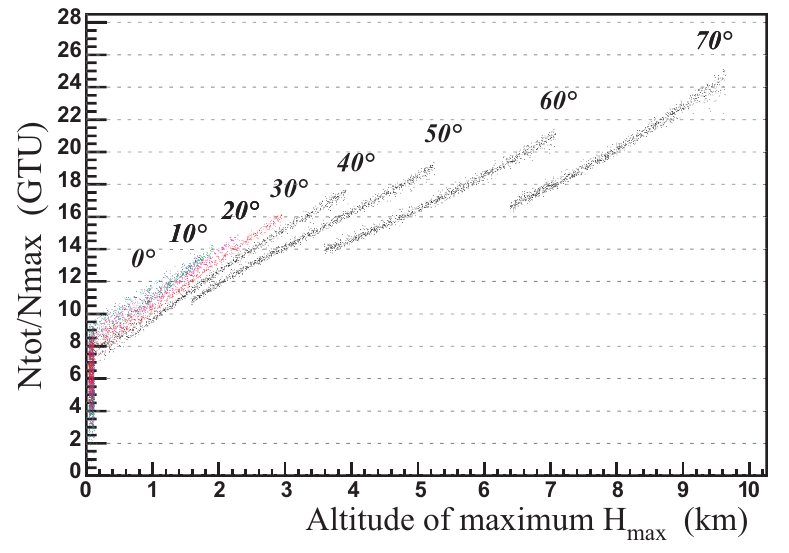}
\caption{Ratio of total over maximum number of fluorescence photons as a function of the shower maximum altitude at different zenith angles. As $N_\mathrm{max}$ is expressed in GTU$^{-1}$, the ratio of $N_\mathrm{tot}$ over $N_\mathrm{max}$ is expressed in GTU. The spread of the curves is due to the random 
first interaction point of UHECP used in the simulation. No correction is applied for the shower distance. }
\label{fig:fluo-length}
\end{center}
\end{figure}

As already mentioned, at higher altitude, where the atmosphere is less dense, the shower is longer than at lower altitude. On the contrary, because the fluorescence yield is almost constant between 0 and 20 km, the number of photons at the maximum depends weakly on the altitude. Therefore at fixed zenith angle the ratio between the total and the maximum number of fluorescence photons is an indicator of the shower altitude as illustrated in Fig.~\ref{fig:fluo-length} and could be used when the reflected Cherenkov signal cannot be observed.

\subsubsection{Ground reflected Cherenkov}\label{sec:pupil-cher}

Cherenkov light that is beamed around shower axis, travels with shower front down to Earth and then back to the detector for the part that is reflected. It provides an extrapolation of shower axis to Earth, therefore the time difference between Cherenkov peak and fluorescence light maximum is an indicator of the shower altitude. 

\begin{figure}[ht]
\begin{center}
\includegraphics[width=0.5\textwidth]{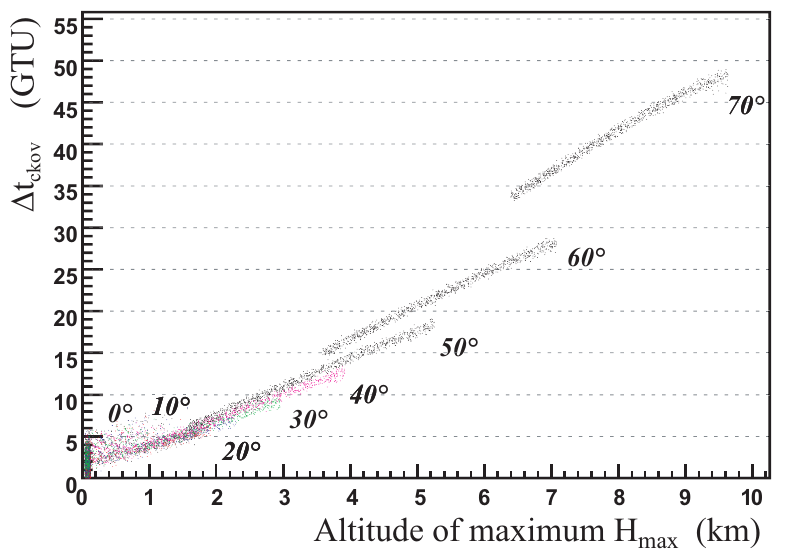}
\caption{Time difference between ground reflected Cherenkov and direct fluorescence peaks as a function of the altitude of the shower maximum at different zenith angles. The spread of the curves is due to the random 
first interaction point of UHECP used in the simulation. No correction is applied for the shower distance.}
\label{fig:cher-deltat-vs-hmax}
\end{center}
\end{figure}

Fig.~\ref{fig:cher-deltat-vs-hmax} shows results obtained at different zenith angles. Nevertheless, this signal can only be used if it is sufficiently intense.

\begin{figure}[ht]
\begin{center}
\includegraphics[width=0.5\textwidth]{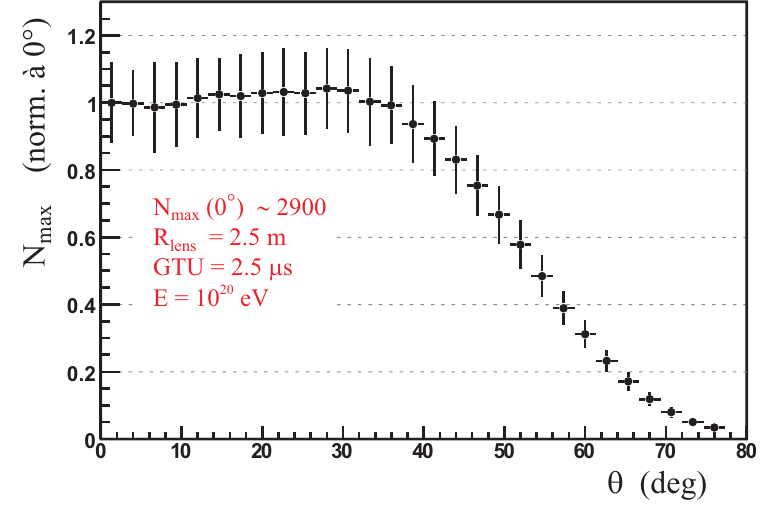}
\caption{Amplitude of the ground reflected Cherenkov peak as a function of shower zenith angle. Numbers of photons are normalized to the number of photons for a vertical shower, which is   around 2900 for a primary energy of 10$^{20}$~eV.}
\label{fig:cher-vs-theta}
\end{center}
\end{figure}

In case of inclined showers, the Cherenkov beam has a longer path to go down to the ground and is therefore more attenuated than in the case of vertical showers. Moreover Cherenkov photons are more dispersed around the shower impact due to the photon angular distribution. As a consequence the Cherenkov peak amplitude decreases when the zenith angle increases (Fig.~\ref{fig:cher-vs-theta}). Being quite constant up to 40\degr the signal is reduced by a factor of 30 at 75\degr.

In most cases Earth surface cannot be modeled by a Lambertian surface and specular reflection has to be used.

Taking into account showers with an impact at nadir, in the case of a specular reflection on oceans with a wind speed of 2 m/s (respectively 9 m/s), one can deduce from Fig.~\ref{fig:brdf} that events with a zenith angle greater than 20\degr (respectively 30\degr) will give a weaker Cherenkov peak than in Lambertian case. On the contrary, for smaller zenith angles, the Cherenkov peak should be greatly enhanced.

The anisotropy of the specular reflection has consequences for the dependence of the reflected Cherenkov light on the shower position in the FoV. 

\begin{figure}[ht]
\begin{center}
\includegraphics[width=0.5\textwidth]{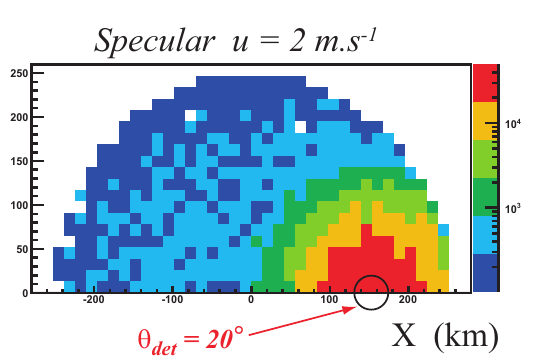}
\caption{Amplitude of the ground reflected Cherenkov peak as a function of the shower impact position in the FoV for showers with a zenith angle of 20\degr in case of specular reflection over ocean with a wind speed of \mbox{2 m$\cdot$s$^{-1}$}.  The total albedo is 8\%. }
\label{fig:specular-fov}
\end{center}
\end{figure}

As an example, Fig.~\ref{fig:specular-fov} shows the amplitude of the ground reflected Cherenkov peak as a
function of the position of the shower impact in the FoV for showers with 20\degr zenith angle. For a wind 
speed of  \mbox{2 m/s}, 76\% of the showers that impact in the FoV have a reflected Cherenkov 
light lower than the one expected in the Lambertian case. 
Furthermore, for showers with zenith angle greater than 30\degr, the 
specular peak does not point toward the telescope, regardless of the shower impact position in the 
FoV.  Over the oceans the hypothesis of a Earth Lambertian surface seems to be too optimistic and the anisotropy 
of the specular reflection will be responsible of a dependence of the reflected Cherenkov light with the 
position in the FoV. 

We conclude, therefore, that in case of specular reflection on Earth, the Cherenkov light of most 
of the showers will not be detected. 
Simulation results have been produced with a constant albedo of a few percents. 
In case of specular reflection, the Cherenkov peak could be enhanced only if, by chance, the shower 
angle and its impact position in the FoV allows the light to be reflected towards the telescope.  

\subsubsection{Scattered light}\label{sec:pupil-diff}
In this section, the term ``scattered light'' represents all photons that have been scattered by Earth surface (assumed lambertian) or by atmosphere in clear sky conditions. The signal component previously named ``ground reflected Cherenkov'' is not included in this scattered light component, but is considered as a component of the basic signal.
As already said, scattered photons are slightly time delayed compared to the direct signal and spread around the main track in the FoV. Therefore they can partly mix with the basic signal as a noise.

\begin{figure}[p]   
\begin{center}
\begin{tabular}{c}
\includegraphics[width=0.46\textwidth]{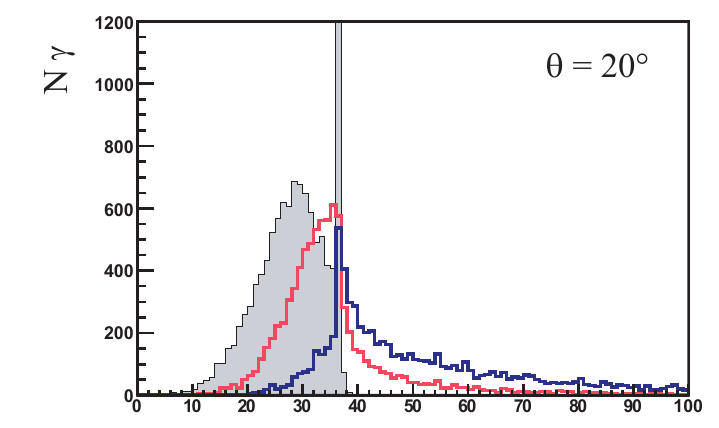}\\
\includegraphics[width=0.46\textwidth]{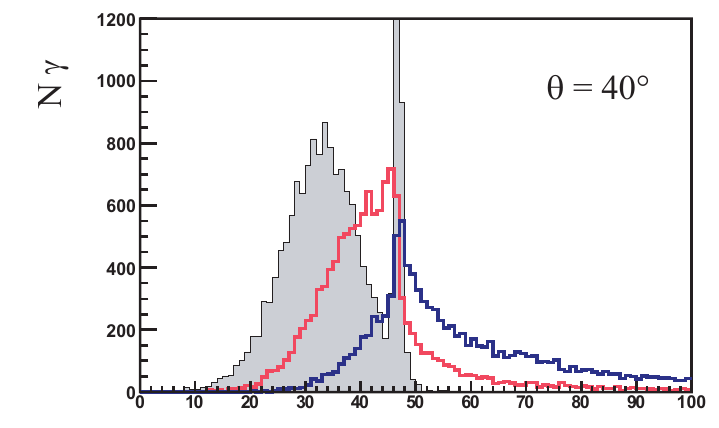}\\
\includegraphics[width=0.46\textwidth]{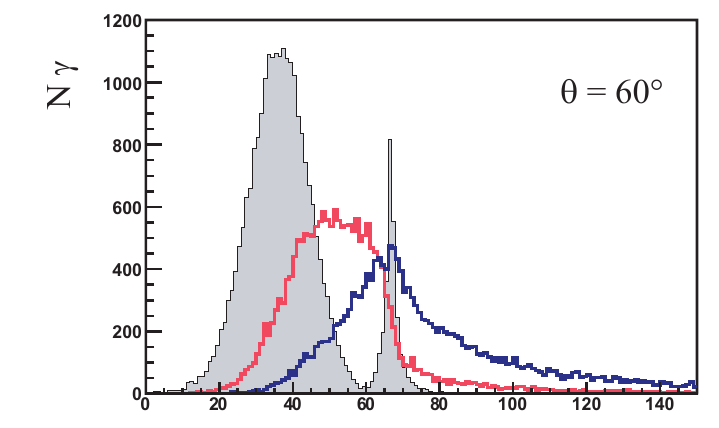}\\
\includegraphics[width=0.46\textwidth]{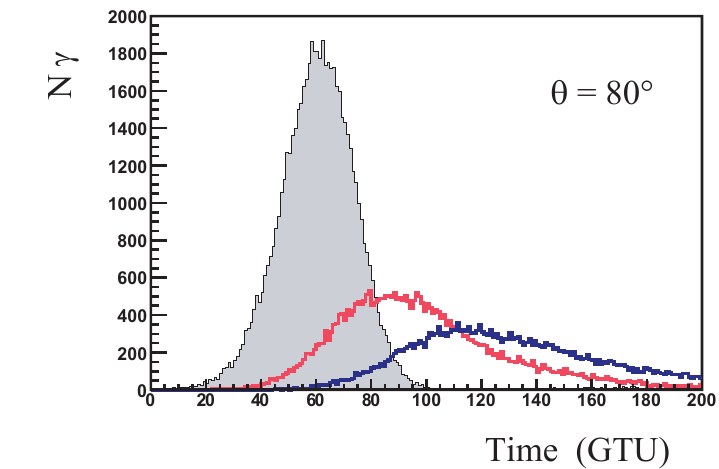}\\
\end{tabular}
\caption{Number of photons reaching the pupil as a function of time at 4 zenith angles. The signal is integrated in the whole FoV. In gray are the basic signals: direct fluorescence and ground reflected Cherenkov.  The single scattered signal is superimposed in light red and the multi-scattered signal in dark blue.}
\label{fig:scattered-time}
\end{center}
\end{figure}
Fig.~\ref{fig:scattered-time} shows time profiles of photons reaching the pupil at different zenith angles. The plotted signal is integrated in the whole FoV. Although the scattered light increases with shower zenith angle (by a factor of 5 for single scattered photons between 0\degr and 80\degr and by a factor of 4 for multi-scattered photons), its contribution is of the same order of magnitude as the direct fluorescence light.
However part of the scattered signal is not correlated with the track in the FoV because most of the photons come from a last scattering position (in the atmosphere) far from the shower axis. Thus most of the scattered photons will be rejected in the same way as background photons.

\begin{figure}[ht]
\begin{center}
\includegraphics[width=0.47\textwidth]{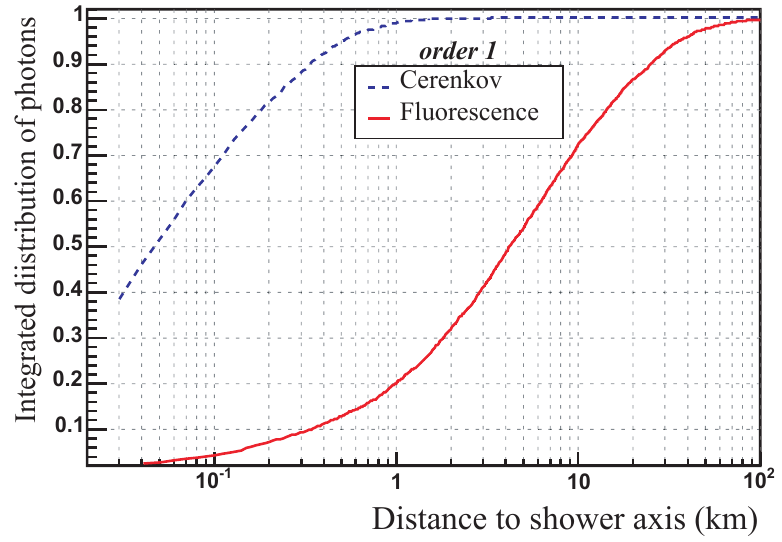}
\caption{Integrated single scattered photon number as a function of the distance to the shower axis for a vertical shower. Fluorescence and Cherenkov photons are compared. Distributions are normalized to 1.}
\label{fig:scattered-lat}
\end{center}
\end{figure}
\begin{figure}[ht]
\begin{center}
\includegraphics[width=0.47\textwidth]{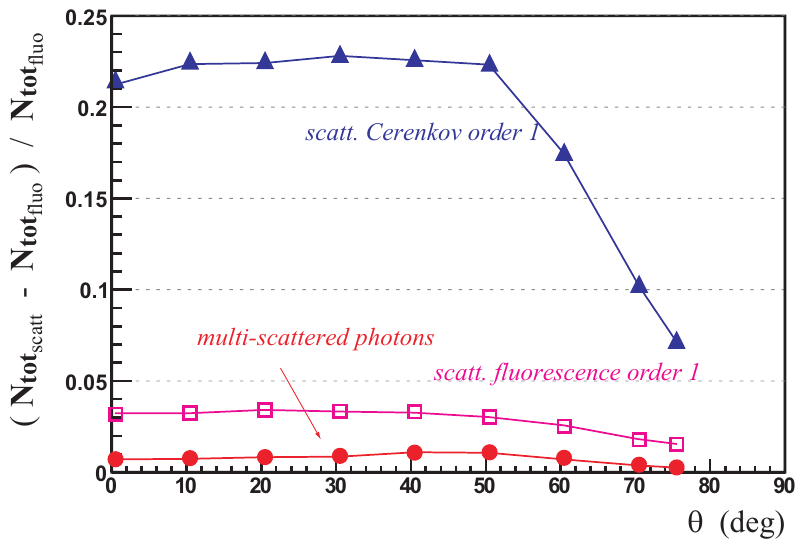}
\caption{Contributions of selected scattered photons from different origins related to direct fluorescence light as a function of shower zenith angle in clear sky conditions. Selected scattered photons are those correlated in space and time with the direct fluorescence signal.}
\label{fig:scattered-contribution}
\end{center}
\end{figure}

As an example Fig.~\ref{fig:scattered-lat} shows the lateral distribution of the last scattering position in the atmosphere for vertical showers. As one can see, due to the fact that Cherenkov light is beamed at emission, 90\% of single scattered Cherenkov photons have their last scattering position in the atmosphere located less than 0.4 km from the shower axis, 
in contrast to only 10\% of single scattered fluorescence photons. 
Simulations performed to study multiple scattering effects prove that only 5\% of the photons scatter more than once before reaching the detector.

If not taken into account during the reconstruction process, the scattered light component would entail systematic errors on the shower parameters. The amount of light is increased and the detected shower profile is distorted. 

To estimate precisely the contribution of scattered light to the detected signal, we have selected the scattered photons, at the photo-electron level, using a simple detector simulation. It uses EUSO configuration (pupil diameter of 2.5 m), applying a global efficiency coefficient (5\%) to convert photons into photo-electrons. The FoV being pixelized and time profile being discretized in GTU, one selects at a given GTU the scattered signal standing in pixels containing the direct fluorescence signal. 

\begin{figure}[t]
\begin{center}
\includegraphics[width=0.5\textwidth]{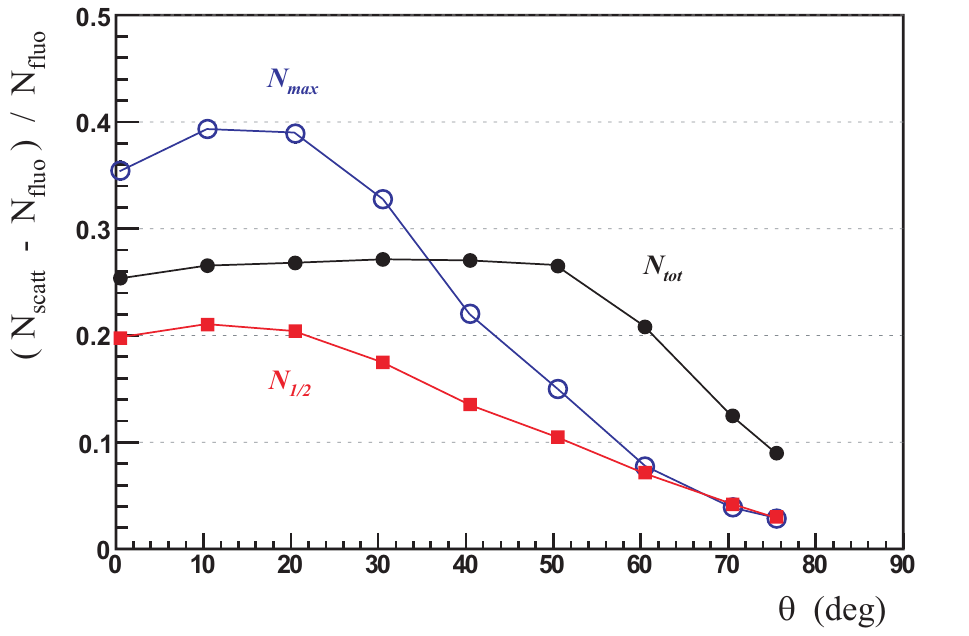}
\caption{Contributions of scattered photons to the integral of the signal $N_\mathrm{tot}$, the integral of the first half part of the signal $N_{1/2}$ and the maximum of the signal $N_\mathrm{max}$.}
\label{fig:scattered-shape}
\end{center}
\end{figure}

It appears that the contribution of scattered photons represents less than 30\% of the direct fluorescence light. This is shown in Fig.~\ref{fig:scattered-contribution} where contributions of scattered photons from different origins are represented as a function of the shower zenith angle. The main contribution is that of single scattered Cherenkov photons that represents slightly more than 20\% of the direct fluorescence photons, in contrast to only less than 4\% for single scattered fluorescence photons and 1\% for multi-scattered photons.

It has to be noticed that presence of scattered photons has consequences not only on the total number of photons reaching the pupil but also on the shape of the reconstructed profile. This can be illustrated by comparison of scattered photon contributions to the integral of the signal, to the integral of the first half part of the signal and to the maximum of the signal as shown in Fig.~\ref{fig:scattered-shape}.

\subsection{Triggering efficiency and energy threshold}\label{sec:trigger}

All the simulations discussed in the present section are based on a standard configuration
which derives from the EUSO Phase A design 
at a typical orbital height $H=\un[400]{km}$. 

From this basic configuration, four different detectors were implemented, identical in all parts but for the diameter of the 
entrance pupil: \un[2.5]{m} (EUSO Phase A configuration), \un[4]{m}, \un[8]{m} and \un[12]{m}. In this way ESAF is able to evaluate the performances of EUSO and of possible larger detectors.
\begin{figure}[t]
\centering
\begin{tabular}{c}
\includegraphics[width=0.45\textwidth]{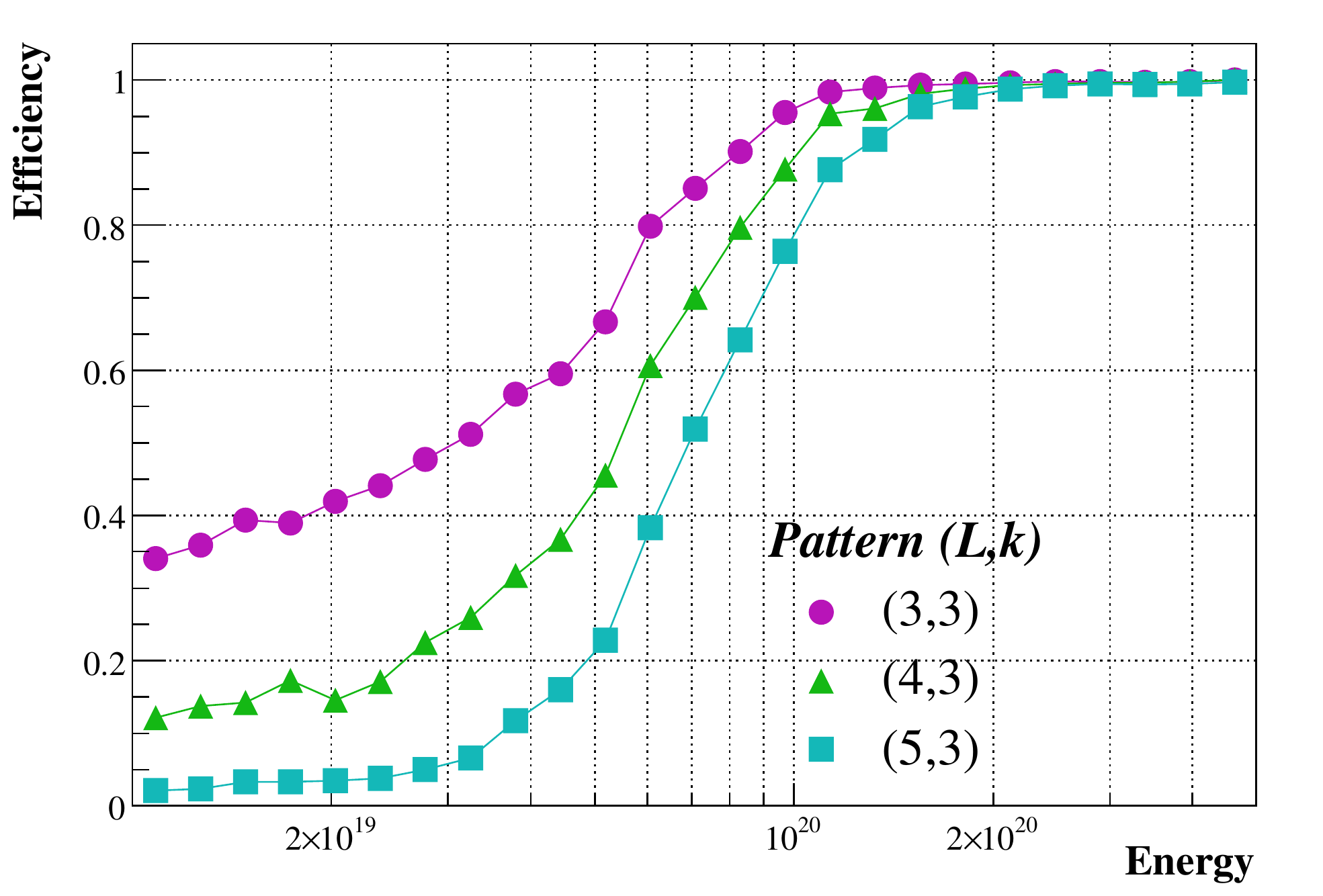} \\
(a) \\
\includegraphics[width=0.45\textwidth]{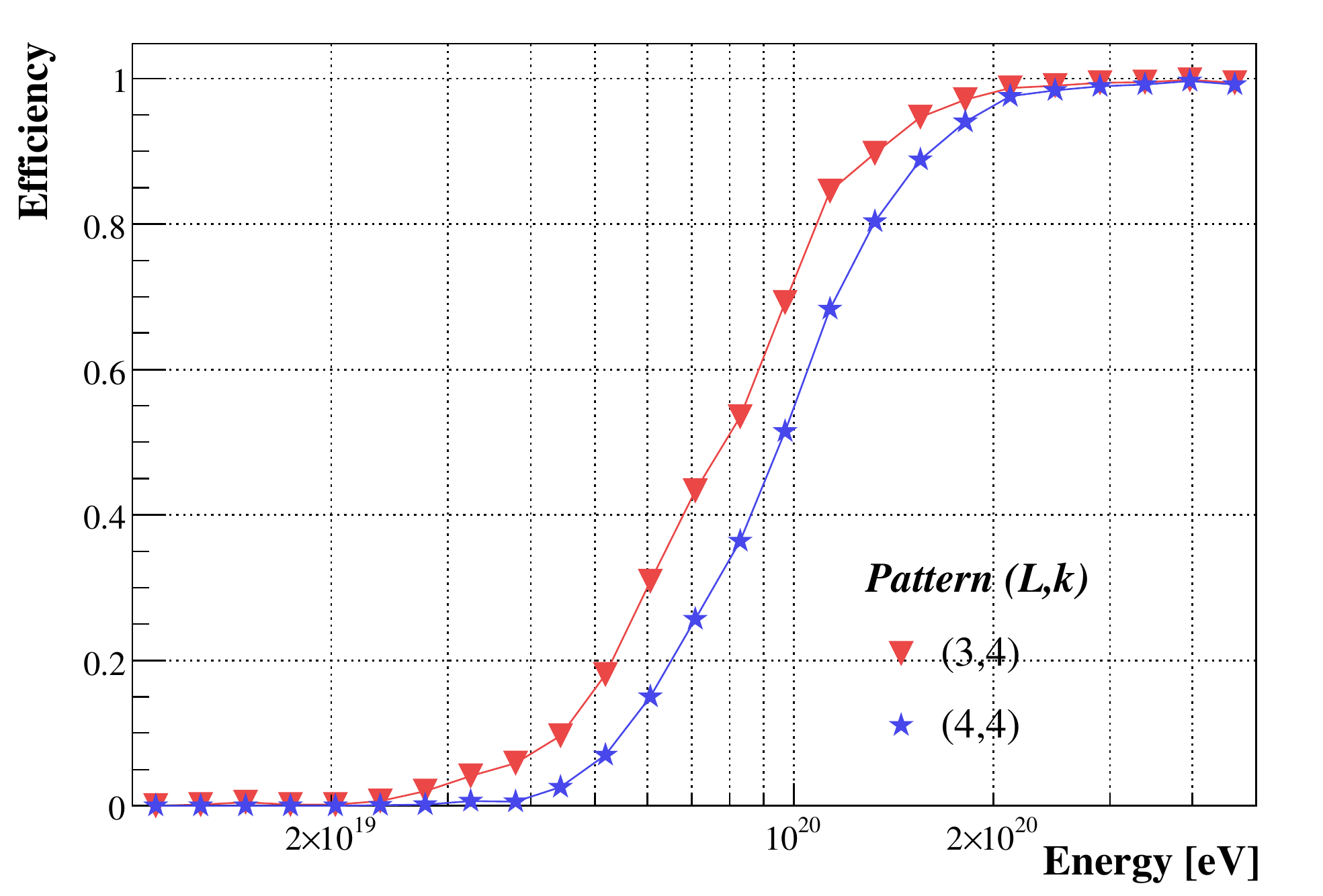} \\
(b) \\
\includegraphics[width=0.45\textwidth]{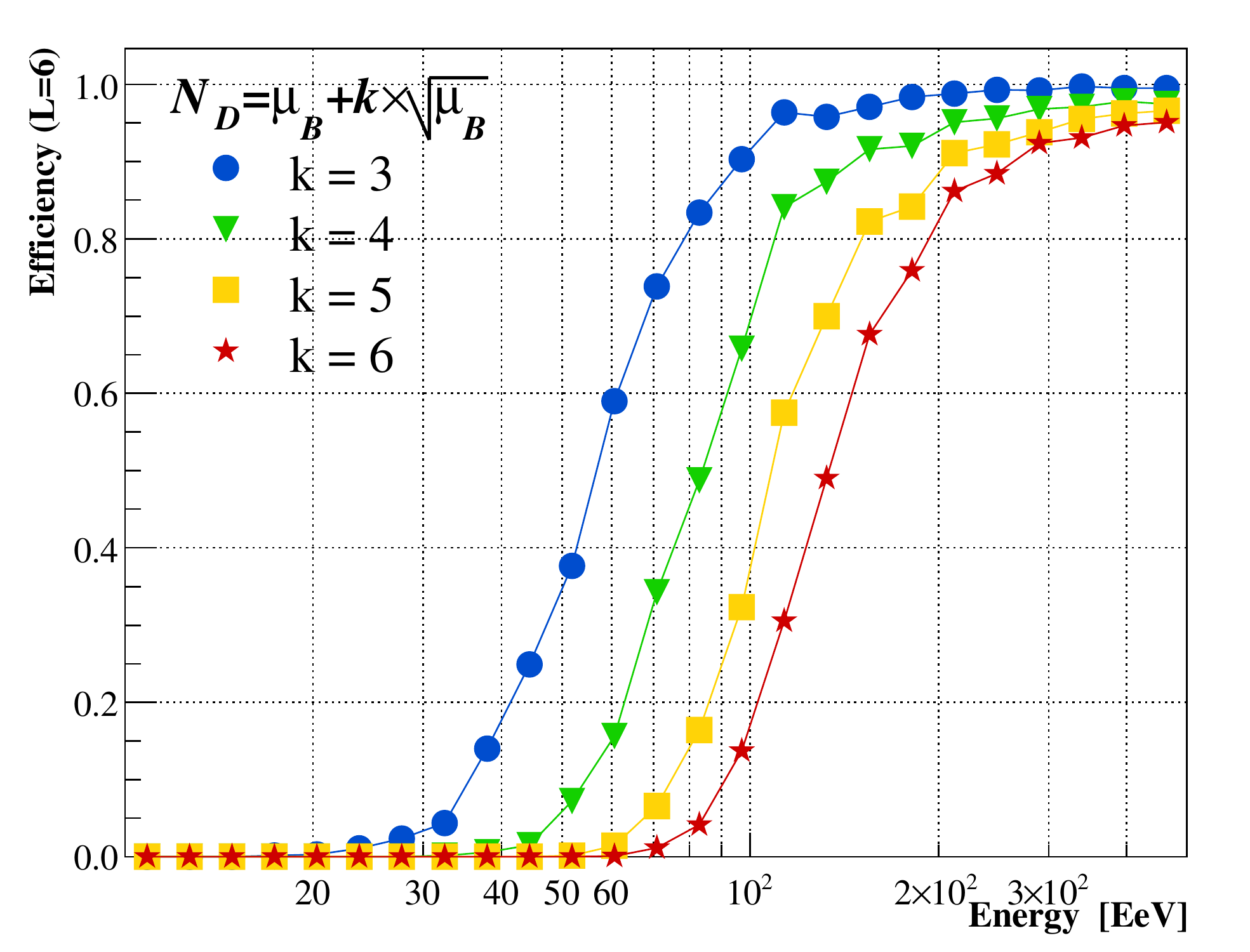} \\
(c) \\
\end{tabular}
\caption{Trigger efficiency curves for various trigger configurations (\un[2.5]{m} detector): (a) FTR above \simun[100]{Hz}; 
(b) FTR below \simun[100]{Hz} (c) Efficiency with fixed $L=6$ and $3\le k \le 6$.}
\label{fig:TrEffFTR}
\end{figure}

A sample of \sci{3}{5} simulated EAS with energy in the range \mbox{$\eVsci{1}{19}<E< \eVsci{5}{20}$} 
(flat energy distribution in a logarithmic scale and angular distribution uniform in solid angle) 
was used for all the configurations. 
\begin{figure}[t]
\centering
\begin{tabular}{c}
\includegraphics[width=0.45\textwidth]{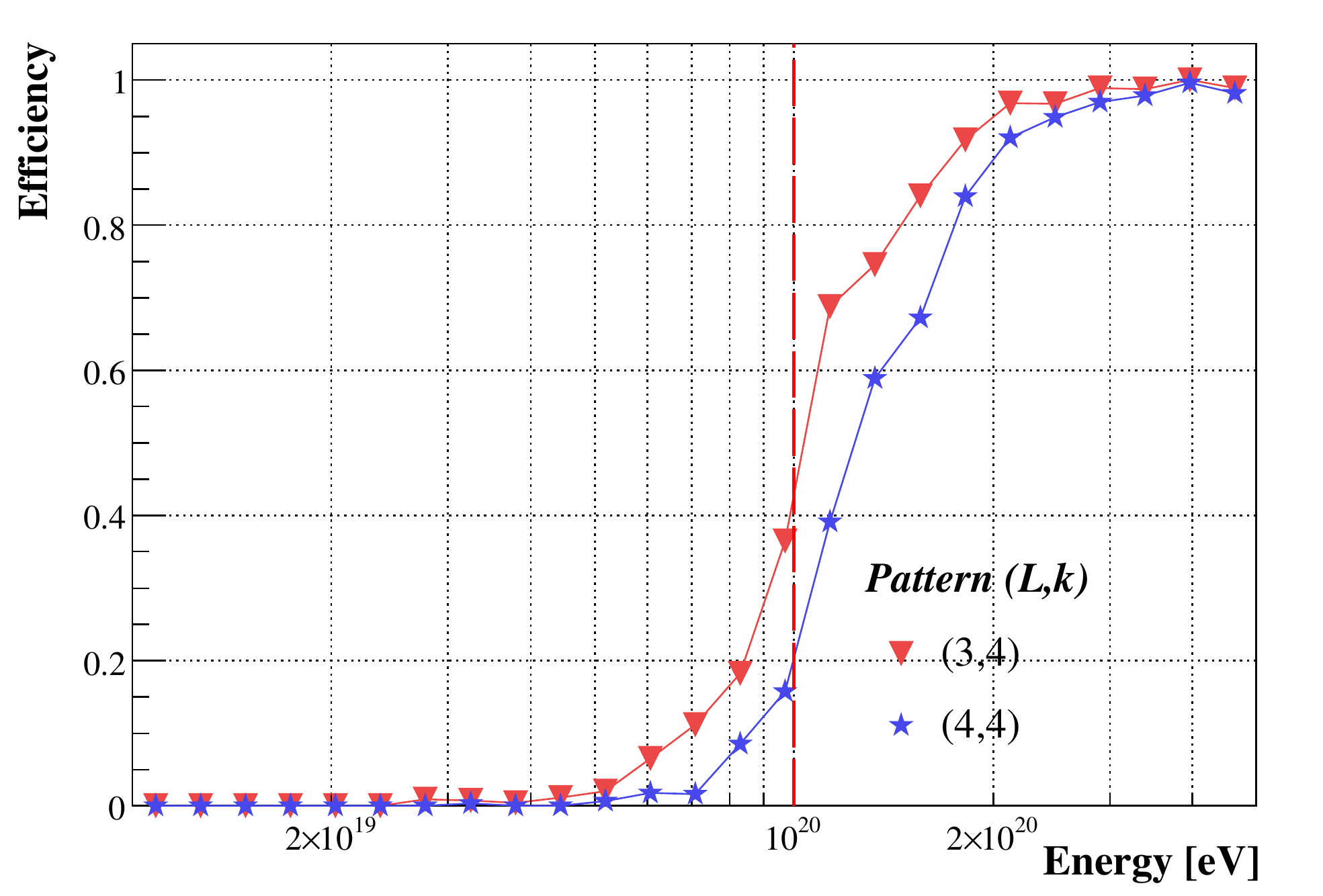} \\	
(a) $\theta \le 30\degr$ \\
\includegraphics[width=0.45\textwidth]{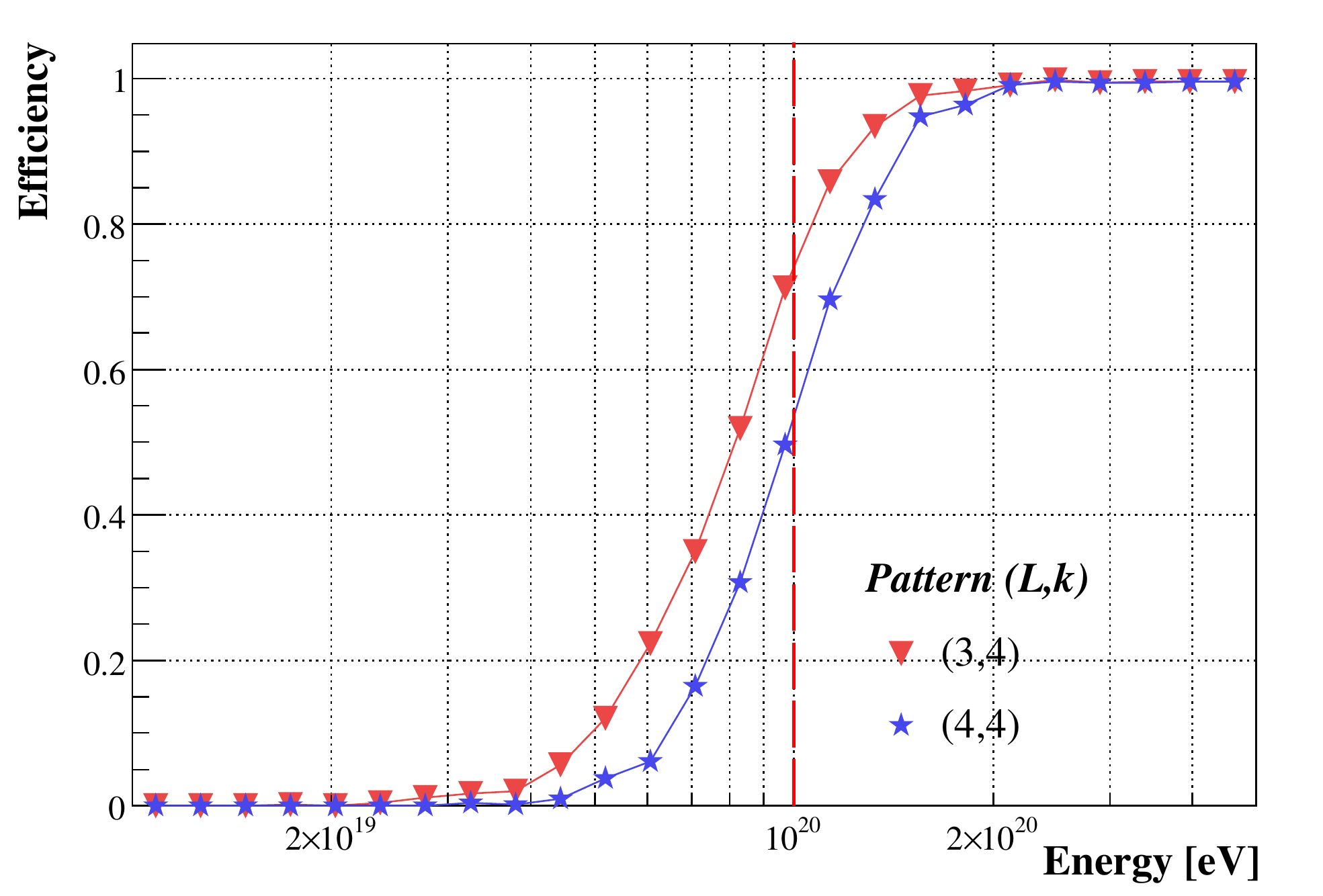} \\
(b) $30 \degr < \theta < 60 \degr$ \\
\includegraphics[width=0.45\textwidth]{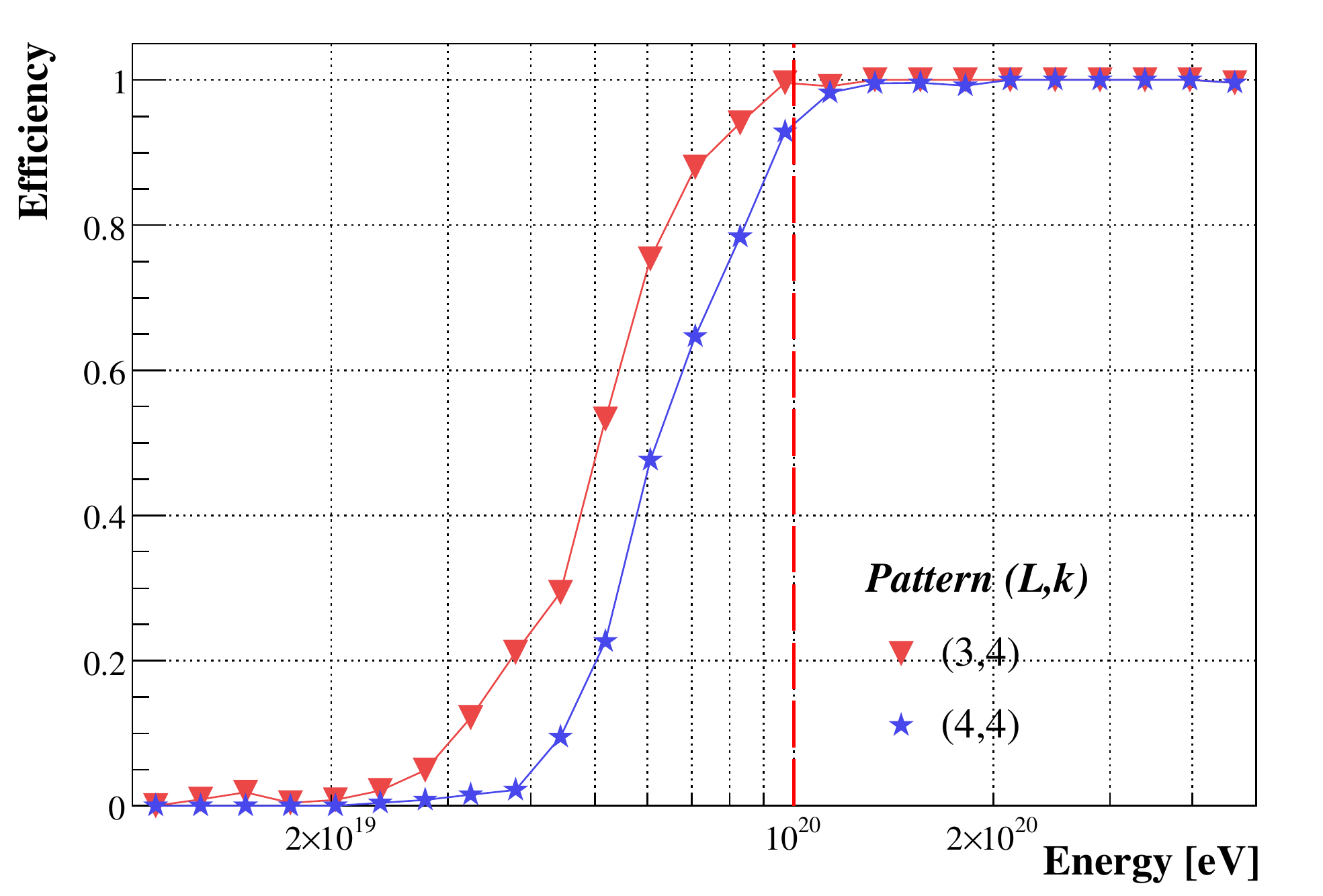} \\
(c) $\theta \ge 60\degr$ \\
\end{tabular}
\caption{Efficiency of the [3;4] and [4;4] trigger patterns on three ranges in $\theta$. The
relative weight of the three samples are (a) $\simeq25\%$, (b) $\simeq50\%$, (c) $\simeq25\%$. The red line
at \eVsci{1}{20} is drawn as a reference.}
\label{fig:TrEffTheta}
\end{figure}

The showers were generated by GIL (see section~\ref{sec:showers}) in clear sky and US-Standard Atmosphere (no clouds). A parameterized optics with a 2D gaussian PSF of \un[5]{mm} RMS, a GTU of $t_\mathrm{GTU}=\un[2.5]{\mu s}$ and an overall photo-detector efficiency of 10\% were used according to the EUSO Phase A design.
\begin{figure}[t]
	\centering
	\includegraphics[width=0.45\textwidth]{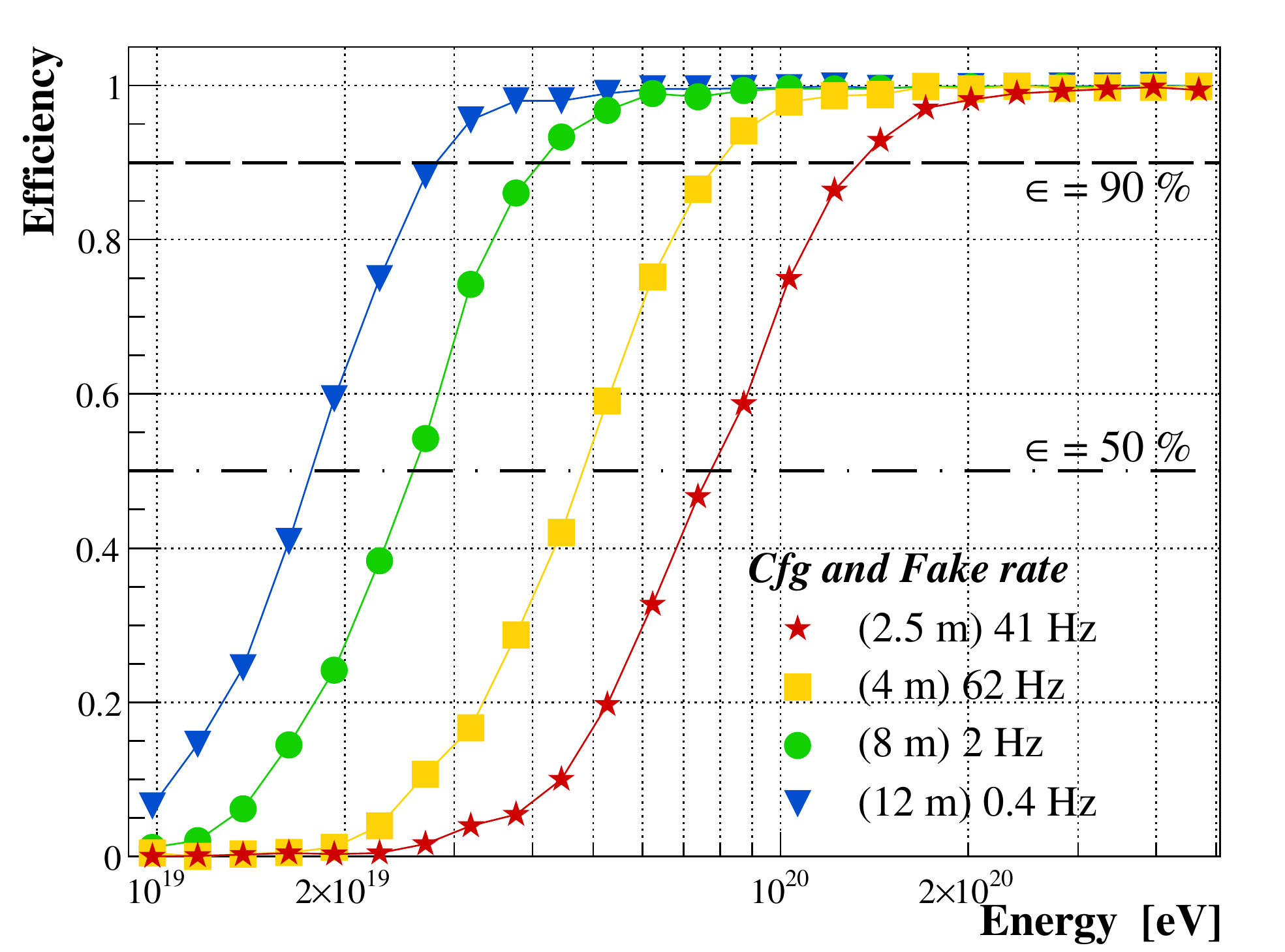}
	\caption{Trigger efficiency curves for various sizes of the detector using the pattern [3;4].}
\label{fig:TrEffDiameters}
\end{figure}

Because of the limits imposed by the available telemetry and computing resources in the EUSO project, 
only the contiguity tracking triggers that satisfy $\nu_\mathrm{FTR}\lesssim\un[100]{Hz}$ were considered useful.
Future detectors may have less stringent requirements.

As shown in Fig.~\ref{fig:TrEffFTR}, if the condition \mbox{$\nu_\mathrm{FTR}\lesssim\un[100]{Hz}$} is not satisfied, 
the efficiency curve flattens at low energies to a value that depends on $[L;k]$ (see section \ref{sec:CTT} for the definition of $[L;k]$), else flattens to zero at low energies. 
Furthermore,  when the value of $L$ is raised to a fixed $k$, the efficiency curve is shifted toward 
higher energies (without changes of shape), i.e. the full efficiency is obtained at a larger value of $E$.

The dependency of the trigger efficiency on the zenith angle $\theta$ is shown in Fig.~\ref{fig:TrEffTheta}. The efficiency curve is shifted toward higher energies for $\theta< 30\degr$ and toward lower energies for $\theta > 60\degr$.

To compare the detectors on the same grounds, the continuity trigger pattern must satisfy
$\nu_\mathrm{FTR}\lesssim\un[100]{Hz}$ for all detectors. $[L;k]=[3;4]$ was the selected 
trigger for the further studies being the less restrictive pattern satisfying this additional condition.

The results are shown in Fig.~\ref{fig:TrEffDiameters}. As expected the curves are shifted 
toward lower energies when increasing the entrance pupil diameter. The energy threshold of the 
experiment scales as the inverse of the diameter or, equivalently, as the inverse of the square root 
of the photo-detection efficiency. 

Simulations with ESAF show that an EUSO like detector has full detection efficiency around \mbox{2. 10$^{20}$ eV} and 50\%
efficiency at \mbox{8. 10$^{19}$ eV}. Better efficiency can be obtained only by enlarging the detector size or
by using better photo-detection devices than commercially available MAPMTs.  We recall, however, that 
these results are obtained in ideal conditions, not taking into account variable 
weather  and dead pixels in the focal surface. 


This result is really very weakly dependent
on the specific triggering algorithm. The lack of efficiency is related to lack of signal at low energy. 

It is worth recalling that these conclusions apply to photo-detector similar to MAPMTs, that have
a global photo-detection efficiency (quantum efficiency $\times$ PMT practical factor $\times$ geometric factors) 
of about 10\%. To our knowledge,
better devices do not exist yet, but new photocathodes with much improved quantum efficiency are already
available and might yield more efficient MAPMTs in the near future. This of course may have a significant
impact on efficiency. However, higher photo-detection efficiency also means higher
nightglow background, so the gain would not be linear. To be more precise, the gain in
photon collection efficiency scales as the pupil area in Fig. \ref{fig:TrEffDiameters}. For instance, 
referring to a 2.5 m diameter pupil (right-most curve in Fig.  \ref{fig:TrEffDiameters}), 
an increase of efficiency of a factor 2.56 (significant improvement!) 
corresponds to 4 m diameter (second curve from right).

\subsection{Angular resolution}\label{sec:track}

\begin{figure}[htb]
\includegraphics[width=0.45\textwidth]{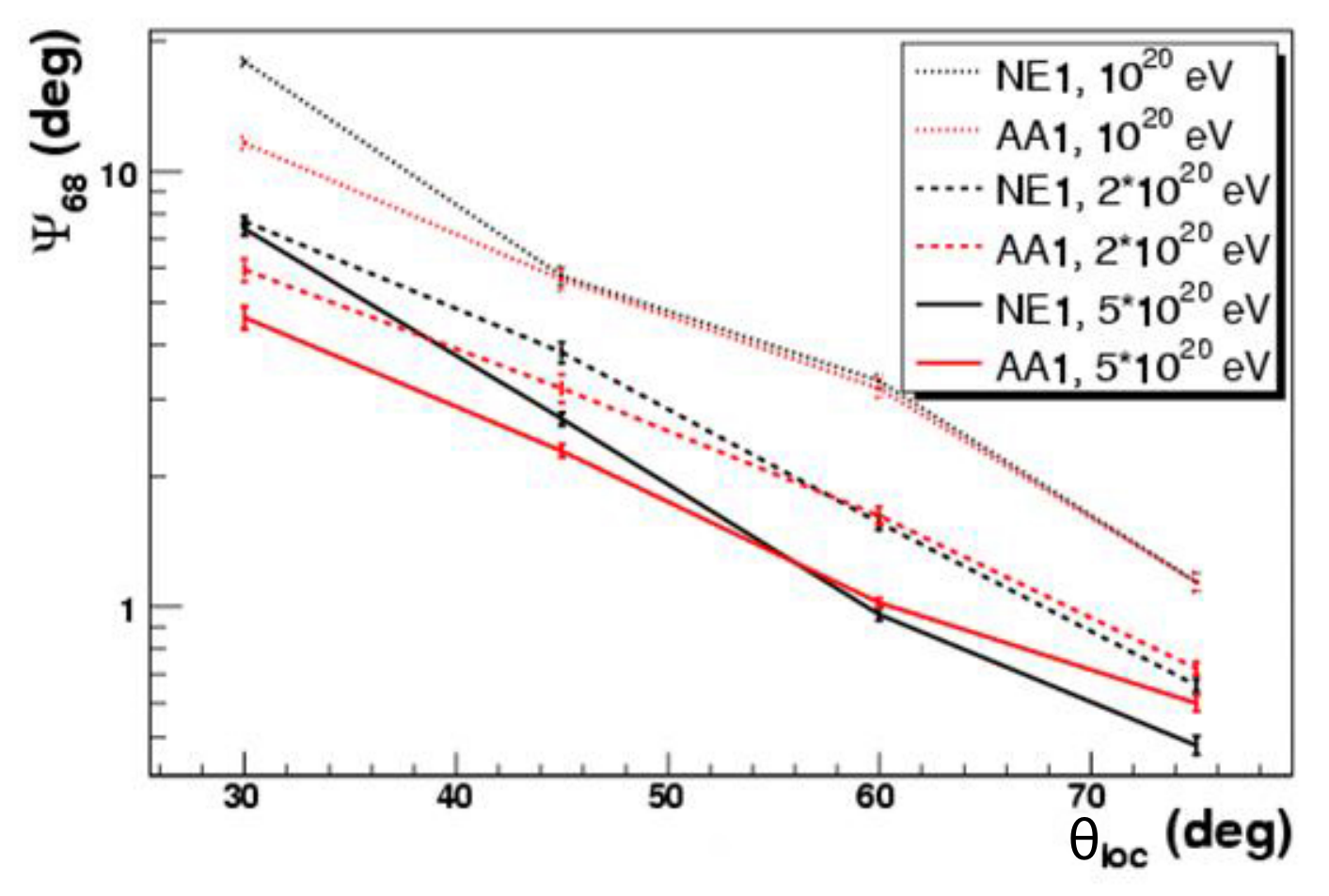}
\caption{Angular resolution.  $\Psi_{68}$ is the angle in space between the simulated and reconstructed 
directions which defines the cone containing 68 \% of the events.}
\label{fig:AngularResolution}
\end{figure}

Fig.~\ref{fig:AngularResolution} shows the angular resolution concerning the identification of the
shower axis direction in space in presence of typical random background of \mbox{500 ph m$^{-2}$ns$^{-1}$ sr$^{-1}$} (see section~\ref{sec:RBSimu})\footnote{We recall here that the typical night glow background in EUSO
is 500 ph m$^{-2}$ns$^{-1}$ sr$^{-1}$, reaching 1000 with 1/4 of moon. With more than 1/4 moon, the observation
of EAS is not possible. }; in the figure,  
$\Psi_{68}$ is the angle in space
between the simulated and reconstructed directions which defines the cone containing
68\% of the events, i.e. one standard deviation in angle; $\theta_\mathrm{loc}$ is the zenith angle of the incoming shower calculated at the ground impact point. Standard EUSO parameters are used for this simulation. The showers are simulated with UNISIM in clear sky conditions neglecting the multiple scattering effects.

The curves labeled $NE1$ refers to exact formulas and numerical methods, while $AA1$ refers to approximated formulas
and analytical methods (see section~\ref{sec:DirectionReco}).
As expected the direction reconstruction is far better for very inclined showers, since for such inclinations
the track length is greater and consequently the total number of detected photons and the number
of hit pixels are higher (see section~\ref{sec:pupil-fluo}). 

At 10$^{20}$ eV the
angular resolution is close to 10\degr for zenith angles smaller
than 40\degr, it goes down to a few degrees in the angular range
50\degr-70\degr and finally approaches about 1\degr for very inclined showers
with zenith angle greater than 80\degr. The angular
resolution reaches 1\degr already at
70\degr zenith angle for 2 10$^{20}$ eV and at 60\degr zenith angle for \mbox{5 10$^{20}$ eV. }

In a conservative way, the detection
of Cherenkov light is not taken into account here and the detector shower distance is set in the 
reconstruction at the expected average value. Shower to shower fluctuations yields approximately 0.5 
km rms error on the altitude. 

The relation between the uncertainty in the estimation of the shower maximum height (and so of the detector-shower distance) and the uncertainty in the reconstructed direction has been studied by Monte-Carlo simulation and analytical calculation.
In average we find an uncertainty of $\sim$0.2\degr in the reconstructed zenith angle for every km of error in the altitude of the EAS maximum.

\subsection{Energy and \Xmax reconstruction performances}\label{sec:energy}
As previously seen, the event reconstruction starts with the determination of the shower track. 
Then a functional form for the shower longitudinal development is assumed and the expected collected light is calculated.
The energy and $X_\mathrm{max}$ of the EAS are deduced by varying the parameters of the longitudinal shower function until an agreement with the measured signal is obtained.

Several sources of uncertainties can be identified at each step of this reconstruction process.
ESAF is a useful tool  to evaluate most of them. Different samples of  proton showers  
have been generated. In each sample, either one shower parameter (the energy, the zenith angle, 
the shower maximum altitude, the ground impact position in the FoV)  
or the atmosphere profile, or one telescope characteristics is varied.  
Possible uncertainty sources can be studied using each of these simulated samples.

\subsubsection{Uncertainties on the shower track altitude}
Beside the angular resolution described in section~\ref{sec:track}, the precision on the shower track geometry 
is related to the knowledge of the shower track altitude. As explained, two different  methods 
may be applied to determine \Hmax, either using the reflected Cherenkov signal, or when the information can not be retrieved,
using the shower track length only. In both cases, the shower track zenith angle $\theta$ must be known to infer  \Hmax. 

At the first order, according to geometrical considerations only, the impact on \Hmax of  an uncertainty on  $\theta$ is estimated.
Whereas an uncertainty \mbox{$\delta\theta = + 1$\degr} induces an underestimation of  \Hmax when the Cherenkov information is used (relatively increasing with $\theta$ but less than 2\% for shower with zenith angle $\theta \leq 60$\degr),
it induces an overestimation of \Hmax when the Cherenkov information is not used
(relatively decreasing with $\theta$ but less than 2\% for shower with zenith angle $\theta \geq 20$\degr).

The results are obtained simulating about 12600 events (of which 11200 reconstructed) in clear sky (US-Standard atmosphere) with \texttt{LOWTRAN} and UNISIM and a telescope aperture of 2.5 m (parametric optics).

ESAF results show that in clear sky and US-Standard Atmosphere,
\Hmax can be reconstructed with an error of \mbox{100-300 m} for events with both fluorescence 
and Cherenkov  echo components for \mbox{$10\degr<\theta<60\degr$}, and with an error of \mbox{400-800 m} 
when the Cherenkov echo is either 
missing or just ignored for \mbox{$10\degr<\theta<90\degr$}.  
The absolute error grows with $\theta$ for both methods, while the 
relative error is rising for the method based on the Cherenkov echo (from 5\% to 8\%) and decreases 
for the method based on the time width of the p.e. distribution (from 22\% to 5\%). 
Therefore these methods can be considered as complementary.

\subsubsection{Effects of geometric uncertainties}
Uncertainties on the shower geometry affect the factors used in the expected light profile calculation.  
At each  shower point $P_i$, the reconstruction process involves the fluorescence yield $Y_i$, the light transmission
$T_i$, and a solid angle factor $(4 \pi R_i^2)^{-1}$. ESAF allows to  evaluate uncertainties induced on them
related to shower geometry.
A $\sim$1~km shift on altitude affects  the solid angle computation by less than 1\%.
Despite the fact that the light transmission is the more affected factor by an uncertain altitude,
an uncertainty on the reconstructed altitude of $\sim$1~km leads to less than 7\% uncertainty on the direct transmission in clear sky.
The fluorescence yield dependence with altitude is almost canceled by the dependence of the ionization energy deposit per meter of track length,
thus an uncertainty  of $\sim$1~km on altitude affects the fluorescence yield per meter by less than 1\%.

Furthermore, the uncertainty on the shower geometry affects the slant depth calculation. 
The integral of the reconstructed profile is thus affected by an uncertainty of 10\% for every km of error in altitude, 
and \Xmax by around $ 100$~g/cm$^{2}$.
In the same way an uncertainty on shower zenith angle leads to an uncertainty on \Xmax of around $-20$~g/cm$^{2}$ for $\delta\theta = + 1$\degr 
but to only 0.2\% for $\delta\theta = + 1$\degr on the integral of the shower profile. 

\subsubsection{Effects of atmospheric uncertainties}
ESAF allows also to  evaluate uncertainties on $Y_i$, $T_i$ and slant depth calculation related to atmosphere knowledge. 
During simulation runs, atmosphere parameters are varied according to all available 
MSISE profiles, and in the reconstruction  process, only US Standard atmosphere 
model is considered. These  studies show that the uncertainty on transmission and on the fluorescence yield per meter 
stays below 2\% (respectively 2.5\%).
They also demonstrate that uncertainties induced by 
slant depth calculation lead to a shift on \Xmax from \gcmsq[-30] to \gcmsq[25], and to an uncertainty on the reconstructed profile integral around 3\% which could reach 4\% for shower detected close to the equator.

\subsection{Energy and \Xmax resolution}\label{sec:energyres}

\begin{figure}[htb]
\centering
\includegraphics[width=0.5\textwidth]{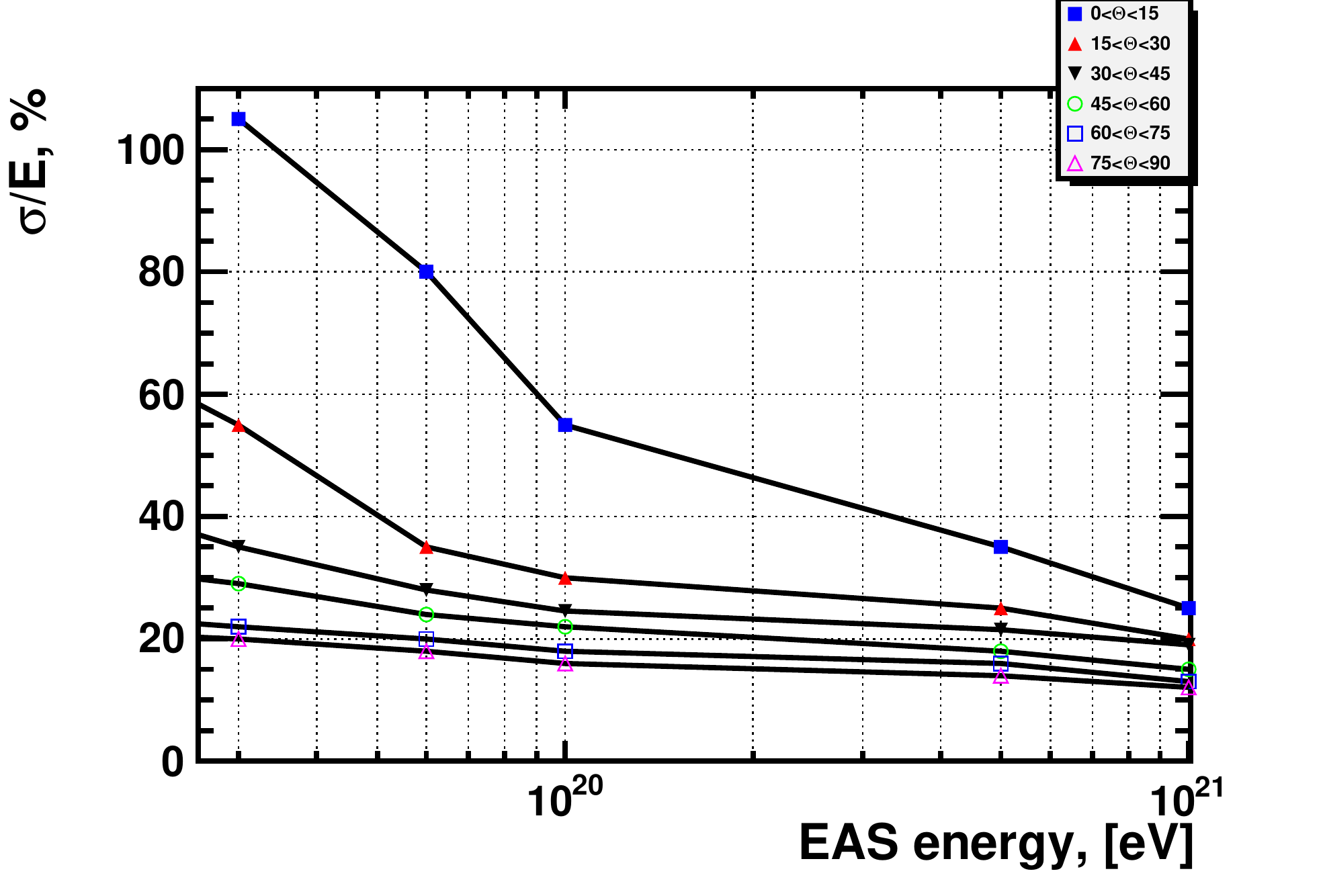} 
\caption{\label{fig:energy_res} Relative energy resolution $\sigma/E$ as a function of energy in various bins of zenith angle $\theta$ for proton initiated EAS.} 
\end{figure}

Apart from the uncertainties on the atmospheric profile,  different contributions described above
are taken into account to perform preliminary studies of the relative resolution of energy $\sigma/E$ for an EUSO detector with 2.5 m entrance pupil diameter. 
Showers are simulated with UNISIM, standard EUSO parameters are used for the detector and simulations have been performed in clear sky conditions with a US-Standard atmosphere, neglecting the multiple scattering effects. 
Results are displayed in Fig.~\ref{fig:energy_res} as a function of UHECP energy in various bins of zenith angle $\theta$ for proton initiated EAS. 
An energy resolution between $20-25$\% is expected at \mbox{$E=10^{20}$ eV}. It worsens at lower energy due to a low signal statistics and improves at higher energies. There is a strong dependence on the zenith angle of developing EAS, also due  to the 
increasing number of collected photons as $\theta$ increases.

As already said, an uncertainty on the shower altitude induces an important uncertainty on the reconstructed \Xmax,
that is why results on \Xmax resolution are presented depending on the method used for the first altitude estimation.
For events with both fluorescence and Cherenkov signals  \Xmax can be reconstructed with resolution 
better than \gcmsq[40] for $\theta<60\degr$. 
At larger angles  the Cherenkov signal becomes suppressed, and the method based on the fluorescence
profile must be used. In this case the resolution on \Xmax is degraded, but it is still in the range 
between \gcmsq[50] and \gcmsq[100], depending on the shower inclination. 
\begin{figure}[htb]
\includegraphics[width=0.5\textwidth]{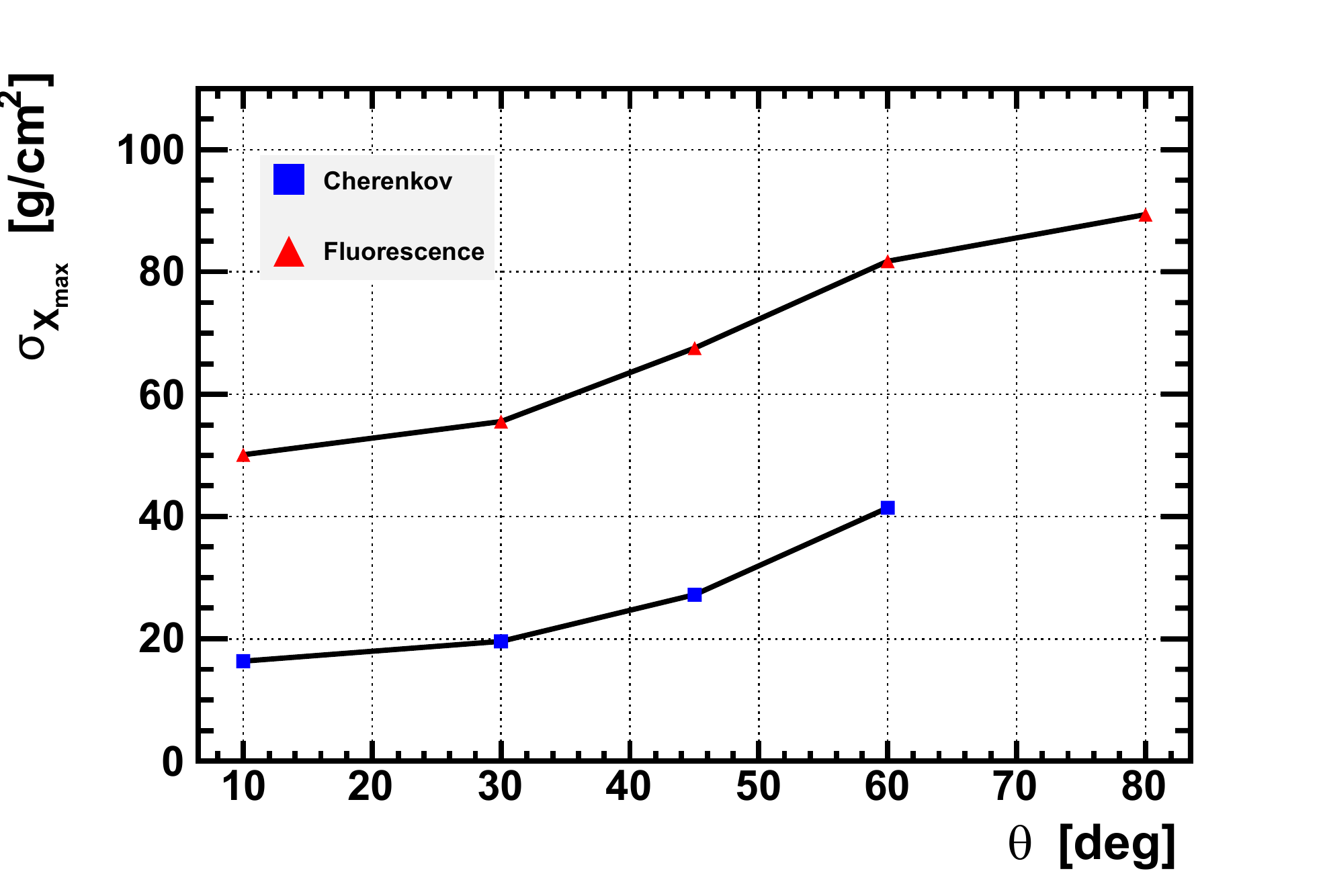}
\caption{The plot shows the error (in g/cm$^2$) with which \Xmax is reconstructed when the Cherenkov pulse is
available (lower curve, blue points) and when only fluorescence profile is used (upper curve, red points). }
\label{fig:xmax_with_cher}
\end{figure}

\subsubsection{Effect of scattering on E and \Xmax resolution}

Simulation results show that photon scattering distorts the fluorescence signal. 
Studies described in section~\ref{sec:pupil-diff} are used to evaluate systematics related to the fluorescence
signal contamination. In case of clear sky conditions, the total number of photons at the entrance 
pupil is increased by 10\% to 25\%, depending on the shower zenith angle (see Fig.~\ref{fig:scattered-contribution}). 
The contamination is mainly due to the scattered Cherenkov light. If this contribution is not
subtracted to the signal, it  leads to an uncertainty on $E$ in the same range, and to a shift of 
the maximum of the profile (in time); \Xmax is then overestimated by around \gcmsq[100].
Scattering of fluorescence photons as well as multi-scattering (more than once) of the Cherenkov photons
have weaker impacts on the energy (less than 5\%) and  \Xmax  (\gcmsq[10]) resolutions.
The subtraction of the diffused Cherenkov light is possible, 
as it was already done in HiRes experiment (see \cite{bib:hires-subtraction}).

\subsection{Effect of clouds on the detected signal}\label{sec:clouds}
\subsubsection{Reflected peak}\label{sec:cloud-peak}
\begin{figure}[ht]
\begin{center}
\includegraphics[width=0.5\textwidth]{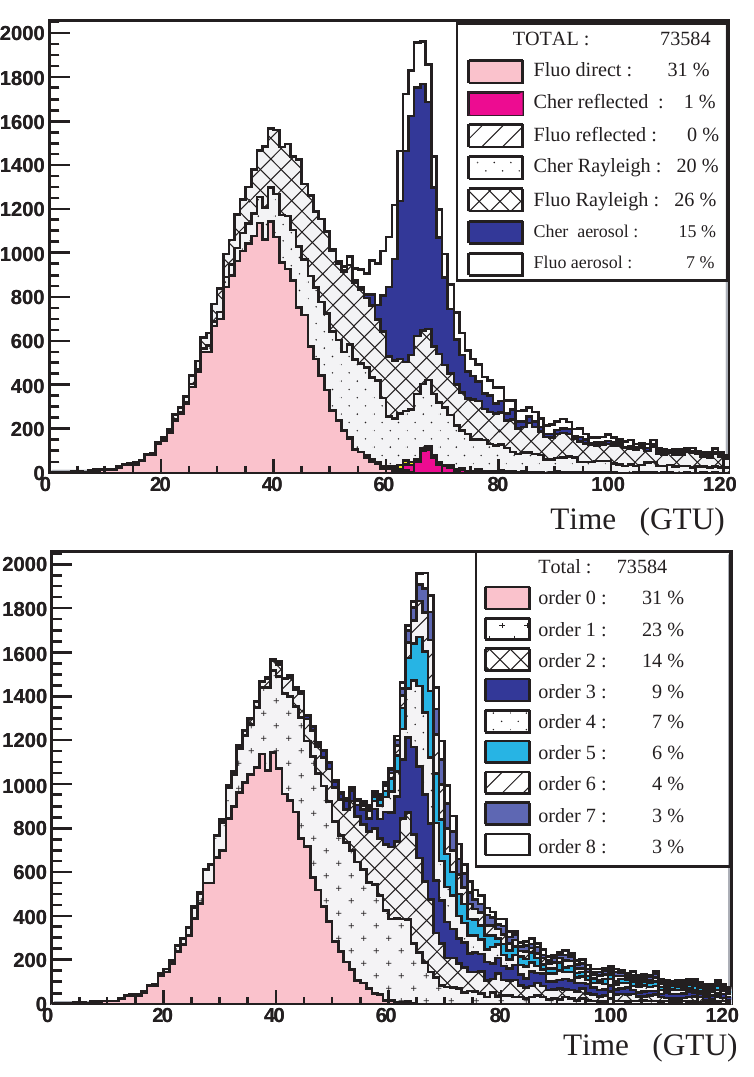}
\caption{Number of photons reaching the pupil as a function of time. Signal integrated in the whole FoV. Top histograms refer to the last photon interaction and bottom histograms refer to the number of scattering interactions before reaching the pupil. The simulation has been carried out with 8 orders of scattering in a US-Standard atmosphere including a rural aerosol model of visibility 5 km. The energy of the shower is 10$^{20}$~eV and the zenith angle is 60\degr.}
\label{fig:AerosolsSignalOrder8}
\end{center}
\end{figure}

\begin{figure}[ht]
\begin{center}
\includegraphics[width=0.5\textwidth]{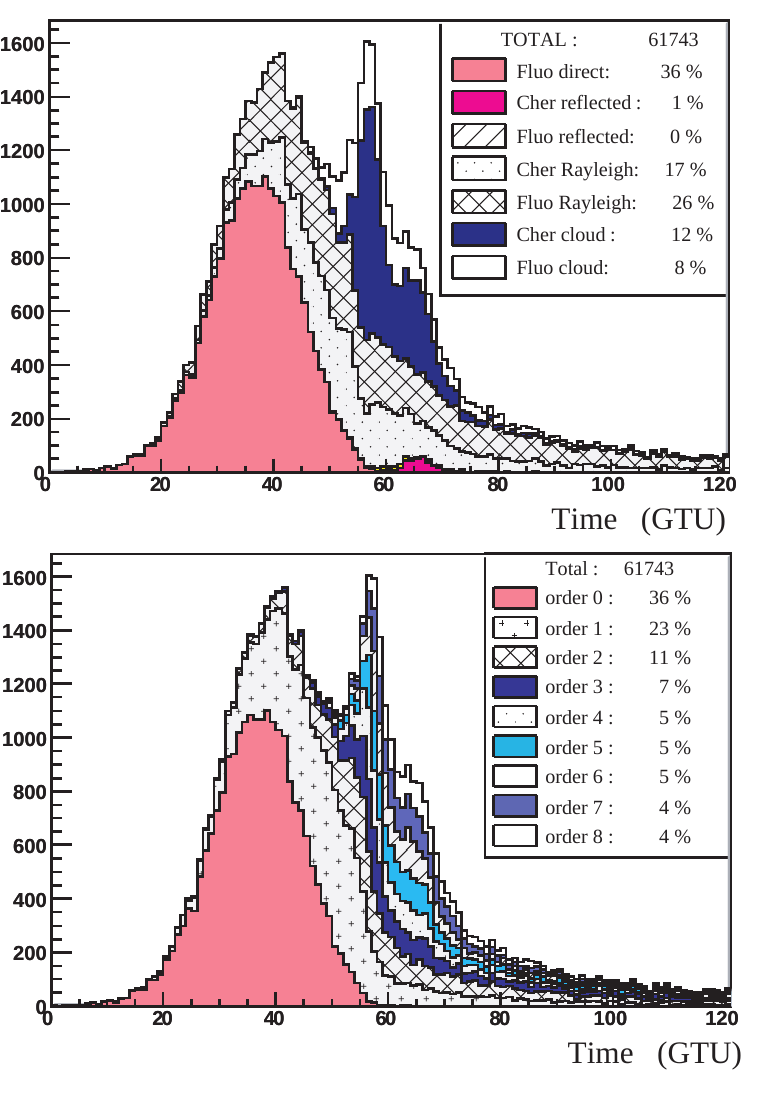}
\caption{Number of photons reaching the pupil as a function of time. Top histograms refer to the last photon interaction and bottom histograms refer to the number of scattering interactions before reaching the pupil. The simulation has been carried out with 8 orders of scattering in a US-Standard atmosphere including an homogeneous cumulus cloud model, characterized by a top altitude of  \mbox{3 km}, a vertical extension of \mbox{1 km} and an optical depth of 2. The  shower energy is 10$^{20}$~eV and the zenith angle is 60\degr.}
\label{fig:CloudSignalOrder8}
\end{center}
\end{figure}

Because clouds and aerosols are responsible for Mie scattering, an adequate simulation including multi-scattering must be carried out. As examples Fig.~\ref{fig:AerosolsSignalOrder8} and~\ref{fig:CloudSignalOrder8} show simulation results at the pupil entrance of the same shower as in section~\ref{sec:pupil} but with presence of rural aerosols or clouds. As one can see, contributions from scattering orders greater than 4 cannot be neglected as in the clear sky case. Therefore we carried out the simulations taking into account up to 8 scattering orders.

In presence of aerosols, fluorescence photons being emitted mostly above aerosols layers, direct fluorescence light is only weakly affected. On the contrary, ground reflected light is strongly attenuated since it has to travel twice through the dense layers of atmosphere before being detected.  Aerosols being mostly a scattering medium, a multi-scattering peak dominated by the contribution of Mie scattered Cherenkov photons appears in the total photon time distribution (Fig.~\ref{fig:AerosolsSignalOrder8}). Its maximum is between time corresponding to ground impact and that corresponding to aerosol top layer impact. However most of these photons are not contiguous with the shower track. To estimate the shape of the multi-scattering peak likely to be detected, an appropriate selection of the scattered signal component at the photo-electron level is applied, using a simple detector simulation (already mentioned in section~\ref{sec:pupil-diff}). Only the scattered signal which is contiguous in space and time with the shower track is selected, keeping in mind that selecting too many pixels reduces the signal to noise ratio.

\begin{figure}[ht]
\begin{center}
\includegraphics[width=0.5\textwidth]{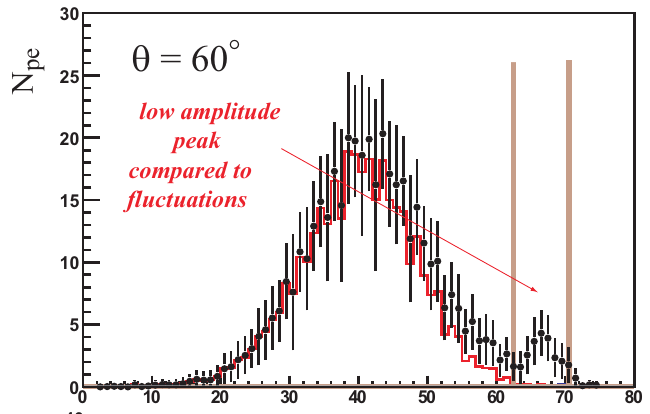}
\caption{Mean photoelectrons number as a function of time after selection of the shower track contiguous pixels. 
The energy of the shower is 10$^{20}$~eV and the zenith angle is 60\degr. The atmosphere model is US-Standard with rural aerosols of visibility 5 km. The vertical bars indicate respectively the impact on aerosols top layer and the impact on ground.}
\label{fig:AerosolsSignalDet}
\end{center}
\end{figure}

This selection has been applied to the considered event and the resulting signal is shown on Fig.~\ref{fig:AerosolsSignalDet}. As one can see the amplitude of the multi-scattering peak is weak compared to statistical fluctuations. Let us notice moreover that the peak maximum after pixel selection in space and time still does not correspond to the impact on the aerosol top layer. In a general way, its position in time should depend on the aerosols properties. Therefore the use of the multi-scattering peak (if detected) instead of ground reflected Cherenkov appears in the reconstruction procedure to be more difficult than expected.

\begin{figure}[ht]
\begin{center}
\includegraphics[width=0.5\textwidth]{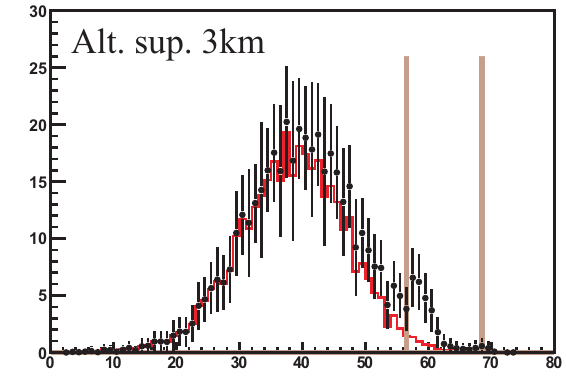}
\caption{Mean photoelectron number as a function of time after selection of shower track contiguous pixels. The shower energy is 10$^{20}$~eV and the zenith angle is 60 degrees. The atmosphere model is US-Standard with an homogeneous cumulus cloud model, characterized by a top altitude of 3 km, a vertical extension of 1 km and an optical depth of 2. The vertical bars indicate respectively the impact on cloud layer and the impact on ground.}
\label{fig:CloudSignalDet}
\end{center}
\end{figure}

Contrary to aerosols, clouds can be found at higher altitudes, and thus distort strongly direct fluorescence light depending on their altitude, optical depth and the shower zenith angle. However, in some cases of low altitude clouds, when major part of fluorescence light (including the maximum of development) is emitted above clouds, direct fluorescence light is only weakly distorted. Nevertheless reflected Cherenkov is strongly attenuated and scattered photons are added to signal. As in case of aerosols, a multi-scattering peak dominated by Mie scattered Cherenkov photons appears in total photon time distribution (Fig.~\ref{fig:CloudSignalOrder8}). Mie scattered  photons are also spread in FoV therefore the amplitude of the multi-scattering peak is greatly reduced and low compared to statistical fluctuations when pixel selection is applied (Fig.~\ref{fig:CloudSignalDet}).  We have studied several configurations as a function of shower zenith angle, cloud top altitude, cloud thickness and cloud optical depth. In some particular situations, for dense clouds located at low altitude and showers weakly inclined ($<40\degr$), the multi-scattering peak may be clearly visible. Two peaks may even appear in very specific situations, the first related to the impact on the cloud layer, the second related to the impact on ground. In a general way, the peak shape and its position in time depend strongly on clouds properties. Therefore, in the reconstruction procedure, the use of the multi-scattering peak (if detected) instead of ground reflected Cherenkov appears to be more difficult than expected.

An on board atmospheric sounding device associated with the telescope is thus required in order to interpret the signal distortions due to aerosols and cloud layers.

\subsubsection{Effect of clouds on acceptance}
Presence of clouds and aerosols in the FoV makes signal more complex to exploit and modifies also detector acceptance. Despite the fact that multi-scattering until order 8 is needed in cloud simulations and because the ``Reduced'' Monte-Carlo algorithm for radiative transfer is time-consuming, a compromise using the ``Bunch'' algorithm has been done for preliminary studies concerning acceptance with clouds. The main effect of clouds on the acceptance comes from the attenuation of the detected fluorescence light when a cloud layer stands between a shower and the detector. For this study, the scattering component has not been calculated, only the attenuation effect has been considered, using a simplified version of the ``Bunch'' algorithm.

\begin{figure}[ht]
\begin{center}
\includegraphics[width=0.5\textwidth]{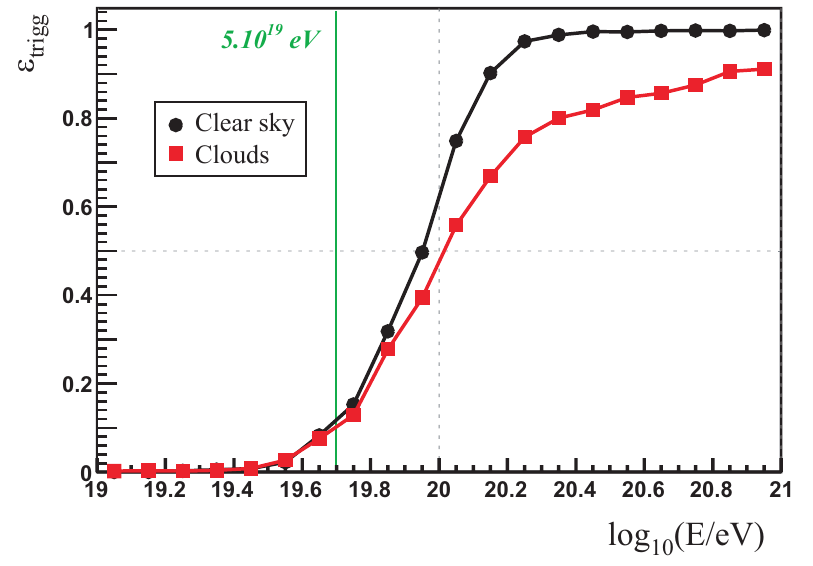}
\caption{Trigger efficiency curves in a modeled atmosphere with and without clouds.  }
\label{fig:trigger-clouds}
\end{center}
\end{figure}

\begin{figure}[t]
\centering
\includegraphics[width=0.48\textwidth]{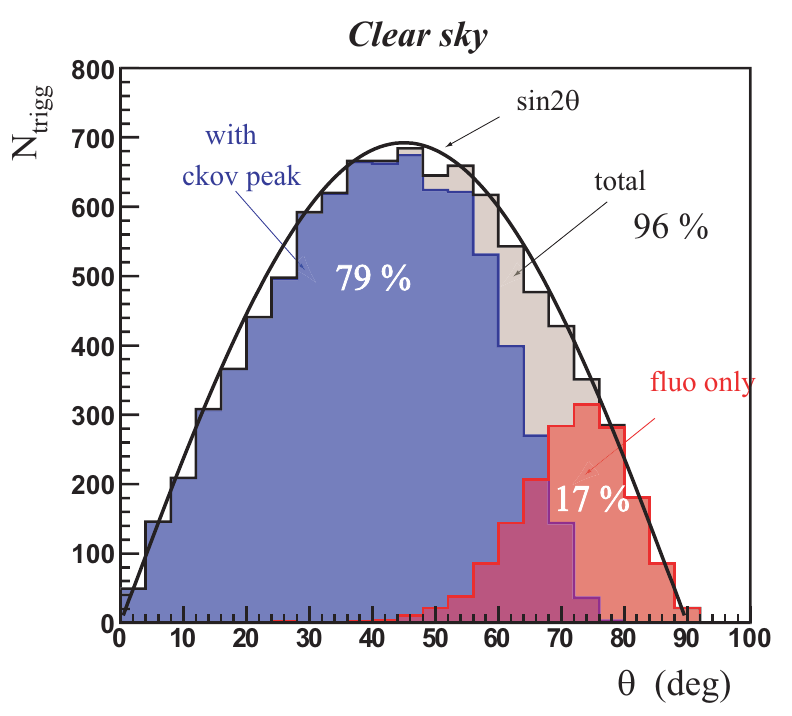} \\
\includegraphics[width=0.48\textwidth]{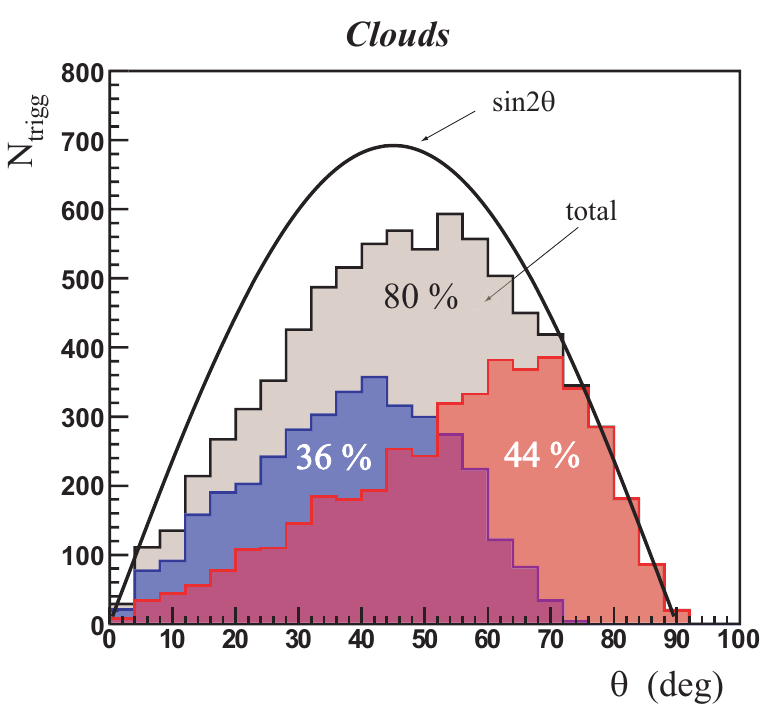}
\caption{Acceptance for triggering showers that have their maximum within the FoV
as a function of the zenith angle $\theta$ in clear sky conditions (top) and cloudy conditions (bottom). 
The plot also shows the contribution of events with (in blue) and without (in red) Cherenkov peak.
The shower energy spectrum is flat (in a logaritmic scale) between 10$^{20}$ eV and 10$^{21}$ eV.  
To guide the eyes, the curves giving an estimate of the geometric acceptance (showers having
their maximum in the FoV)  are also superimposed.
Percentage are related to the simulated geometric acceptance whose integrated value is 633 000 km$^2$ sr 
(the estimated one gives a 4\% lower value than the simulated one). }
\label{fig:trigg-acc-theta}
\end{figure}

Two samples of 2$\cdot$10$^{4}$ proton EAS with flat energy distribution (in a logarithmic scale between 10$^{19}$~eV and 10$^{21}$~eV), isotropic  angular distribution at top of atmosphere and a maximum of shower development in FoV have been used. Showers were generated with GIL parametrization in an atmosphere distribution obtained from MSISE model combined with the International Space Station Trajectory data. TOVS distribution and description of clouds (probability of presence, opacity and altitude) have been included in one of the two samples. The Earth surface is considered as a lambertian one, with an albedo of 5\%. The fluorescence yield is based on the Kakimoto model completed with Bunner rays. The detector is derived from the EUSO Phase A design and the trigger efficiency is computed using the pattern [4;4].

In Fig.~\ref{fig:trigger-clouds} the effect of clouds  on the trigger efficiency is illustrated: it decreases when presence of clouds is taken into account. At the highest energies, between 10$^{20}$eV and 10$^{21}$eV, the mean efficiency is only 80\% when clouds are taken into account instead of  96\% in clear sky conditions. 

In Fig. \ref{fig:trigg-acc-theta} we also show the trigger acceptance as a function of the shower zenith
angle in clear sky conditions and in the presence of an atmosphere with clouds. In clear sky conditions
the data nicely agree with the expectation that is at first order proportional to sin(2$\theta$). 
The contribution to the total number of events of near horizontal proton initiated showers is negligibly small; however, 
the possibility to reconstruct and identify such showers might possibly open a new window for the detection of 
high energy neutrinos. We have not considered in this paper the problems related to neutrino detection.

The main effect of clouds, 
as already said, is that the global trigger acceptance is decreased from 96\% to 80\%, but also that events having a detectable Cherenkov peak is reduced from 79\% down to 36\%. To obtain this last result we have defined a ``detectable Cherenkov peak'' as a peak with maximum amplitude greater than 4 standard deviations above background.

\begin{figure}[ht]
\begin{center}
\includegraphics[width=0.5\textwidth]{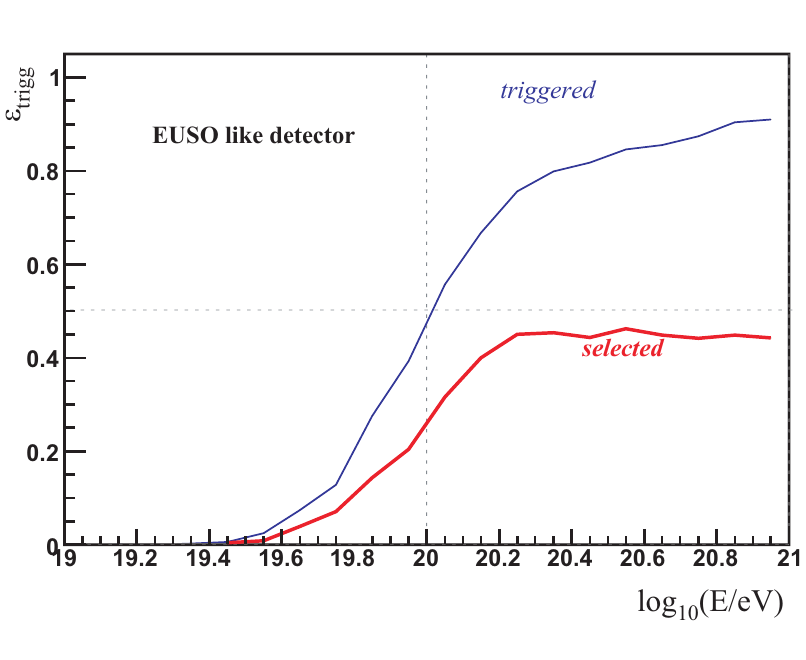}
\caption{Trigger efficiency in a MSISE modeled atmosphere with TOVS clouds. Superimposed in red:  triggered events which have part of their longitudinal profile (between ${s=0.9}$ and ${s=1.1}$) located above clouds.}
\label{fig:profile-clouds}
\end{center}
\end{figure}

In case of clouds presence, a triggered event may result in a strongly distorted signal. To estimate the rate of showers with a maximum of development standing above cloud layer, if any, a condition on shower age has been applied to simulated events:  only events for which the portion of the longitudinal profile between ${s=0.9}$ and ${s=1.1}$  develops above potential clouds are kept. As shown in Fig.~\ref{fig:profile-clouds}, this condition reduces the efficiency by a factor of 2. This result is a mean value assuming a realistic cloud coverage.

To estimate the event rates for 3 years of data taking of an EUSO-like experiment, 
13316 protons have been simulated between 4.10$^{19}$ and 10$^{21}$ eV. 
The energy spectrum is given by \cite{bib:hires} and normalized to a flux of 3. 10$^{-33}$ 
m$^{-2}$ s$^{-1}$ sr$^{-1}$ eV$^{-1}$ at the energy of 10$^{19}$ eV. 
The behaviour of this spectrum was recently confirmed by the Pierre Auger collaboration. The number of generated protons takes into account a duty cycle of 18\% and 
the acceptance of 633000 km$^2$ sr. 
Simulation conditions are the same as previously mentioned. Results are presented in Table \ref{tab:rates}.
In Table \ref{tab:rates2} we also report the same data as in Table \ref{tab:rates} but using a different
energy spectrum, namely the one that was suggested by Agasa data at the time when 
the EUSO telescope was studied.


\begin{table}
\begin{center}
\begin{tabular}{|c|c|c|c|c|} \hline 
Hires     & \multicolumn{2}{c|}{Triggering} & \multicolumn{2}{c|}{$0.9< s< 1.1$} \\
Spectrum   & \multicolumn{2}{c|}{Events} & \multicolumn{2}{c|}{above clouds} \\ \hline
Energy   & ~EUSO~ & ~EUSOx4~ & ~EUSO~ & ~EUSOx4~ \\ \hline
~E$>4.~10^{19}$  eV~ & 2135 & 5872 & 1101 & 3206 \\ 
~E$>5.~10^{19}$ eV~ & 1731 & 4040 & 896   & 2254 \\ 
~E$>1.~10^{20}$ eV~ & 406   & 418    & 228   & 310   \\ 
~E$>5.~10^{20}$ eV~ & 6        & 6         & 3       & 3  \\     \hline
\end{tabular}
\vspace{2mm}
\end{center}
\caption{Events collected in 3 years of data taking for various energy intervals by an EUSO like 
detector and by a detector with a pupil 4 times larger with the assumption of a cosmic ray spectrum as measured by HiRes 
collaboration \cite{bib:hires}. The first two columns report the triggering events. 
The second two report the events that have their maximum well above a cloud layer with average world cloud 
coverage. The conditions and the assumption made to obtain these numbers are explained in text. }
\label{tab:rates}
\end{table}

\begin{table}
\begin{center}
\begin{tabular}{|c|c|c|c|c|} \hline 
 E$^{-2.75}$  &  \multicolumn{2}{c|}{Triggering} & \multicolumn{2}{c|}{$0.9< s< 1.1$} \\
 Spectrum      &  \multicolumn{2}{c|}{Events} & \multicolumn{2}{c|}{above clouds} \\ \hline
Energy   & ~EUSO~ & ~EUSOx4~ & ~EUSO~ & ~EUSOx4~ \\ \hline
~E$>4.~10^{19}$  eV~ & 4324 & 8667 & 2283 & 4803     \\ 
~E$>5.~10^{19}$ eV~ & 3982 & 7093 & 2111   & 3989      \\ 
~E$>1.~10^{20}$ eV~ & 2235   & 2699    & 1252   & 1521        \\ 
~E$>5.~10^{20}$ eV~ & 117        & 119         & 67       & 67             \\     \hline
\end{tabular}
\vspace{2mm}
\end{center}
\caption{Same as in Table \ref{tab:rates} but numbers are obtained for a cosmic ray spectrum modeled 
by a single power law with spectral index equal to  2.75}
\label{tab:rates2}
\end{table}

\section{Conclusions}\label{sec:conclu}
In order to study the scientific potential of an observatory in orbit around the Earth, a full simulation of the 
detection of EAS resulting of UHECP interaction in atmosphere has been performed. Although this work has 
been done within the framework of the ESA EUSO project, the results are quite general and apply to a large
class of possible telescopes. The code also implements a complete reconstruction chain of direction, energy and
shower \Xmax . 

The simulation of the radiative transfer in the atmosphere up to the telescope emphasizes the importance 
of multiple scattering in atmosphere and Earth. The contamination of the direct fluorescence signal by 
the scattered light is enhanced, particularly in presence of clouds or aerosols: in these cases the longitudinal 
profile is distorted. The behaviour of the reflected Cherenkov light, according to the ground and atmospheric 
parameters, has been studied. The simulation results show that the effects of the atmosphere on the EAS 
light signal have to be considered during the reconstruction procedure.

The studies reported in this paper show that the detection energy threshold (defined
as the energy above which detection efficiency exceeds 50\%) cannot be significantly less that 10$^{20}$ eV,
unless the aperture is much larger than the one originally foreseen for EUSO (2.5 m in diameter). However, 
it is shown that a large EUSO-like telescope of the order of 8 to 12 m in diameter might perform very well 
around the GZK region. In fact, while a triggering threshold of 2. 10$^{19}$ eV can be obtained only with a 
pupil diameter of at least 8 m with current photo-detection technology.

The measurement of the arrival direction of the UHECP depends mostly on the focal surface granularity 
and on the electronics time resolution, but
with a small detector like EUSO the angular resolution is also limited by lack of signal. Below 10$^{20}$ eV the 
angular resolution is always worse than 1\degr even for very inclined showers and is about 4\degr on average at
10$^{20}$ eV between 30\degr and 70\degr. 

Significant improvements are possible only with a detector larger than EUSO  (larger signal with smaller
fluctuations) and possibly with a high granularity focal surface.


The ESAF code, the tool that we have developed to make this study, is available upon request and is
suitable for future developments in this field.

\section*{Acknowledgements}
We acknowledge the support of the italian Istituto Nazionale di Fisica Nucleare, the italian Agenzia Spaziale Italiana (ASI), the universities of Firenze and Genova, Italy, 
and the European Space Agency (ESA). D.N. thanks Russian Fund for Basic Research 07-02-00215-a and 
Joint Institute for Nuclear Research fellowship for young candidate of science.

We are grateful to M. Dattoli, S. Biktemerova and M. Gonchar for their help in developing 
some parts of the ESAF code.

We dedicate this paper to the memory of our friends and teachers John Linsley and Livio Scarsi.

\end{document}